\documentclass[aps,prd,showpacs,eqsecnum,nofootinbib,floatfix,twocolumn,a4paper]{revtex4-1}

\usepackage{amsmath,amsfonts,amssymb}
\usepackage{graphicx}

\allowdisplaybreaks

\newcommand{\be}{\begin{equation}}
\newcommand{\ee}{\end{equation}}

\newcommand{\da}{\delta_a}
\newcommand{\xa}{\mathbf{x}_a}
\newcommand{\pa}{\mathbf{p}_a}
\newcommand{\x}{\mathbf{x}}
\newcommand{\p}{\mathbf{p}}
\newcommand{\rv}{\mathbf{r}}
\newcommand{\n}{\mathbf{n}}

\newcommand{\Bvi}[2]{B_{(6)#1#2}}
\newcommand{\BTTvi}[2]{B^\mathrm{TT}_{(6)#1#2}}

\newcommand{\gradf}[1]{\phi_{,#1}}
\newcommand{\ggradf}[2]{\phi_{,#1#2}}

\newcommand{\fii}{\phi_{(2)}}
\newcommand{\gradfii}[1]{\phi_{(2),#1}}
\newcommand{\ggradfii}[2]{\phi_{(2),#1#2}}

\newcommand{\fiv}{\phi_{(4)}}
\newcommand{\gradfiv}[1]{\phi_{(4),#1}}
\newcommand{\ggradfiv}[2]{\phi_{(4),#1#2}}

\newcommand{\fvib}{\bar{\phi}_{(6)}}

\newcommand{\ggradfvib}[2]{\bar{\phi}_{(6),#1#2}}

\newcommand{\fvinoTT}{\phi_{(6)1}}
\newcommand{\fvibTT}{\bar{\phi}_{(6)2}}
\newcommand{\fviTT}{\phi_{(6)2}}

\newcommand{\fviiib}{\bar{\phi}_{(8)}}

\newcommand{\fxb}{\bar{\phi}_{(10)}}

\newcommand{\hn}{H_\mathrm{N}}
\newcommand{\hi}{H_\mathrm{1PN}}
\newcommand{\hii}{H_\mathrm{2PN}}
\newcommand{\hiii}{H_\mathrm{3PN}}

\newcommand{\CTTvi}[2]{C^\mathrm{TT}_{(6)#1#2}}
\newcommand{\CTTdotvi}[2]{\dot{C}^\mathrm{TT}_{(6)#1#2}}
\newcommand{\gradCTTvi}[3]{C^\mathrm{TT}_{(6)#1#2,#3}}

\newcommand{\dhTT}[2]{\delta h^\mathrm{TT}_{#1#2}}
\newcommand{\graddhTT}[3]{\delta h^\mathrm{TT}_{#1#2,#3}}
\newcommand{\dhTTdot}[2]{\delta \dot{h}^\mathrm{TT}_{#1#2}}

\newcommand{\hTT}[2]{{h^\mathrm{TT}_{#1#2}}}
\newcommand{\gradhTT}[3]{h^\mathrm{TT}_{#1#2,#3}}
\newcommand{\ggradhTT}[4]{h^\mathrm{TT}_{#1#2,#3#4}}

\newcommand{\hTTdot}[2]{\dot{h}^\mathrm{TT}_{#1#2}}
\newcommand{\hTTddot}[2]{\ddot{h}^\mathrm{TT}_{#1#2}}

\newcommand{\hTTivpo}[2]{h^\mathrm{TT}_{(4)0#1#2}}
\newcommand{\hTTivpii}[2]{h^\mathrm{TT}_{(4)2#1#2}}
\newcommand{\hTTivpoddot}[2]{\ddot{h}^\mathrm{TT}_{(4)0#1#2}}
\newcommand{\hTTivpiiddot}[2]{\ddot{h}^\mathrm{TT}_{(4)2#1#2}}

\newcommand{\hTTiv}[2]{h^\mathrm{TT}_{(4)#1#2}}
\newcommand{\gradhTTiv}[3]{h^\mathrm{TT}_{(4)#1#2,#3}}
\newcommand{\hTTivdot}[2]{\dot{h}^\mathrm{TT}_{(4)#1#2}}
\newcommand{\hTTivddot}[2]{\ddot{h}^\mathrm{TT}_{(4)#1#2}}
\newcommand{\ilhTTivddot}[2]{\Deltad^{-1}\ddot{h}^\mathrm{TT}_{(4)#1#2}}
\newcommand{\gradilhTTivddot}[3]{\Deltad^{-1}\ddot{h}^\mathrm{TT}_{(4)#1#2,#3}}
\newcommand{\ilhTTivdddot}[2]{\Deltad^{-1}\dddot{h}^\mathrm{TT}_{(4)#1#2}}

\newcommand{\hTTv}[2]{h^\mathrm{TT}_{(5)#1#2}}

\newcommand{\hTTvii}[2]{h^\mathrm{TT}_{(7)#1#2}}

\newcommand{\hTTvi}[2]{h^\mathrm{TT}_{(6)#1#2}}
\newcommand{\hTTvidot}[2]{\dot{h}^\mathrm{TT}_{(6)#1#2}}

\newcommand{\md}{\mathrm{d}}

\newcommand{\ore}[1]{\mathcal{O}(\epsilon^{#1})}

\newcommand{\ora}[1]{\mathcal{O}\big(r^{#1}\big)}

\newcommand{\na}{\nabla}

\newcommand{\nipi}{(\mathbf{n}_1\cdot\mathbf{p}_1)}
\newcommand{\niipi}{(\mathbf{n}_2\cdot\mathbf{p}_1)}
\newcommand{\nipii}{(\mathbf{n}_1\cdot\mathbf{p}_2)}
\newcommand{\niipii}{(\mathbf{n}_2\cdot\mathbf{p}_2)}

\newcommand{\npi}{(\mathbf{n}_{12}\cdot\mathbf{p}_1)}
\newcommand{\npii}{(\mathbf{n}_{12}\cdot\mathbf{p}_2)}

\newcommand{\piipii}{{\bf p}_2^2}
\newcommand{\piipiip}{({\bf p}_2^2)}
\newcommand{\pipi}{{\bf p}_1^2}
\newcommand{\pipii}{({\mathbf{p}}_1\cdot{\mathbf{p}}_2)}
\newcommand{\pipip}{({\bf p}_1^2)}

\newcommand{\nainaii}{\left(\na_1\cdot\na_2\right)}
\newcommand{\pinai}{\left(\mathbf{p}_1\cdot\nabla_1\right)}
\newcommand{\pinaii}{\left(\mathbf{p}_1\cdot\nabla_2\right)}
\newcommand{\piinai}{\left(\mathbf{p}_2\cdot\nabla_1\right)}
\newcommand{\piinaii}{\left(\mathbf{p}_2\cdot\nabla_2\right)}

\newcommand{\papa}{\mathbf{p}_a^2}
\newcommand{\pbpb}{\mathbf{p}_b^2}
\newcommand{\papap}{(\mathbf{p}_a^2)}
\newcommand{\napa}{(\mathbf{n}_a\cdot\mathbf{p}_a)}
\newcommand{\nabpa}{(\mathbf{n}_{ab}\cdot\mathbf{p}_a)}

\newcommand{\pp}{\mathbf{p}^2}
\newcommand{\ppn}{(\mathbf{p}^2)}
\newcommand{\np}{(\mathbf{n}\cdot\mathbf{p})}

\def\papb{\left({\bf p}_a\cdot{\bf p}_b\right)}
\def\nanb{\left(\na_a\cdot\na_b\right)}
\def\pana{\left({\bf p}_a\cdot\na_a\right)}
\def\pbnb{\left({\bf p}_b\cdot\na_b\right)}
\def\panb{\left({\bf p}_a\cdot\na_b\right)}
\def\pbna{\left({\bf p}_b\cdot\na_a\right)}

\def\pan{\left({\bf p}_a\cdot\na\right)}
\def\pbn{\left({\bf p}_b\cdot\na\right)}

\def\nna{\left(\na\cdot\na_a\right)}
\def\nnb{\left(\na\cdot\na_b\right)}

\def\uapa{\left({\bf n}_a\cdot{\bf p}_a\right)}

\newcommand{\piTT}[2]{\pi^{#1#2}_\mathrm{TT}}
\newcommand{\piTTdot}[2]{{\dot{\pi}}^{#1#2}_\mathrm{TT}}

\newcommand{\dpiTT}[2]{\delta\pi^{#1#2}_\mathrm{TT}}

\newcommand{\pit}[2]{\tilde{\pi}^{#1#2}}
\newcommand{\gradpit}[3]{{\tilde{\pi}^{#1#2}}_{,#3}}

\newcommand{\pitiii}[2]{\tilde{\pi}^{#1#2}_{(3)}}
\newcommand{\gradpitiii}[3]{{\tilde{\pi}^{#1#2}}_{(3),#3}}

\newcommand{\pitv}[2]{\tilde{\pi}^{#1#2}_{(5)}}
\newcommand{\gradpitv}[3]{{\tilde{\pi}^{#1#2}}_{(5),#3}}

\newcommand{\pitviib}[2]{\tilde{\bar{\pi}}^{#1#2}_{(7)}}
\newcommand{\gradpitviib}[3]{{\tilde{\bar{\pi}}}^{#1#2}_{(7),#3}}

\newcommand{\pitixb}[2]{\tilde{\bar{\pi}}^{#1#2}_{(9)}}
\newcommand{\gradpitixb}[3]{{\tilde{\bar{\pi}}}^{#1#2}_{(9),#3}}

\newcommand{\STT}[2]{S^\mathrm{TT}_{#1#2}}
\newcommand{\dSTT}[2]{\delta S^\mathrm{TT}_{#1#2}}

\newcommand{\Siv}[2]{S_{(4)#1#2}}
\newcommand{\gradSiv}[3]{S_{(4)#1#2,#3}}
\newcommand{\gradiiiSiv}[5]{S_{(4)#1#2,#3#4#5}}
\newcommand{\STTiv}[2]{S^\mathrm{TT}_{(4)#1#2}}

\newcommand{\Svi}[2]{S_{(6)#1#2}}
\newcommand{\STTvi}[2]{S^\mathrm{TT}_{(6)#1#2}}

\newcommand{\SivI}{S_{(4)1}}
\newcommand{\gradSivI}[1]{S_{(4)1,#1}}
\newcommand{\ggradSivI}[2]{S_{(4)1,#1#2}}

\newcommand{\SivII}{S_{(4)2}}
\newcommand{\gradSivII}[1]{S_{(4)2,#1}}
\newcommand{\ggradSivII}[2]{S_{(4)2,#1#2}}

\newcommand{\Vv}[1]{V_{(5)}^#1}
\newcommand{\gradVv}[2]{V_{(5),#2}^#1}

\newcommand{\Viii}[1]{V_{(3)}^#1}
\newcommand{\gradViii}[2]{V_{(3),#2}^#1}

\newcommand{\ve}{\varepsilon}

\newcommand{\eldt}[2]{\big[\big[#1\big]\big]_#2}

\newcommand{\infctiii}{\tilde{c}_\infty}
\newcommand{\infciii}{c_\infty}

\newcommand{\mc}{\mathfrak{c}}
\newcommand{\ms}{\mathfrak{s}}

\newcommand{\bxi}{\boldsymbol{\xi}}

\newcommand{\omk}{\omega_\textrm{k}}
\newcommand{\oms}{\omega_\textrm{s}}

\newcommand{\Deltad}{\Delta_d}

\begin{document}

\title{Derivation of local-in-time fourth post-Newtonian ADM Hamiltonian\\ for spinless compact binaries}

\author{Piotr Jaranowski}
\email{p.jaranowski@uwb.edu.pl}
\affiliation{Faculty of Physics,
University of Bia{\l}ystok,
Cio{\l}kowskiego 1L, 15--245 Bia{\l}ystok, Poland}

\author{Gerhard Sch\"afer}
\email{gos@tpi.uni-jena.de}
\affiliation{Theoretisch-Physikalisches Institut,
Friedrich-Schiller-Universit\"at Jena,
Max-Wien-Pl.\ 1, 07743 Jena, Germany}

\date{\today}

\begin{abstract}

The paper gives full details of the computation
within the canonical formalism of Arnowitt, Deser, and Misner of the local-in-time part of the fourth post-Newtonian,
i.e.\ of power eight in one over speed of light, conservative Hamiltonian of spinless compact binary systems.
The Hamiltonian depends only on the bodies' positions and momenta.
Dirac delta distributions are taken as source functions. Their full control is furnished by dimensional continuation,
by means of which the occurring ultraviolet (UV) divergences are uniquely regularized.
The applied near-zone expansion of the time-symmetric Green function leads to infrared (IR) divergences.
Their analytic regularization results in one single ambiguity parameter.
Unique fixation of it was successfully performed in T.~Damour, P.~Jaranowski, and G.~Sch\"afer, Phys.\ Rev.\ D \textbf{89}, 064058 (2014) through far-zone matching.
Technically as well as conceptually (backscatter binding energy), the level of the Lamb shift in quantum electrodynamics is reached.
In a first run a computation of all terms is performed in three-dimensional space using analytic Riesz-Hadamard regularization techniques.
Then divergences are treated locally (i.e., around particles' positions for UV and in the vicinity of spatial infinity for IR divergences)
by means of combined dimensional and analytic regularization.
Various evolved analytic expressions are presented for the first time. 
The breakdown of the Leibniz rule for distributional derivatives is addressed
as well as the in general nondistributive law when regularizing value of products of functions evaluated at their singular point.

\end{abstract}

\maketitle

\section{Introduction}

In the early days of Einstein's theory of gravity, the theory of general relativity as Einstein coined it, 
the weak-field and slow-motion expansion of the field equations, nowadays known as post-Newtonian (PN) expansion, played a crucial role already.
The first post-Newtonian (1PN) approximation revealed a convincing understanding of the perihelion advance of Mercury \cite{AE15a}.
Later on, in 1919, the light bending at the limb of the Sun was measured in agreement with the 1PN prediction of Einstein's theory \cite{AE15b,DED20}.
A rich 1PN scenario came up with the discovery of the Hulse-Taylor binary pulsar PSR B1913+16 in 1974,
where after four years of observation even the dissipative two-and-a-half post-Newtonian (2.5PN) effect could be measured in agreement with general relativity
in the form of energy loss through gravitational radiation damping \cite{TFC79}.
Also the precession of the proper angular momentum (spin) of the pulsar PSR B1913+16 from spin-orbit coupling,
a 1PN or 1.5PN effect according to the counting of the spin as of 0PN or 0.5PN order, respectively, could be seen \cite{WRT89}.
The discovery of the double pulsar system J0737$-$3039 in 2003 allowed measurements of even more 1PN effects
than with the Hulse-Taylor binary pulsar \cite{KW09}.
Also in the solar system and in the gravitational field of the Earth the 1PN approximation is crucial \cite{CW14,KEK2011}.
In the observation of the Hulse-Taylor pulsar, even 2PN measurement accuracy has been achieved in the periastron advance \cite{WNT10},
deduced for the first time in \cite{DS87} with full details given in \cite{DS88}.  
In the long-term run, the observation of the double pulsar system is likely to allow
detailed measurements of conservative 2PN and dissipative 3.5PN effects \cite{KW09,NW2014,BS1989,JS1997}.
On the other side, in the coming years gravitational-wave astronomy will come to operation 
through advanced (or second-generation) LIGO and Virgo observatories \cite{LSC2015,VC2015}
and the cryogenic KAGRA detector (formerly known as LCGT) \cite{Akutsu2015}.
Even more sensitive third-generation detectors, like ET \cite{Punturo2010}, are already under consideration.
Then still higher-order PN effects will become important.
Great efforts have already been undertaken in the precise data analysis of possible wave forms
(see, e.g., \cite{JK2009,SS2009,JK2012,BS2015}),
and as well as in the computation of the orbital and spin dynamics through higher PN orders including the gravitational wave emission
(see, e.g., \cite{FI07,Damour2012,BDGNR11,BBLT2012,JS2011,HSS13,LB14,LeviSteinhoff2015}).

The complete conservative third post-Newtonian (3PN) binary dynamics has been fully achieved for the first time in 2001 \cite{DJS2001},
based on earlier work \cite{JS98,JS99,JS00a,DJS00}, rooted in the canonical formalism of Arnowitt, Deser, and Misner (ADM)
with application of a coordinate system of maximal isotropy in three-dimensional space \cite{ADM62}.
Around 2004 and later in 2011 the final results of \cite{JS98,JS99,JS00a,DJS00,DJS2001} became fully confirmed \cite{IF03,YI04,BDEF04,FS11},
in using completely different techniques but yet all with the employment of harmonic coordinate conditions.
Let us stress that the rather tedious (but much easier than reported in the present paper)
calculations performed in \cite{JS98,JS99,JS00a,DJS00,DJS2001}
were not just a matter of straightforward computational efforts.
The standard regularization Hadamard ``partie-finie'' technique
together with the Hadamard ``partie-finie'' based Riesz analytic continuation
supplemented by the Schwartz distribution theory
in (3+1)-dimensional space-time resulted in ambiguities \cite{JS98,JS99},
which yet could be parametrized by two variables only, $\oms$ and $\omk$ \cite{JS00a}
(for a detailed presentation of the mentioned techniques, see \cite{J1997} and Appendix \ref{Regularization} of the present paper).
Matching to the Brill-Lindquist initial value solution for a two-black-hole system \cite{BL63},
a solution which was shown derivable from ``fictitious'' point-mass sources of Dirac $\delta$-function type (one for each black hole) \cite{JS00b},
brought out (by the very definition of $\oms$) $\oms$ = 0 \cite{JS99,JS00a},
and implementation of Lorentz invariance in the form of fulfillment of the Poincar\'e algebra resulted in $\omk=41/24$
(see \cite{DJS00} and Sec.\ IX of the present paper).
Another procedure to obtain $\oms = 0$ could have been to insist on the ``tweedling of products'' structure
in the course of computation of regularized value of the product of singular functions
(see \cite{JS99,IP60} and Appendix \ref{AHPF} of the current paper)
or, connected herewith, by taking just the finite terms (without limiting procedure)
in the two-body restriction of the many-body static potential derived in \cite{KT72}.
Only by making use of the technique of dimensional regularization a uniform treatment was achieved in Ref.\ \cite{DJS2001},
with confirmation of the numerical values for $\oms$ and $\omk$ found earlier
(a summary of the applied dimensional-regularization techniques is presented in \cite{DJS2008} and in Appendix \ref{Regularization} below).
For extensive application of the dimensional regularization in the generalized ADM formalism 
for spinning compact binaries the reader may consult \cite{HSS13}.

It might be worth comparing the derivation of the 3PN ADM Hamiltonian in \cite{JS98,JS99,JS00a,DJS00,DJS2001}
with the history of the harmonic-coordinate-based calculations,
which led to the results achieved in \cite{BDEF04}, where finally also dimensional regularization played the crucial role.
In the papers \cite{BF00a,BF01} a manifest Lorentz-invariant ``extended'' Hadamard regularization procedure was developed,
which allowed the calculation of the 3PN binary dynamics with only one ambiguity parameter $\lambda$.  
The comparison, for circular motion, of energy in terms of orbital angular frequency
(which is a coordinate-invariant or gauge-invariant relation)
revealed the relation $\lambda=-3\oms/11-1987/3080$ \cite{BF00b}.
Evidently, the application of the Poincar\'e algebra in \cite{DJS00} is the equivalent of the manifest Lorentz invariance in \cite{BF00a,BF01}. 
However, as shown in \cite{BDEF04}, the extended Hadamard regularization procedure, 
being incompatible with Schwartz distribution theory
[see, e.g., Eqs.\ (7.17) and (8.34) in \cite{BF00a},
which should be compared with Eqs.\ \eqref{e41a2} and \eqref{e41c2}, respectively, of our paper],
could not be made compatible with dimensional regularization [see, e.g., Sec.\ III~D in \cite{BDEF04}].
Therefore all terms of the extended Hadamard procedure
different from the standard Hadamard (``pure Hadamard-Schwartz'' in \cite{BDEF04}) ones,
had to be traced back to their origins and eliminated [see, e.g., Eq.\ (3.55) in \cite{BDEF04}]
before dimensional generalization and regularization could be employed.
There exists another derivation of the 3PN binary dynamics in harmonic coordinates made in \cite{FS11},
using an effective field theory approach with space-Fourier transformed fields,
which was from the very beginning put on a ($d$+1)-dimensional space-time footing with dimensional regularization.

It is interesting to note that the still other derivation of the 3PN binary dynamics in \cite{IF03,YI04},
based on the Einstein-Infeld-Hoffmann surface integral technique \cite{EIH38}, is a purely (3+1)-dimensional one.
The used surface-integral technique allowed a manifest Lorentz-invariant calculation with surface integrals defined in the smooth vacuum regime.
Divergent integrals entered only on technical reasons in simplifying computations.
(In the paper by Einstein, Infeld, and Hoffmann \cite{EIH38} the field singularities were already seen in correspondence with Dirac $\delta$-functions,
though their treatment as full field sources had had to await for later developments in the form of Infeld's ``good $\delta$-functions,''
see, e.g., the book by Infeld and Pleba{\'n}ski \cite{IP60}.) 
Remarkably, the Brill-Lindquist initial value solution in (3+1)-dimensional space-time mentioned above
has originally been obtained from the pure vacuum Einstein field equations
without facing any physical or geometrical divergences \cite{BL63}.

The fourth post-Newtonian (4PN) conservative dynamics of two-point-mass systems has been completed only quite recently \cite{DJS14},
based on previous calculations \cite{JS12,JS13}.
The results of Refs.\ \cite{JS12,JS13,DJS14} were in part confirmed by \cite{LSB08,TD2010,BDTW10,FS13a} (see also \cite{FS13b,GRR14}).
The present paper, announced in \cite{DJS14}, delivers the details of the involved computations
of the 4PN conservative two-point-mass ADM Hamiltonian,
where ultraviolet (UV) and infrared (IR) divergences have to be tackled simultaneously
and where a mixed dimensional-analytic regularization treatment is needed to be applied (like in \cite{Blanchet&others2005}).
Pure dimensional regularization allows one to control UV divergences only         
and Refs.\ \cite{JS12,JS13} have uniquely regularized all UV divergences.
Regularization of IR divergences turned out to be ambiguous (what is discussed in the present paper)
and this ambiguity was resolved in Ref.\ \cite{DJS14}
by taking into account the breakdown of a usual PN scheme
(based on a formal near-zone expansion) due to infinite-range tail-transported temporal correlations
found in Ref.\ \cite{BD88}. Reference \cite{DJS14} showed that the total 4PN conservative Hamiltonian
is the sum of instantaneous (local-in-time) near-zone Hamiltonian and time-symmetric but nonlocal-in-time tail Hamiltonian.
The present paper is devoted to details of computation of the 4PN near-zone local-in-time Hamiltonian.
A rather nontrivial application of Ref.\ \cite{DJS14}
to 4PN-accurate generalization of effective-one-body approach to compact binary dynamics
has been worked out in Ref.\ \cite{DJS15} already.

Since the seminal work by 't Hooft and Veltman \cite{tHV1972},
dimensional regularization (for history see \cite{BP2012})
has become very popular in quantum field theory (see, e.g., \cite{JCC1984}).
Even the famous Lamb shift with simultaneously occurring UV and IR divergences
found its elegant computation by applying dimensional regularization \cite{LSB1992}.
However, whereas UV divergences are nicely controlled by dimensional regularization,
IR divergences can pose problems (see, e.g., Refs.\ \cite{CT1992,HG1993}).
This also happens in our approach with the final solution indicated above.
It is worth pointing out that also the Lamb shift calculation of Ref.\ \cite{LSB1992}
shows up an undefined constant in the IR sector, which gets fixed by some dimensional matching.

A short overview of the various sections will give the reader some help in orientation. 
Section II presents the ADM formalism for two-point-mass systems developed in $(d+1)$-dimensional space-time.
It also gives the transition to the Routhian functional
which is crucial for the obtention of Hamiltonian which depends on matter variables only.
Section III expands the nonpropagating part of the ADM structure
for two-point-mass systems through 4PN order.
It ends with the formulas for the 4PN-accurate reduced Hamiltonian.
Section IV is concerned with the derivation of field equations
for the propagating degrees of freedom valid to the next-to-leading order
(which is enough for our purposes).
It also includes the formal near-zone (and thus PN) expansion of the time-symmetric solution of field equations.
Section V develops the 4PN-accurate Routhian functional,
without expanding propagating degrees of freedom into the PN series.
Section VI presents the 4PN-accurate conservative Hamiltonian dependent on matter variables only,
with still non-PN-expanded propagating degrees of freedom.
Section VII gives the 4PN-exact near-zone conservative matter Hamiltonian
with fully PN-expanded propagating degrees of freedom.
Section VIII delivers some details of the computation and regularization of integrals,
taking into account the various appendices,
and presents the fully explicit result for the total 4PN-accurate matter conservative Hamiltonian.
Section IX is devoted to checks with the aid of the Poincar\'e algebra.
All technical calculational details are shifted to three appendices.
Appendix A is devoted to all regularization procedures used.
Appendix B gives a variety of needed inverse Laplacians.
Finally, Appendix C presents field functions in fully explicit forms.

We employ the following notation:
$\x=(x^i)$ ($i=1,\ldots,d$) denotes a point in the $d$-dimensional
Euclidean space $\mathbb{R}^d$ endowed with a standard Euclidean metric
and a scalar product (denoted by a dot).
Spatial latin indices run through $1,\ldots,d$,
and space-time greek indices vary from 0 to $d$.
Letters $a$ and $b$ ($a,b=1,2$) are particle labels,
so $\xa=(x_a^i)\in\mathbb{R}^d$ denotes the position of the $a$th point mass.
We also define ${\bf r}_a:=\x-\xa$, $r_a:=|{\bf r}_a|$, ${\bf n}_a:={\bf r}_a/r_a$;
and for $a\ne b$, ${\bf r}_{ab}:=\xa-\x_b$, $r_{ab}:=|{\bf r}_{ab}|$, ${\bf n}_{ab}:={\bf r}_{ab}/r_{ab}$;
$|\cdot|$ stands here for the Euclidean length of a vector. 
The linear momentum vector of the $a$th particle is denoted by $\pa=(p_{ai})$,
and $m_a$ denotes its mass parameter.
An overdot, as in $\dot{\x}_a$, means differentiation with respect to time coordinate $t$.
The partial differentiation with respect to $x^i$ is denoted by $\partial_i$ or by a comma, i.e., $\partial_i\phi\equiv\phi_{,i}$,
and the partial differentiation with respect to $x_a^i$ is denoted by $\partial_{ai}$.
We abbreviate Dirac delta distribution $\delta(\x-\xa)$ by $\da$ (both in $d$ and in 3 dimensions);
it fulfills the condition $\int\md^dx\,\da=1$.
\emph{Flat} $d$-dimensional Laplacian is denoted by $\Deltad$,
whereas $\Delta$ (without any subscript) is reserved for Laplacian in $d=3$ dimensions.
Extensive use has been made of the computer algebra system \textsc{Mathematica}.

\section{The ADM canonical formalism in~$d$~space dimensions}

We consider here a system of two point masses, i.e.\ monopolar, pointlike bodies,
which interact gravitationally according to general relativity theory.
We model point masses by means of Dirac delta distributions $\da$.
Let $D:={d+1}$ denote the (analytically continued) spacetime dimension.
The ADM canonical approach \cite{ADM62} uses a $d+1$ split of the coupled gravity-plus-matter
dynamics and works with the pairs of the canonical variables:
positions $\xa$ and momenta $\pa$ of point masses,
and for the gravitational field the space metric $\gamma_{ij}=g_{ij}$
induced by the full space-time metric $g_{\mu\nu}$ on the hypersurface $t=\textrm{const}$,
and its conjugate momentum $\pi^{ij}=\pi^{ji}$.

The full Einstein field equations in $D$ dimensions
in an asymptotically flat space-time and in an asymptotically Minkowskian coordinate system,
written in the canonical variables introduced above,  
are derivable from the Hamiltonian \cite{regge74}
(in units where $16\pi\,G_D=c=1$, with $G_D$ denoting
the generalized Newtonian gravitational constant and $c$ the speed of light)
\begin{align}
\label{totham}
H &= \int \md^dx\,(N {\cal{H}} - N^i{\cal{H}}_i)
\nonumber\\&\quad
+ \oint_{i^0}\md^{d-1}S_i\,\partial_j(\gamma_{ij}-\delta_{ij}\gamma_{kk})\,,
\end{align}
where $N$ and $N^i$ are called \emph{lapse} and \emph{shift} functions
and the super-Hamiltonian ${\cal{H}}$ and super-momentum ${\cal{H}}_i$ are defined as follows:
\begin{subequations}
\begin{align}
\mathcal{H}(\xa,\pa,\gamma_{ij},\pi^{ij}) &:= \sqrt{\gamma}N^2\left(T^{00} - 2G^{00}\right),
\\[1ex]    
\mathcal{H}_i(\xa,\pa,\gamma_{ij},\pi^{ij}) &:= \sqrt{\gamma}N\left(T^0_i - 2G^0_i\right).
\end{align}
\end{subequations}
Here $T^{\mu\nu}$ and $G^{\mu\nu}$ denote the energy-momentum and the Einstein tensor, respectively,
and $\gamma$ is the determinant of the $d$-dimensional matrix $(\gamma_{ij})$.
In the surface integral in \eqref{totham} $i^0$ denotes spacelike infinity
and $\md^{d-1}S_i$ is the $(d-1)$-dimensional out-pointing surface element there.
In terms of lapse and shift functions, the space-time line element reads
(we employ signature $d-1$, $x^0=t$)
\begin{align}
\md s^2 &= g_{\mu\nu}\md x^{\mu}\md x^{\nu}
\nonumber\\[1ex]
&= -N^2 (\md x^0)^2 + \gamma_{ij}(\md x^i + N^i\md x^0)(\md x^j + N^j \md x^0).
\end{align}

The lapse and shift functions are Lagrangian multipliers only and deliver the famous 
Hamiltonian and momentum constraint equations of the Einstein theory,
\be
\label{CE0}
\mathcal{H} = 0, \quad \mathcal{H}_i = 0.
\ee
The fulfillment of the constraint equations together with $d+1$ additional (independent) coordinate conditions,
see Eqs.\ \eqref{coADMTT} below, 
reduces the total Hamiltonian of Eq.\ \eqref{totham} to the reduced one, $H_{\rm red}$, which obviously is a pure surface expression.
This is the ADM Hamiltonian which evolves the independent dynamical degrees of freedom of the total system.
Here the lapse and shift functions are not involved at all, only the space-asymptotic numerical value 1 of the lapse function enters. 
The lapse and shift functions are to be calculated by making use of the coordinate conditions
within the remaining set of the Einstein field equations. 

The dimensionally continued constraint equations \eqref{CE0}
written for the two-point-mass system read
\begin{subequations}
\label{CE}
\begin{align}
\label{CEh}
\sqrt{\gamma}\,R &= \frac{1}{\sqrt{\gamma}} \left(\gamma_{ik}\,\gamma_{j\ell}\,\pi^{ij} \, 
\pi^{k\ell} - \frac{1}{d-1}(\gamma_{ij}\,\pi^{ij})^2 \right)
\nonumber\\[1ex]&\quad
+ \sum_a (m_a^2 + \gamma_a^{ij}\,p_{ai}\,p_{aj})^{\frac{1}{2}} \, \da,
\\[2ex]
\label{CEm}
-2\,D_j\,\pi^{ij} &= \sum_a \gamma_a^{ij} \, p_{aj} \, \da. 
\end{align}
\end{subequations}
Here $R$ denotes the space curvature of the hypersurface $t=\text{const}$,
$\gamma_a^{ij}:=\gamma_\textrm{reg}^{ij}(\xa)$ is the finite part of the inverse metric $\gamma^{ij}$
($\gamma^{ij}\gamma_{jk}=\delta^i_k$) evaluated at the particle position
(which can be perturbatively and unambiguously defined, see Appendix \ref{AHPF} of the current paper),
and $D_j$ is the $d$-dimensional covariant derivative (acting on a tensor density of weight one).
Let us note that the source terms of both constraint equations
are proportional to Dirac $\delta$-functions.

We employ the following ADM transverse-traceless (TT) coordinate conditions,
resulting in irreducible canonical field variables,
\be
\label{coADMTT}
\displaystyle \gamma_{ij} = \left(1 + \frac{d-2}{4(d-1)}\phi\right)^{4/(d-2)} \delta_{ij} + \hTT{ij},
\quad \pi^{ii}=0,
\ee
where the metric function ${\hTT ij}$ is a symmetric TT quantity,
i.e.
\be
\label{hTTproperties}
{\hTT ii} = 0, \quad \partial_j{\hTT ij} = 0,
\ee
and the field momentum $\pi^{ij}$ is split into its longitudinal and TT parts, 
respectively
\be
\pi^{ij} = {\pit ij} + {\piTT ij}.
\ee
The longitudinal part of the field momentum can be expressed in terms of a vectorial function $V^i$,
\be
\label{pitbyv}
\tilde{\pi}^{ij} = \partial_i V^j +\partial_j V^i - \frac{2}{d}\delta^{ij}\partial_k V^k,
\ee
and the TT part satisfies conditions analogous to \eqref{hTTproperties},
\be
{\piTT ii} = 0, \quad \partial_j{\piTT ij} = 0.
\ee

After solving [with the usage of the coordinate conditions \eqref{coADMTT} and by a perturbative expansion]
the constraints \eqref{CE} with respect to the longitudinal variables $\phi$ and $\pit ij$,
we plug these solutions (expressed in terms of $\xa$, $\pa$ and ${\hTT ij}$, ${\piTT ij}$)
into the right-hand side of Eq.\ \eqref{totham}.
This way we get the reduced ADM Hamiltonian of the total matter-plus-field system,
which can be written in the form
\begin{align}
\label{redH}
H_\text{red}&\big[\xa,\pa,{\hTT ij},{\piTT ij}\big]
\nonumber\\[1ex]
&= -\int\!\md^dx\,\Deltad\phi\big[\xa,\pa,{\hTT ij},{\piTT ij}\big].
\end{align}
This Hamiltonian describes the evolution of the matter and independent gravitational field variables.
The equations of motion of the bodies read
\be
\label{EOMm}
\dot{\mathbf{x}}_a = \frac{\partial H_\text{red}}{\partial\pa},
\quad
\dot{\mathbf{p}}_a = -\frac{\partial H_\text{red}}{\partial\xa},
\ee
and the field equations for the independent degrees of freedom have the form
\be
\label{EOMf}
\frac{\partial}{\partial t}{\hTT ij} = \delta^{\textrm{TT}kl}_{ij}\frac{\delta H_\text{red}}{\delta {\piTT kl}},
\quad
\frac{\partial}{\partial t}{\piTT ij} = -\delta^{\textrm{TT}ij}_{kl}\frac{\delta H_\text{red}}{\delta {\hTT kl}},
\ee
where the $d$-dimensional TT-projection operator is defined by
\begin{align}
\label{defTT}
\delta^{\textrm{TT}ij}_{kl} &:= \frac{1}{2}(\delta_{ik}\delta_{jl}+\delta_{il}\delta_{jk})
- \frac{1}{d-1}\delta_{ij}\delta_{kl}
\nonumber\\[1ex]&\quad
-\frac{1}{2}(\delta_{ik}\partial_{jl}+\delta_{jl}\partial_{ik}+\delta_{il}\partial_{jk}+\delta_{jk}\partial_{il})\Deltad^{-1}
\nonumber\\[1ex]&\quad
+ \frac{1}{d-1}(\delta_{ij}\partial_{kl}+\delta_{kl}\partial_{ij})\Deltad^{-1}
\nonumber\\[1ex]&\quad
+ \frac{d-2}{d-1}\partial_{ijkl}\Deltad^{-2}.
\end{align}

In the current paper we are interested only in \emph{conservative} dynamics of the matter-plus-field system
and we want to describe this dynamics in terms of \emph{only matter variables $\xa$ and $\pa$}.
An autonomous (thus conservative) matter Hamiltonian
can be obtained through the transition to a Routhian description.
By means of the first field equation \eqref{EOMf}
one expresses the TT part ${\piTT ij}$ of the field momentum
as a function of matter variables $\xa$, $\pa$
and field variables $\hTT ij$, $\hTTdot ij$.
Then the Routhian is defined as
\begin{align}
\label{Rdef}
R\big[\xa,\pa,{\hTT ij},{\hTTdot ij}\big]
&:= H_\text{red}\big[\xa,\pa,{\hTT ij},{\piTT ij}\big]
\nonumber\\[1ex]
&\quad - \int\md^dx\,{\piTT ij}{\hTTdot ij}.
\end{align}
Finally the matter Hamiltonian reads
\be
\label{mH}
H(\xa,\pa) := R\big[\xa,\pa,{\hTT ij}(\xa,\pa),{\hTTdot ij}(\xa,\pa)\big],
\ee 
where the field variables $\hTT ij$, $\hTTdot ij$ are replaced by solutions of field equations,
i.e.\ they are expressed in terms of the matter variables and eventually their time derivatives,
which are in turn eliminated through lower-order equations of motion
(this procedure is equivalent to performing a higher-order contact transformation \cite{GS1984,DS1991}).

The dependence in the right-hand side of Eq.\ \eqref{mH} of the field functions
on the same position and momentum variables as those being located outside the field expressions vitiates dissipation.
For dissipation (radiation damping) to be obtained,
another set of (primed, say) variables has to be used in the field expressions, cf.\ \cite{JS1997}.
Because of our restriciton to the conservative dynamics,
we can iteratively solve the field equations by means of the {\em time-symmetric}
half-retarded plus half-advanced Green function (see Sec.\ \ref{FEqs} below),
which from the very beginning excludes dissipation.

\section{The 4PN-accurate reduced Hamiltonian}

To compute the 4PN-accurate reduced Hamiltonian given by Eq.\ \eqref{redH}
we have to perturbatively solve the constraint equations \eqref{CE} through 4PN order.
To do this we expand these equations into the PN series,
i.e., into series in powers in the inverse velocity of light, $\epsilon:={1/c}$.
We take into account that
\begin{alignat}{2}
m_a &\sim \mathcal{O}(\epsilon^2),
&\qquad
\pa &\sim \mathcal{O}(\epsilon^3),
\nonumber\\[1ex]
\phi &\sim \mathcal{O}(\epsilon^2),
&\qquad
{\hTT ij} &\sim \mathcal{O}(\epsilon^4),
\nonumber\\[1ex]
{\pit ij} &\sim \mathcal{O}(\epsilon^3),
&\qquad
{\piTT ij} &\sim \mathcal{O}(\epsilon^5).
\end{alignat}
To compute the 4PN-accurate Hamiltonian
we need to expand the Hamiltonian constraint equation \eqref{CEh}
up to the order $\epsilon^{12}$. It is convenient to put this equation
in the form solved for the (flat) Laplacian of the metric function $\phi$.
After long calculation we obtain
(from here the numbers written in subscripts within parentheses denote the formal orders in $\epsilon$)
\begin{align}
\label{Delta-phi}
\Deltad\phi = \sum_{n=1}^6\Phi_{(2n)} + \mathcal{O}(\epsilon^{14}),
\end{align}
where $\Phi_{(2)}$, \ldots, $\Phi_{(12)}$ are given by
\begin{widetext}
\begin{subequations}
\label{Delta-phi-o}
\begin{align}
\Phi_{(2)} &= -\sum_a m_a\da,
\\[2ex]
\Phi_{(4)} &= -\sum_a\frac{\papa}{2m_a}\da - \frac{d-2}{4(d-1)}\phi\,\Deltad\phi,
\\[2ex]
\Phi_{(6)} &= \sum_a\left( \frac{\papap^2}{8m_a^3} + \frac{\papa}{2(d-1)m_a}\phi \right)\da
- ({\pit ij})^2
+ \frac{d-2}{d-1}{\ggradf ij}{\hTT ij},
\\[2ex]
\Phi_{(8)} &= \sum_a\left( -\frac{\papap^3}{16m_a^5}
- \frac{\papap^2}{4(d-1)m_a^3}\phi
- \frac{(d+2)\papa}{16(d-1)^2 m_a}\phi^2 \right)\da
- \frac{4-d}{2(d-1)}\phi({\pit ij})^2
\nonumber\\[1ex]&\kern-3ex
+ \bigg\{\sum_a\frac{p_{ai}p_{aj}}{2m_a}\da-\frac{d-2}{4(d-1)^2}\big((6-d)\phi\phi_{,ij}+3\phi_{,i}\phi_{,j}\big)\bigg\}{\hTT ij}
+ \frac{3}{4}({\gradhTT ijk})^2 - \frac{1}{2}{\gradhTT ijk}{\gradhTT ikj} + {\hTT ij}\Deltad{\hTT ij} - 2{\pit ij}{\piTT ij},
\\
\Phi_{(10)} &= \sum_a\left( \frac{5\papap^4}{128m_a^7}
+\frac{3 \papap^3}{16 (d-1) m_a^5}\phi
+\frac{(d+6) \papap^2}{32 (d-1)^2 m_a^3} \phi^2
+\frac{(d+2)d\,\papa}{96 (d-1)^3 m_a} \phi^3 \right)\da
-\frac{(10-3 d) (4-d)}{16(d-1)^2} \phi ^2 ({\pit ij})^2
\nonumber\\[1ex]&\kern-3ex
+ \bigg\{ \sum_a\left(-\frac{\papa}{4m_a^3}-\frac{1}{(d-1)m_a}\phi\right)p_{ai}p_{aj}\da
+ \frac{3 (d-2)}{4 (d-1)^3} \phi{\gradf i} {\gradf j}
+ \frac{(d-2) (6-d)}{8 (d-1)^3} \phi ^2 {\ggradf ij}
- 2 {\pit ik} {\pit jk} \bigg\}{\hTT ij}
\nonumber\\[1ex]&\kern-3ex
-\frac{5-d}{4(d-1)}(\Deltad\phi)({\hTT ij})^2
-\frac{d-3}{d-1}{\ggradf ij} {\hTT ik}{\hTT jk}
-\frac{8-d}{2(d-1)} {\gradf k} {\hTT ij} {\gradhTT ijk}
+\frac{4-d}{d-1} {\gradf k} {\hTT ij}{\gradhTT ikj}
\nonumber\\[1ex]&\kern-3ex
- \frac{6-d}{2(d-1)}\phi\left(\frac{3}{4}({\gradhTT ijk})^2-\frac{1}{2}{\gradhTT ijk}{\gradhTT ikj}+{\hTT ij}\Deltad{\hTT ij}\right)
- \frac{4-d}{d-1}\phi{\pit ij}{\piTT ij}
- ({\piTT ij})^2,
\\[2ex]
\Phi_{(12)} &= \sum_a\left( -\frac{7\papap^5}{256m_a^9}
-\frac{5 \papap^4}{32 (d-1) m_a^7}\phi
-\frac{3 (d+10) \papap^3}{128 (d-1)^2m_a^5}\phi^2
-\frac{(d+6) (d+2) \papap^2}{192 (d-1)^3 m_a^3}\phi^3
-\frac{(d+2) d (3 d-2) \papa}{1536 (d-1)^4m_a} \phi^4 \right)\da
\nonumber\\[1ex]&\kern-3ex
+\frac{(d-3)(10-3 d)(4-d)}{48(d-1)^3} \phi^3 ({\pit ij})^2
+ \bigg\{ \sum_a\left(\frac{3\papap^2}{16m_a^5}+\frac{3\papa}{4(d-1)m_a^3}\phi+\frac{d+6}{8(d-1)^2m_a}\phi^2\right)p_{ai}p_{aj}\da
+ \frac{d-2}{d-1}\phi{\pit ik}{\pit jk}
\nonumber\\[1ex]&\kern-3ex
- \frac{d^2-4}{32(d-1)^4}\,\phi^2\left(\frac{1}{3}(6-d)\phi{\ggradf ij}+3{\gradf i}{\gradf j}\right) \bigg\}{\hTT ij}
+ \left(\frac{(5-d)(10-d)}{16(d-1)^2}\phi\,\Deltad\phi+\frac{7-d}{4(d-1)^2}({\gradf k})^2\right)({\hTT ij})^2
\nonumber\\[1ex]&\kern-3ex
+ \bigg\{-\sum_a\frac{p_{ai}p_{aj}}{2 m_a}\da+\frac{d-3}{2(d-1)^2}\left(3{\gradf i}\,{\gradf j}+\frac{1}{2}(10-d)\phi\,{\ggradf ij}\right)\bigg\}{\hTT ik}{\hTT jk}
\nonumber\\[1ex]&\kern-3ex
+ \frac{10-d}{4(d-1)^2}\phi{\gradf k}{\hTT ij}\left(\frac{1}{2}(8-d){\gradhTT ijk}-(4-d){\gradhTT ikj}\right)
+ \frac{(6-d)(10-d)}{16(d-1)^2}\phi^2\left(\frac{3}{4}({\gradhTT ijk})^2-\frac{1}{2}{\gradhTT ijk}{\gradhTT ikj}+{\hTT ij}\Deltad{\hTT ij}\right)
\nonumber\\[1ex]&\kern-3ex
+ {\hTT ij}\left(\frac{1}{2}{\gradhTT ikl}{\gradhTT jlk}-\frac{3}{2}{\gradhTT ikl}{\gradhTT jkl}+{\gradhTT ikl}{\gradhTT klj}-\frac{3}{4}{\gradhTT kli}{\gradhTT klj}\right)
- {\hTT ik} {\hTT kj} \Deltad{\hTT ij}
+ {\hTT ij} {\hTT kl}\left({\ggradhTT ikjl}-{\ggradhTT ijkl}\right)
\nonumber\\[1ex]&\kern-3ex
- \frac{(10-3 d)(4-d)}{8(d-1)^2} \phi^2 {\pit ij} {\piTT ij}
- 4{\pit ik}{\hTT ij}{\piTT jk}
- \frac{4-d}{2(d-1)}\phi({\piTT ij})^2.
\end{align}
\end{subequations}
We also need to expand the momentum constraint equation \eqref{CEm} up to the order $\epsilon^9$.
This expansion reads
\begin{align}
\label{grad-pit}
{\gradpit ijj} = \Pi^i_{(3)} + \Pi^i_{(5)} + \Pi^i_{(7)} + \Pi^i_{(9)} + \mathcal{O}(\epsilon^{11}),
\end{align}
where $\Pi^i_{(3)}$, \ldots, $\Pi^i_{(9)}$ are equal to
\begin{subequations}
\label{grad-pit-e}
\begin{align}
\Pi^i_{(3)} &= -\frac{1}{2} \sum_a p_{ai}\da,
\\[2ex]
\Pi^i_{(5)} &= \frac{1}{d-1}\left(\frac{1}{2}\phi \sum_a p_{ai}\da - {\gradf j}{\pit ij}\right),
\\[2ex]
\Pi^i_{(7)} &= -\frac{d+2}{16(d-1)^2}\phi^2 \sum_a p_{ai}\da
+ \frac{d-2}{4(d-1)^2}\phi{\gradf j}{\pit ij}
+ \frac{1}{2}{\hTT ij}\sum_a p_{aj}\da
+ {\pit jk}\left(\frac{1}{2}{\gradhTT jki}-{\gradhTT ijk}\right)
- \frac{1}{d-1}{\gradf j}{\piTT ij},
\\[2ex]
\Pi^i_{(9)} &= \frac{(d+2)d}{96(d-1)^3}\phi^3 \sum_a p_{ai}\da
- \frac{(d-2)^2}{16(d-1)^3}\phi^2{\gradf j}{\pit ij}
- \frac{1}{d-1}\phi{\hTT ij}\sum_a p_{aj}\da
\nonumber\\[1ex]&\quad
+ \frac{1}{d-1}{\pit jk} \left\{{\gradf k}{\hTT ij}
  + \phi\left({\gradhTT ijk}-\frac{1}{2}{\gradhTT jki}\right)\right\}
+ \frac{d-2}{4(d-1)^2}\phi\,{\gradf j}{\piTT ij}
+ \left(\frac{1}{2}{\gradhTT jki}-{\gradhTT ijk}\right){\piTT jk}.
\end{align}
\end{subequations}

We use the ADM canonical approach in an asymptotically flat space-time
and we employ asymptotically Minkowskian coordinates.
Therefore we have to assume that the functions which enter the formalism
have the following asymptotic behavior for $r\to\infty$
(see \cite{regge74} for discussion of asymptotics in the $d=3$ case):
\be
\phi \sim \frac{1}{r^{d-2}},\quad
{\hTT ij} \sim \frac{1}{r^{d-2}},\quad
{\pit ij} \sim \frac{1}{r^{d-1}},\quad
{\piTT ij} \sim \frac{1}{r^{d-1}}.
\ee
Making use of the above asymptotics and the expansion \eqref{Delta-phi}--\eqref{Delta-phi-o} of the Laplacian $\Deltad\phi$,
after dropping many total divergences which decay fast enough at spatial infinity (so they do not contribute to the Hamiltonian),
the 4PN-accurate reduced Hamiltonian can be written as
\begin{align}
\label{hrede1}
H^\mathrm{red}_\mathrm{\le 4PN}\big[\xa,\pa,{\hTT ij},{\piTT ij}\big]
= \int \md^dx \bigg( \sum_a m_a\da + \sum_{n=2}^{6} h^\mathrm{red}_{(2n)}\big(\x;\xa,\pa,{\hTT ij},{\piTT ij}\big)\bigg),
\end{align}
where the Hamiltonian densities $h^\mathrm{red}_{(4)}$, $h^\mathrm{red}_{(6)}$, \ldots, $h^\mathrm{red}_{(12)}$ read
\begin{subequations}
\label{hrede2}
\begin{align}
h^\mathrm{red}_{(4)} &= \sum_a\frac{\papa}{2m_a}\da + \frac{d-2}{4(d-1)}\phi\,\Deltad\phi,
\\[2ex]
h^\mathrm{red}_{(6)} &= \sum_a \left( -\frac{\papap^2}{8 m_a^3}
- \frac{\papa}{2(d-1)m_a}\phi \right)\da
+ ({\pit ij})^2,
\\[2ex]
h^\mathrm{red}_{(8)} &= \sum_a \left( \frac{\papap^3}{16 m_a^5}
+\frac{\papap^2}{4 (d-1) m_a^3} \phi
+\frac{(d+2)\papa}{16 (d-1)^2 m_a} \phi^2 \right)\da
+\frac{4-d}{2(d-1)} \phi\,({\pit ij})^2
\nonumber\\[1ex]&\quad
+ \bigg(-\sum_a\frac{p_{ai}p_{aj}}{2m_a}\da + \frac{(d-2)(d-3)}{4(d-1)^2}{\gradf i}\,{\gradf j}\bigg) {\hTT ij}
+ \frac{1}{4}({\gradhTT ijk})^2,
\\[2ex]
h^\mathrm{red}_{(10)} &= \sum_a \left( -\frac{5\papap^4}{128m_a^7} - \frac{3\papap^3}{16(d-1) m_a^5}\phi
-\frac{(d+6)\papap^2}{32(d-1)^2 m_a^3} \phi^2
-\frac{(d+2)d\,\papa}{96(d-1)^3 m_a} \phi^3 \right)\da
+ \frac{(10-3d)(4-d)}{16(d-1)^2} \phi^2\,({\pit ij})^2
\nonumber\\[1ex]&\quad
+ \bigg\{ \sum_a\Big(\frac{\papa}{4m_a^3}+\frac{1}{(d-1)m_a}\phi\Big)p_{ai}p_{aj}\da
- \frac{(d-2)(d-3)}{4(d-1)^3}\phi{\gradf i}\,{\gradf j}
+ 2{\pit ik}{\pit jk}\bigg\} {\hTT ij}
\nonumber\\[1ex]&\quad
- \frac{d-3}{4(d-1}(\Deltad\phi)\,({\hTT ij})^2 + \frac{d-2}{4(d-1)}{\ggradf ij}{\hTT ik}{\hTT jk}
- \frac{6-d}{8(d-1)} \phi ({\gradhTT ijk})^2
+ \frac{4-d}{d-1}\phi\,{\pit ij}{\piTT ij} + ({\piTT ij})^2,
\\[2ex]
h^\mathrm{red}_{(12)} &= \sum_a \left( \frac{7\papap^5}{256m_a^9} + \frac{5\papap^4}{32(d-1)m_a^7}\phi + \frac{3(d+10)\papap^3}{128(d-1)^2 m_a^5}\phi^2
+ \frac{(d+6)(d+2)\papap^2}{192(d-1)^3 m_a^3} \phi^3
+ \frac{(d+2)d\,(3d-2)\papa}{1536(d-1)^4 m_a} \phi^4 \right)\da
\nonumber\\[1ex]&\quad
- \frac{(d-3)(10-3d)(4-d)}{48(d-1)^3} \phi^3\,({\pit ij})^2
+ \bigg\{ \sum_a\Big(-\frac{3\papap^2}{16m_a^5}-\frac{3\papa}{4(d-1)m_a^3}\phi-\frac{d+6}{8(d-1)^2m_a}\phi^2\Big)p_{ai}p_{aj}\da
\nonumber\\[1ex]&\quad
+ \frac{(d+2)(d-2)(d-3)}{32(d-1)^4}\phi^2{\gradf i}\,{\gradf j}
- \frac{d-2}{d-1}\phi\,{\pit ik}{\pit jk} \bigg\}{\hTT ij}
+ \bigg(\frac{(d-3)(10-d)}{16(d-1)^2}\phi\,\Deltad\phi - \frac{4-d}{8(d-1)^2}({\gradf k})^2\bigg)({\hTT ij})^2 
\nonumber\\[1ex]&\quad
+ \bigg( \sum_a \frac{p_{ai}p_{aj}}{2m_a}\da + \frac{3(4+4d-d^2)}{16(d-1)^2}{\gradf i}\,{\gradf j}
+ \frac{(6-d)(10-d)}{16(d-1)^2}\phi\,{\ggradf ij} \bigg){\hTT ik}{\hTT jk}
+ \frac{(6-d)(10-d)}{64(d-1)^2} \phi^2 ({\gradhTT ijk})^2
\nonumber\\[1ex]&\quad
-\frac{1}{2}{\hTT ij}\Big({\gradhTT ikl}({\gradhTT jkl} + {\gradhTT jlk}) +  \frac{1}{2}{\gradhTT kli}{\gradhTT klj}\Big)
+ 4{\pit ik}{\hTT ij}{\piTT jk}
+ \frac{(10-3d)(4-d)}{8(d-1)^2}\phi^2\,{\pit ij}{\piTT ij} + \frac{4-d}{2(d-1)}\phi\,({\piTT ij})^2.
\end{align}
\end{subequations}
\end{widetext}

In the next step we perform the PN expansion of the field functions $\phi$ and ${\pit ij}$.
From Eqs.\ \eqref{hrede2} it follows that to get the 4PN-accurate Hamiltonian
we need to expand the function $\phi$ up to $\ore{10}$
and the function ${\pit ij}$ up to $\ore{9}$,
\begin{subequations}
\label{phi-pit-e}
\begin{align}
\label{phi-e}
\phi &= \fii + \fiv + \fvib + \fviiib + \fxb + \ore{12},
\\[1ex]
\label{pit-e}
{\pit ij} &= {\pitiii ij} + {\pitv ij} + {\pitviib ij} + {\pitixb ij} + \ore{11}.
\end{align}
\end{subequations}
Equations \eqref{Delta-phi-o} imply that the functions $\fii$ and $\fiv$ depend only on matter variables $\xa$, $\pa$,
the function $\fvib$ depends on matter variables and on $\hTT ij$,
whereas the functions $\fviiib$ and $\fxb$ depend both on $\xa$, $\pa$ and on ${\hTT ij}$, ${\piTT ij}$.
The leading-order and next-to-leading-order functions ${\pitiii ij}$ and ${\pitv ij}$
depend on matter variables only; the functions ${\pitviib ij}$ and ${\pitixb ij}$
depend on both matter and TT field variables.
The overbar in the functions $\fvib$, $\fviiib$, $\fxb$, ${\pitviib ij}$, and ${\pitixb ij}$ means that they depend
on the non-PN-expanded TT variables ${\hTT ij}$, ${\piTT ij}$.

To obtain the equations fulfilled by the functions $\fii$ up to $\fxb$
we substitute Eqs.\ \eqref{phi-pit-e} into Eqs.\ \eqref{Delta-phi-o} and reexpand them with respect to $\epsilon$.
This way we first obtain the Poisson equations for the functions $\fii$ and $\fiv$, which read
\begin{subequations}
\label{delta-fii-fiv}
\begin{align}
\label{delta-fii}
\Deltad\fii &= -\sum_a m_a\da,
\\[1ex]
\label{delta-fiv}
\Deltad\fiv &=  \sum_a \bigg(-\frac{\papa}{2m_a}+\frac{(d-2)m_a}{4(d-1)}\,\fii\bigg) \da,
\end{align}
\end{subequations}
where in the right-hand side of \eqref{delta-fiv} we have used \eqref{delta-fii}.
We have found it useful to split the function $\fiv$ into two pieces,
\be
\label{phi4byS}
\fiv = -\frac{1}{2}\SivI + \frac{d-2}{4(d-1)}\SivII,
\ee
where $\SivI$ and $\SivII$ fulfill the following Poisson equations
\be
\label{deltaSivI/II}
\Deltad\SivI = \sum_a \frac{\papa}{m_a}\delta_a,
\quad
\Deltad\SivII =  \fii \sum_a m_a\da.
\ee
In the case of the function $\fvib$ it is also convenient to split it into two pieces,
\begin{align}
\fvib[\x;\xa,\pa,{\hTT ij}] &= \fvinoTT(\x;\xa,\pa)
\nonumber\\[1ex]&\quad
+ \fvibTT[\x;\xa,{\hTT ij}],
\end{align}
where the first piece $\fvinoTT$ depends only on matter variables $\xa$, $\pa$,
whereas the second piece $\fvibTT$ depends on $\xa$ and (functionally) on ${\hTT ij}$.
The Poisson equation for the functions $\fvinoTT$ and $\fvibTT$ read
[in their source terms the Laplacians $\Deltad\fii$ and $\Deltad\fiv$ were eliminated by means of Eqs.\ \eqref{delta-fii-fiv}]
\begin{widetext}
\begin{subequations}
\label{lapphi6b}
\begin{align}
\label{lapphi61}
\Deltad\fvinoTT &= \sum_a\ \bigg\{ \frac{\papap^2}{8m_a^3}
+ \frac{(d+2)\papa}{8(d-1)m_a} \fii
- \frac{(d-2)m_a}{8(d-1)} \Big(\SivI+\frac{d-2}{2(d-1)}\big(\fii^2-\SivII\big)\Big) \bigg\}\da
- ({\pitiii ij})^2,
\\[1ex]
\label{lapphi62b}
\Deltad\fvibTT &= \frac{d-2}{d-1}{\ggradfii ij}{\hTT ij}.
\end{align}
\end{subequations}

The Poisson equations fulfilled by the functions $\fviiib$ and $\fxb$ we present in the form where the lower-order
Laplacians $\Deltad\fii$, $\Deltad\fiv$, and $\Deltad\fvib$ are not replaced by their source terms. The equations read
\begin{subequations}
\begin{align}
\Deltad\fviiib &= - \sum_a\left\{ \frac{\papap^3}{16m_a^5}
+ \frac{\papap^2}{4(d-1)m_a^3}\fii
+ \frac{\papa}{2(d-1)m_a}\bigg(\frac{d+2}{8(d-1)}\fii^2-\fiv\bigg) \right\}\da
- \frac{4-d}{2(d-1)}\fii({\pitiii ij})^2
\nonumber\\[1ex]&\kern-3ex
-2{\pitiii ij}{\pitv ij}
+ \left\{ \sum_a\frac{p_{ai}p_{aj}}{2m_a}\da - \frac{d-2}{4(d-1)^2}\big(3{\gradfii i}{\gradfii j}+(6-d)\fii{\ggradfii ij}\big)
+ \frac{d-2}{d-1}{\ggradfiv ij} \right\}{\hTT ij}
\nonumber\\[1ex]&\kern-3ex
+ \frac{3}{4}({\gradhTT ijk})^2
- \frac{1}{2}{\gradhTT ijk}{\gradhTT ikj}
+ {\hTT ij}\Deltad{\hTT ij}
- 2{\pitiii ij}{\piTT ij}
-\frac{d-2}{4(d-1)}\big(\fii\Deltad\fvib+\fiv\Deltad\fiv+\fvib\Deltad\fii\big),
\\[2ex]
\Deltad\fxb &= \sum_a\Bigg\{ \frac{5 \papap^4}{128 m_a^7}
+\frac{3 \papap^3}{16 (d-1) m_a^5}\fii
+\frac{\papap^2}{4(d-1)m_a^3} \bigg(\frac{d+6}{8(d-1)}\fii^2-\fiv\bigg)
\nonumber\\[1ex]&\kern-3ex
+\frac{\papa}{2(d-1)m_a} \bigg(\frac{(d+2)d}{48(d-1)^2}\fii^3-\frac{d+2}{4(d-1)}\fii\fiv+\fvib\bigg) \Bigg\}\da
\nonumber\\[1ex]&\kern-3ex
-\frac{4-d}{d-1}\bigg(\frac{10-3d}{16(d-1)}\fii^2({\pitiii ij})^2 + \frac{1}{2}\fiv({\pitiii ij})^2 + \fii{\pitiii ij}{\pitv ij}\bigg)
- ({\pitv ij})^2 - 2{\pitiii ij}{\pitviib ij}
\nonumber\\[1ex]&\kern-3ex
+ \Bigg\{ -\sum_a\frac{p_{ai}p_{aj}}{m_a}\bigg(\frac{\papa}{4 m_a^2}+\frac{1}{d-1}\fii\bigg)\da
+ \frac{d-2}{d-1}\bigg(
\frac{3}{4(d-1)}\Big(\frac{1}{d-1}\fii{\gradfii i}{\gradfii j} - {\gradfii i}{\gradfiv j} - {\gradfii j}{\gradfiv i}\Big)
\nonumber\\[1ex]&\kern-3ex
+ \frac{6-d}{4(d-1)}\Big(\frac{1}{2(d-1)}\fii^2{\ggradfii ij}-{\ggradfii ij}\fiv-\fii{\ggradfiv ij}\Big)
+ {\ggradfvib ij} \bigg)
- 2 {\pitiii ik} {\pitiii jk} \Bigg\}{\hTT ij}
\nonumber\\[1ex]&\kern-3ex
- \frac{5-d}{4(d-1)}(\Deltad\fii)({\hTT ij})^2
- \frac{d-3}{d-1}{\ggradfii ij} {\hTT ik}{\hTT jk}
- \frac{8-d}{2 (d-1)}{\gradfii k}{\hTT ij}{\gradhTT ijk}
+ \frac{4-d}{d-1} {\gradfii k} {\hTT ij}{\gradhTT ikj}
\nonumber\\[1ex]&\kern-3ex
+ \frac{6-d}{2(d-1)}\fii\Big(\frac{1}{2}{\gradhTT ijk} {\gradhTT ikj} - \frac{3}{4} ({\gradhTT ijk})^2 - {\hTT ij}\Deltad{\hTT ij}\Big)
- \bigg(\frac{4-d}{d-1}\fii{\pitiii ij}+2{\pitv ij}\bigg) {\piTT ij}
- ({\piTT ij})^2
\nonumber\\[1ex]&\kern-3ex
-\frac{d-2}{4(d-1)}\big(\fii\Deltad\fviiib+\fiv\Deltad\fvib+\fvib\Deltad\fiv+\fviiib\Deltad\fii\big).
\end{align}
\end{subequations}

The equations satisfied by the longitudinal field momenta $\pitiii ij$, $\pitv ij$, $\pitviib ij$, and $\pitixb ij$
we obtain by substituting expansions \eqref{phi-pit-e} into Eqs.\ \eqref{grad-pit}--\eqref{grad-pit-e}
and reexpanding them with respect to $\epsilon$. The result is
\begin{subequations}
\label{grad-pitiii-ixb}
\begin{align}
\label{grad-pitiii}
{\gradpitiii ijj} &= -\frac{1}{2}\sum_a p_{ai}\da,
\\[2ex]
\label{gradpitv}
{\gradpitv ijj} &= -\frac{1}{d-1}\partial_j\big(\fii{\pitiii ij}\big),
\\[2ex]
{\gradpitviib ijj} &= \partial_j\bigg\{
- \frac{1}{d-1}\big(\fiv{\pitiii ij} + \fii{\pitv ij} + \fii{\piTT ij}\big)
- \frac{6-d}{8(d-1)^2}\fii^2{\pitiii ij} 
- {\pitiii jk}{\hTT ik} + {\Viii k}{\gradhTT jki} \bigg\},
\\[2ex]
{\gradpitixb ijj} &= \partial_j\bigg\{
- \frac{1}{d-1}\big(\fvib{\pitiii ij} + \fiv{\pitv ij} + \fii{\pitviib ij} + \fiv{\piTT ij}\big)
- \frac{6-d}{4(d-1)^2}\fii\fiv{\pitiii ij}
- \frac{(4-d)(6-d)}{48(d-1)^3}\fii^3{\pitiii ij}
\nonumber\\[1ex]&\quad
- \frac{6-d}{8(d-1)^2}\fii^2\big({\pitv ij}+{\piTT ij}\big)
+ \frac{1}{d-1}\fii{\pitiii jk}{\hTT ik}
+ {\Vv k}{\gradhTT jki} - {\hTT ik}{\piTT jk} \bigg\}
\nonumber\\[1ex]&\quad
- \frac{1}{d-1}\big(\fii{\pitiii jk}+(d-1){\pitv jk}){\gradhTT ijk}
+ \frac{1}{2}{\gradhTT jki}{\piTT jk}.
\end{align}
\end{subequations}
\end{widetext}

By making use of Eqs.\ \eqref{delta-fii-fiv}--\eqref{grad-pitiii-ixb}
and performing very many integrations by parts in space,
it is possible to rewrite the 4PN-accurate reduced Hamiltonian \eqref{hrede1} in the form
in which its density depends on momenta $\pa$ and on the following functions:
$\fii$, $\SivI$, $\SivII$, $\Viii i$, $\Vv i$, $\fvib$, $\hTT ij$, $\piTT ij$
[for convenience we also use $\pitiii ij$ and $\pitv ij$,
which can be expressed by $\Viii i$ and $\Vv i$, respectively, by means of Eq.\ \eqref{pitbyv}].
We will display now this form.
The 4PN-accurate Hamiltonian is the sum of pieces related with different PN orders,
\begin{align}
\label{hrede3}
H^\mathrm{red}_\mathrm{\le 4PN}\big[\xa,\pa,{\hTT ij},{\piTT ij}\big]&
\nonumber\\[1ex]&\kern-18ex
= \int\md^dx\,h^\mathrm{red}_\mathrm{\le 4PN}\big[\x;\xa,\pa,{\hTT ij},{\piTT ij}\big],
\end{align}
where
\begin{align}
h^\mathrm{red}_\mathrm{\le 4PN}\big[\x;\xa,\pa,{\hTT ij},{\piTT ij}\big] &= \sum_a m_a\da
+ h_{(4)}\big(\x;\xa,\pa\big)
\nonumber\\&\kern-15ex
+ h_{(6)}\big(\x;\xa,\pa\big)
+ h_{(8)}\big(\x;\xa,\pa,{\hTT ij}\big)
\nonumber\\[1ex]&\kern-15ex
+ h_{(10)}\big(\x;\xa,\pa,{\hTT ij},{\piTT ij}\big)
\nonumber\\[1ex]&\kern-15ex
+ h_{(12)}\big[\x;\xa,\pa,{\hTT ij},{\piTT ij}\big],
\end{align}
where the Newtonian $h_{(4)}$ and the 1PN $h_{(6)}$ densities depend only on matter variables $\xa$, $\pa$ only,
the 2PN densitity $h_{(8)}$ depends on matter variables and on $\hTT ij$,
whereas the 3PN $h_{(10)}$ and 4PN $h_{(12)}$ densities
depend on matter variables and on field variables $\hTT ij$, $\piTT ij$.
The dependence of $h_{(12)}$ on $\hTT ij$ is both pointwise and functional
and this is why we have used square brackets for arguments of $h_{(12)}$.
The explicit forms of the Newtonian $h_{(4)}$ and the 1PN-level $h_{(6)}$ densities are as follows:
\begin{subequations}
\label{hred-h4-h6}
\begin{align}
h_{(4)}\big(\x;\xa,\pa\big) &= \sum_a\bigg(\frac{\papa}{2m_a}-\frac{(d-2)m_a}{4(d-1)}\,\fii\bigg)\da,
\\[2ex]
h_{(6)}\big(\x;\xa,\pa\big) &= \sum_a\bigg( -\frac{\papap^2}{8m_a^3} - \frac{(d+2)\papa}{8(d-1)m_a}\fii
\nonumber\\[1ex]&\kern-10ex
+ \frac{(d-2)m_a}{8(d-1)}\Big(\SivI-\frac{d-2}{2(d-1)}\big(\SivII-\fii^2\big)\Big)
\nonumber\\[1ex]&\kern-10ex
+ {\Viii i}\,p_{ai} \bigg)\da.
\end{align}
\end{subequations}
For displaying the 2PN-level density $h_{(8)}$ we define two auxiliary functions
which depend on matter variables only,
\begin{widetext}
\begin{subequations}
\label{chi8-S4-def}
\begin{align}
\varkappa_{(8)}\big(\x;\xa,\pa\big) &:= \sum_a\Bigg\{ \frac{\papap^3}{16m_a^5}
+ \frac{(d+2)\papap^2}{16(d-1)m_a^3}\,\fii
+ \frac{(d+2)\papa}{16(d-1)m_a} \left(\fii^2+\SivI-\frac{d-2}{2(d-1)}\SivII\right)
\nonumber\\[1ex]&\qquad
- \frac{(d-2)^2 m_a}{32(d-1)^2} \left( \frac{d-2}{d-1}\fii^2 + 3\,\SivI - \frac{3(d-2)}{2(d-1)}\SivII \right)\fii \Bigg\}\da
- \fii({\pitiii ij})^2,
\\[2ex]
\label{Siv}
{\Siv ij}(\x;\xa,\pa) &:= -\sum_a \frac{p_{ai} p_{aj}}{m_a}\da - \frac{d-2}{2(d-1)}{\gradfii i}{\gradfii j}.
\end{align}
\end{subequations}
By means of these functions the density $h_{(8)}$ can be written as
\begin{align}
\label{hred-h8}
h_{(8)}&\big(\x;\xa,\pa,{\hTT ij}\big) = \varkappa_{(8)}\big(\x;\xa,\pa\big)
+ \frac{1}{2}{\Siv ij}\big(\x;\xa,\pa\big){\hTT ij} + \frac{1}{4}({\gradhTT ijk})^2.
\end{align}
To display the 3PN density $h_{(10)}$ we define another two auxiliary functions
\begin{subequations}
\begin{align}
\varkappa_{(10)}\big(\x;\xa,\pa\big) &:= \sum_a \bigg\{
- \frac{5\papap^4}{128m_a^7}
- \frac{(d+4)\papap^3}{32(d-1)m_a^5}\fii
\nonumber\\[1ex]&\quad
- \frac{\papap^2}{32(d-1)m_a^3} \left(\frac{(d+6)d}{2(d-1)}\fii^2+(d+4)\SivI-\frac{d(d-2)}{d-1}\SivII\right)
\nonumber\\[1ex]&\quad
- \frac{(d+2)\papa}{16(d-1)^2m_a} \left(\frac{d(3d-4)}{12(d-1)}\fii^2+d\,\SivI-\frac{(3d-4)(d-2)}{4(d-1)}\SivII\right)\fii
\nonumber\\[1ex]&\quad
+ \frac{(d-2)m_a}{32(d-1)^2} \bigg(\frac{(d-2)^3}{4(d-1)^2}\fii^4+\frac{(3d-4)(d-2)}{2(d-1)}\SivI\fii^2-\frac{(d-2)^3}{(d-1)^2}\SivII\fii^2
\nonumber\\[1ex]&\quad
+ d\,\SivI^2 - \frac{(3d-4)(d-2)}{2(d-1)}\SivI\SivII + \frac{(d-2)^3}{2(d-1)^2}\SivII^2\bigg)
\nonumber\\[1ex]&\quad
+ \frac{{\Viii i}p_{ai}}{2(d-1)} \left(\frac{(3d-2)(3d-4)}{8(d-1)}\fii^2+d\,\SivI-\frac{(3d-4)(d-2)}{4(d-1)}\SivII\right)  \bigg\}\da
\nonumber\\[1ex]&\quad
+ \bigg( \frac{3d-4}{4(d-1)^2}\big((d-2){\gradSivII i}-(3d-2)\fii{\gradfii i}\big)
- \frac{d}{d-1}{\gradSivI i} \bigg){\Viii j}{\pitiii ij},
\\[2ex]
\label{B6def}
{\Bvi ij}(\x;\xa,\pa) &:= \sum_a\bigg( \frac{\papa\,p_{ai}p_{aj}}{4m_a^3}
+ \frac{(d+2)p_{ai}p_{aj}}{4(d-1)m_a}\fii \bigg)\da
- 4 {\gradViii ik}{\gradViii kj}
- 2 {\gradViii ik}{\gradViii jk} + 2 {\gradViii ki}{\gradViii kj}
\nonumber\\[1ex]&\quad
 + \frac{4(d-2)}{d} {\gradViii ij}{\gradViii kk}
+ \frac{(d+1)(d-2)}{4(d-1)^2} {\gradfii i}{\gradSivI j}
- \frac{(2d-3)(d-2)^2}{8(d-1)^3} {\gradfii i}{\gradSivII j}
\nonumber\\[1ex]&\quad
+ \frac{(d-2)(3d-4)}{8(d-1)^3} \fii{\gradfii i}{\gradfii j}.
\end{align}
\end{subequations}
The 3PN density $h_{(10)}$ then reads
\begin{align}
\label{hred-h10}
h_{(10)}\big(\x;\xa,\pa,{\hTT ij},{\piTT ij}\big) &= \varkappa_{(10)}\big(\x;\xa,\pa\big)
+ \frac{2(2d-3)}{(d-1)^2}{\gradfii i}{\Viii j}\big(\fii{\pitiii ij}\big)^\mathrm{TT}
+ {\Bvi ij}(\x;\xa,\pa){\hTT ij}
\nonumber\\[1ex]&\qquad
+ \frac{1}{2(d-1)}\fii{\hTT ij}\Deltad{\hTT ij}
- \frac{2(d-2)}{d-1}\fii{\pitiii ij}{\piTT ij}
+ ({\piTT ij})^2,
\end{align}
where, to diminish number of terms, we have introduced the TT projection of the product $\fii{\pitiii ij}$,
which, by virtue of Eqs.\ \eqref{pitbyTT} and \eqref{gradpitv}, can be written as
\be
\label{fi2pi3TT}
\big(\fii{\pitiii ij}\big)^\mathrm{TT} = \fii{\pitiii ij} + (d-1){\pitv ij}.
\ee
This TT projection should be treated as a function of matter variables only.

To display the very large formula for the 4PN-order density $h_{(12)}$ we introduce three auxiliary quantities
$\varkappa_{(12)}^1$, $\varkappa_{(12)}^2$, and $\varkappa_{(12)}^3$.
We first define the function $\varkappa_{(12)}^1$ which depends only on matter variables $\xa$, $\pa$,
\begin{align}
\varkappa_{(12)}^1&(\x;\xa,\pa) := \sum_a \bigg\{ \frac{7\papap^5}{256m_a^9}
+ \frac{5(d+6)\papap^4}{256(d-1)m_a^7}\,\fii
\nonumber\\[1ex]&
+ \frac{\papap^3}{64(d-1)m_a^5} \bigg[ \frac{d+1}{d-1}\left(\frac{1}{2}(d+10)\fii^2-(d-2)\SivII\right) + (d+6)\SivI \bigg]
\nonumber\\[1ex]&
+ \frac{(d+6)\papap^2}{32(d-1)m_a^3} \bigg[ \frac{(d+2)(3d-2)}{24(d-1)^2}\fii^3
+ \frac{1}{2(d-1)} \bigg((d+1)\SivI-\frac{(3d-2)(d-2)}{4(d-1)}\SivII\bigg) \fii + \fvinoTT \bigg]
\nonumber\\[1ex]&
+ \frac{(d+2)\papa}{32(d-1)^2m_a} \bigg[
  \frac{d(3d-2)(2d-3)}{48(d-1)^2}\fii^4
+ \frac{d}{4(d-1)} \left( (3d-2)\SivI - \frac{(d-2)(2d-3)}{d-1}\SivII \right) \fii^2
\nonumber\\[1ex]&
+ \frac{1}{2} \left((d+1)\SivI^2 - \frac{(d-2)(3d-2)}{2(d-1)}\SivI\SivII + \frac{(d-2)^2(2d-3)}{4(d-1)^2}\SivII^2 \right)
+ (3d-2)\fii\fvinoTT \bigg]
\nonumber\\[1ex]&
+ \frac{(d-2)m_a}{32(d-1)^2} \bigg[
- \frac{(d-2)^4}{16(d-1)^3}\fii^5
- \frac{(d-2)^2}{4(d-1)^2}\left((2d-3)\SivI-\frac{5(d-2)^2}{4(d-1)}\SivII\right)\fii^3
\nonumber\\[1ex]&
- \frac{d-2}{2(d-1)} \left( \frac{1}{2}(3d-2)\SivI^2 - \frac{(d-2)(2d-3)}{d-1}\SivI\SivII + \frac{5(d-2)^3}{8(d-1)^2}\SivII^2 \right) \fii
\nonumber\\[1ex]&
+ \left(\frac{5(d-2)^2}{2(d-1)}\big(\SivII-\fii^2\big) - (3 d-2)\SivI\right) \fvinoTT \bigg] \bigg\}\da
- \frac{3d-2}{4(d-1)}\fvinoTT(\pitiii ij)^2
\nonumber\\[1ex]&
- \frac{3d-2}{8(d-1)^2} \bigg[ \frac{1}{3} (2 d-3) \fii^3 + \frac{1}{2} \left( (3d-2)\SivI - \frac{(d-2)(2d-3)}{d-1}\SivII \right) \fii \bigg] (\pitiii ij)^2.
\end{align}
The function $\varkappa_{(12)}^2$ depends on matter variables $\xa$, $\pa$ and on the field function ${\hTT ij}$
(it contains terms linear, quadratic, and cubic in ${\hTT ij}$). It reads
\begin{align}
\varkappa_{(12)}^2(&\x;\xa,\pa,{\hTT ij}) := \bigglb\{
-\sum_a \frac{p_{ai}p_{aj}}{8m_a} \bigg(
\frac{3\papap^2}{2m_a^4}
+ \frac{(d+4)\papa}{(d-1)m_a^2}\,\fii
+ \frac{(d+6)d}{2(d-1)^2}\fii^2 
+ \frac{d+4}{d-1}\SivI
- \frac{d(d-2)}{(d-1)^2}\SivII \bigg)\da
\nonumber\\[1ex]&
+ \frac{d(d-2)}{4(d-1)^2}{\ggradfii ij}\fvinoTT
- \frac{d-2}{8(d-1)^3} \bigg\{ \bigg[\frac{2d-3}{2(d-1)}\bigg(\frac{d+2}{2}\fii^2-(d-2)\SivII\bigg)+\frac{3d-2}{2}\SivI\bigg]{\gradfii i}{\gradfii j}
\nonumber\\[1ex]&
+ \bigg(3d\,{\gradSivI i}
- \frac{(5d-8)(d-2)}{2(d-1)}{\gradSivII i}\bigg)\fii{\gradfii j} \bigg\}
- \frac{d-2}{16(d-1)^2} \bigg( (d+3)\,{\gradSivI i}{\gradSivI j}
\nonumber\\[1ex]&
- \frac{(2d-1)(d-2)}{d-1}{\gradSivI i}{\gradSivII j}
+ \frac{(3d-5)(d-2)^2}{4(d-1)^2}{\gradSivII i}{\gradSivII j} \bigg)
+ \frac{2(d-2)}{d-1} \bigg\{ \bigg[ {\gradViii ki}{\gradViii kj}
\nonumber\\[1ex]&
- {\gradViii ik}{\gradViii jk}
- 2{\gradViii ik}{\gradViii kj}
+ 2\bigg(1-\frac{2}{d}\bigg){\gradViii ij}{\gradViii kk} \bigg]\fii
+ 2{\Viii k}\big({\gradViii ij}{\gradfii k}-{\gradViii ik}{\gradfii j}\big) \bigg\}
\nonumber\\[1ex]&
+ 4(d-1) \bigg[ {\gradViii ki}{\gradVv kj}
- {\gradViii ik}{\gradVv jk}
- {\gradViii ik}{\gradVv kj}-{\gradViii ki}{\gradVv jk}+\Big(1-\frac{2}{d}\Big)\big({\gradViii kk}{\gradVv ij}+{\gradViii ij}{\gradVv kk}\big) \bigg]
\biggrb\} {\hTT ij}
\nonumber\\[1ex]&
+ \bigg( \sum_a\frac{p_{ai}p_{aj}}{2m_a}\da - \frac{4+8d-3d^2}{16(d-1)^2}{\gradfii i}{\gradfii j}
- \frac{(d-2)^2}{16(d-1)^2}\big(\fii{\ggradfii ij}+{\ggradSivII ij}\big)
+ \frac{1}{4(d-1)}{\ggradSivI ij} \bigg){\hTT ik}{\hTT jk}
\nonumber\\[1ex]&
+ \bigg(-\sum_a\frac{3\papa}{16(d-1)m_a}\da
+ \frac{20-d^2}{64(d-1)^2} ({\gradfii k})^2\bigg) ({\hTT ij})^2
- \frac{1}{8(d-1)} \bigg(\frac{(d+2)(10-d)}{8(d-1)} \fii^2 + 3 \SivI
\nonumber\\[1ex]&
- \frac{d^2-4}{4(d-1)}\SivII \bigg){\hTT ij}\Deltad{\hTT ij}
- \frac{1}{4} \Big( 2 {\gradhTT ikl}\left({\gradhTT jkl}+{\gradhTT jlk}\right)
  + {\gradhTT kli}{\gradhTT klj} \Big){\hTT ij}.
\end{align}
The function $\varkappa_{(12)}^3$ is proportional to the second part $\fvibTT$ of the function $\fvib$.
It is a function of matter variables and it depends on ${\hTT ij}$ both pointwisely and, through the function $\fvibTT$, functionally.
It equals
\begin{align}
&\varkappa_{(12)}^3[\x;\xa,\pa,{\hTT ij}] := \bigg\{
\sum_a\bigg[\frac{(d+6)\papap^2}{32(d-1)m_a^3} + \frac{(d+2)(3d-2)\papa}{32(d-1)^2m_a}\fii
\nonumber\\[1ex]&
+ \frac{(d-2)m_a}{32(d-1)^2}\bigg(\frac{5(d-2)^2}{2(d-1)}\big(\SivII-\fii^2\big)-(3d-2)\SivI\bigg)\bigg]\da
- \frac{3d-2}{4(d-1)}({\pitiii ij})^2 + \frac{d(d-2)}{4(d-1)^2}{\ggradfii ij}{\hTT ij}
\bigg\}\fvibTT.
\end{align}
The 4PN-order Hamiltonian density $h_{(12)}$ finally equals
\begin{align}
\label{hred-h12}
h_{(12)}\big[&\x;\xa,\pa,{\hTT ij},{\piTT ij}\big]
= \varkappa_{(12)}^1(\x;\xa,\pa) + \varkappa_{(12)}^2(\x;\xa,\pa,{\hTT ij}) + \varkappa_{(12)}^3[\x;\xa,\pa,{\hTT ij}]
\nonumber\\[1ex]&
+ \bigg(\frac{d(11d-18)}{8(d-1)^3}\fii^2 + \frac{2}{d-1}\SivI - \frac{(5d-8)(d-2)}{4(d-1)^3}\SivII\bigg){\pitiii ij}(\fii{\pitiii ij})^\mathrm{TT}
- \frac{3d-5}{(d-1)^3}\fii\Big((\fii{\pitiii ij})^\mathrm{TT}\Big)^2
\nonumber\\[1ex]&
+ 2\Big(2{\pitiii ik}{\hTT ij}+{\Viii i}(2{\gradhTT ijk}-{\gradhTT jki})\Big)\bigg({\piTT jk}-\frac{d-2}{d-1}(\fii{\pitiii jk})^\mathrm{TT}\bigg)
- \frac{d-3}{d-1}\fii({\piTT ij})^2
\nonumber\\[1ex]&
+ \bigg( \frac{(9d-14)(4-d)}{4(d-1)^2}\fii{\gradfii j}{\Viii i} - \frac{d}{d-1}{\gradSivI j}{\Viii i} + \frac{(3d-4)(d-2)}{4(d-1)^2}{\gradSivII j}{\Viii i}
+ \frac{2(3d-5)}{d-1}{\gradfii j}{\Vv i} \bigg){\piTT ij}.
\end{align}
\end{widetext}

\section{Field equations}
\label{FEqs}

Dynamical degrees of freedom of gravitational field
described by the functions $\hTT ij$ and $\piTT ij$
are solutions of the field equations \eqref{EOMf}.
From the 4PN-accurate reduced Hamiltonian given in Eq.\ \eqref{hrede3}
one can derive 4PN-accurate approximate field equations.
In the rest of the paper we will only need to use field equations
which follow from the 3PN-accurate part of the Hamiltonian \eqref{hrede3}.
It reads
\begin{multline}
\label{h3rede}
H^\mathrm{red}_\mathrm{\le 3PN}\big[\xa,\pa,{\hTT ij},{\piTT ij}\big]
\\
= \int\md^dx\,h_\mathrm{\le 3PN}\big(\x;\xa,\pa,{\hTT ij},{\piTT ij}\big),
\end{multline}
where
\begin{align}
h_\mathrm{\le 3PN}\big(\x;\xa,\pa,{\hTT ij},{\piTT ij}\big) &=
\sum_a m_a\da + h_{(4)}\big(\x;\xa,\pa\big)
\nonumber\\&\kern-15ex
+ h_{(6)}\big(\x;\xa,\pa\big)
+ h_{(8)}\big(\x;\xa,\pa,{\hTT ij}\big)
\nonumber\\[1ex]&\kern-15ex
+ h_{(10)}\big(\x;\xa,\pa,{\hTT ij},{\piTT ij}\big).
\end{align}
For this Hamiltonian the field equations \eqref{EOMf} take the form
\begin{subequations}
\label{FE}
\begin{align}
{\hTTdot ij} &= \delta^{\textrm{TT}kl}_{ij}\,\frac{\partial h_\mathrm{\le 3PN}}{\partial {\piTT kl}}
+ {\ore 7},
\\[2ex]
{\piTTdot ij} &= -\delta^{\textrm{TT}ij}_{kl} \bigg\{
\frac{\partial h_\mathrm{\le 3PN}}{\partial {\hTT kl}}
- \bigg(\frac{\partial h_\mathrm{\le 3PN}}{\partial {\gradhTT klm}}\bigg)_{\!\!,m}
\nonumber\\[1ex]&\qquad
+ \bigg(\frac{\partial h_\mathrm{\le 3PN}}{\partial h^\mathrm{TT}_{kl,mn}}\bigg)_{\!\!,mn} \bigg\}
+ {\ore 8},
\end{align}
\end{subequations}
or, more explicitly,
\begin{subequations}
\label{FE2}
\begin{align}
\label{FE2a}
{\hTTdot ij} &= \delta^{\textrm{TT}kl}_{ij}
\bigg\{2{\piTT kl}-\frac{2(d-2)}{d-1}\fii{\pitiii kl}\bigg\}
+ {\ore 7},
\\
\label{FE2b}
{\piTTdot ij} &= -\delta^{\textrm{TT}kl}_{ij}
\bigg\{\frac{1}{2} {\Siv kl} - \frac{1}{2}\Deltad{\hTT kl}
+ {\Bvi kl}
\nonumber\\&\qquad
+ \frac{1}{2(d-1)}\bigg(\fii\Deltad{\hTT kl} + \Deltad\Big(\fii{\hTT kl}\Big)\bigg)\bigg\}
\nonumber\\[1ex]&\qquad
+ {\ore 8}.
\end{align}
\end{subequations}
By combining these two equations one gets the equation fulfilled by the function $\hTT ij$.
It can be written in the form of the wave equation,
\begin{align}
\label{FEhTT}
\Box_{d+1}{\hTT ij} = S^{\rm TT}_{ij},
\end{align}
where $\Box_{d+1}$ is d'Alembertian in $(d+1)$-dimensional flat space-time,
\be
\Box_{d+1} := -\partial_t^2 + \Deltad,
\ee
and where the source term is
\begin{align}
S^{\rm TT}_{ij} &= \delta^{\textrm{TT}kl}_{ij}\bigg\{
{\Siv kl} + 2{\Bvi kl}
\nonumber\\[1ex]&\quad
+ \frac{1}{d-1}\bigg(\fii\Deltad{\hTT kl} + \Deltad\Big(\fii{\hTT kl}\Big)\bigg)
\nonumber\\[1ex]&\quad
+ \frac{2(d-2)}{d-1}\partial_t\big(\fii{\pitiii kl}\big)
\bigg\} + {\ore 8}.
\end{align}
After solving field equation \eqref{FEhTT} for $\hTT ij$
one can obtain the TT field momentum $\piTT ij$ from Eq.\ \eqref{FE2a}:
\be
\label{piTTbyhTT}
{\piTT ij} = \frac{1}{2}{\hTTdot ij} + \frac{d-2}{d-1}\delta^{\textrm{TT}kl}_{ij}\Big(\fii{\pitiii kl}\Big)
+ {\ore 7}.
\ee

In the current paper we are interested only in conservative dynamics of the matter-plus-field system;
therefore, we use a time-symmetric (half-retarded half-advanced) formal solution of the wave equation \eqref{FEhTT}.
We then expand this solution with respect to the retardation.
This way we obtain the \emph{near-zone} PN expansion of $\hTT ij$ of the form
\be
\label{timesymhTT}
{\hTT ij} = \left(\Deltad^{-1} + \Deltad^{-2}\partial_t^2 + \Deltad^{-3}\partial_t^4 + \cdots \right) S^{\rm TT}_{ij}.
\ee
By making the expansion \eqref{timesymhTT} we exclude from the 4PN-level near-zone metric
the nonlocal-in-time contribution coming from tail effects (what was found in Ref.\ \cite{BD88}).
Expansion of $\hTT ij$ to the order required in our computations reads
\be
\label{hTTe}
{\hTT ij} = {\hTTiv ij} + {\hTTvi ij} + {\ore 8};
\ee
we thus omit the non-time-symmetric parts $\hTTv ij$ and $\hTTvii ij$.
After plugging the expansion \eqref{hTTe} into Eq.\ \eqref{FEhTT}
one obtains equations fulfilled by the functions $\hTTiv ij$ and $\hTTvi ij$.
The leading-order function $\hTTiv ij$ is the solution of the Poisson equation
\be
\label{feqhTT4}
\Deltad{\hTTiv ij} = {\STTiv ij},
\ee
where the source term $\Siv ij$ is defined in Eq.\ \eqref{Siv}.
The next-to-leading-order function $\hTTvi ij$ fulfills equation
\be
\label{feqhTT6}
\Deltad{\hTTvi ij} = {\STTvi ij} + {\hTTivddot ij},
\ee
where the source function $\Svi ij$ equals
\begin{align}
\label{STT6}
{\Svi ij} &= 2{\Bvi ij} + \frac{1}{d-1}\bigg(\fii\Deltad{\hTTiv ij} + \Deltad\Big(\fii{\hTTiv ij}\Big)\bigg)
\nonumber\\[1ex]&\qquad
+ 2\frac{d-2}{d-1}\partial_t\big(\fii{\pitiii ij}\big),
\end{align}
and the source function ${\Bvi ij}$ is defined in Eq.\ \eqref{B6def}.

\section{4PN-accurate Routhian}

Introducing Routhian description is an intermediate step on our route
to the Hamiltonian which depends only on matter variables.
The 4PN-accurate Routhian $R_\mathrm{\le 4PN}$ we obtain
from the 4PN-accurate reduced Hamiltonian $H^\mathrm{red}_\mathrm{\le 4PN}$ of Eq.\ \eqref{hrede3}
by performing the Legendre transformation with respect to the field momentum ${\piTT ij}$ [see Eq.\ \eqref{Rdef}].
The result is
\begin{align}
\label{Routhian2}
R_\mathrm{\le 4PN}\big[\xa,\pa,{\hTT ij},{\hTTdot ij}\big]
&= H^\mathrm{red}_\mathrm{\le 4PN}\big[\xa,\pa,{\hTT ij},{\piTT ij}\big]
\nonumber\\[1ex]&\quad
- \int\md^dx\,{\hTTdot ij}{\piTT ij},
\end{align}
where on the right-hand side the field momentum $\piTT ij$ is expressed in terms of $\hTT ij$, $\hTTdot ij$, and matter variables.
The Legendre transformation is to leading order realized by Eq.\ \eqref{piTTbyhTT}.
We will show now that this leading-order formula is enough to get the 4PN-accurate Routhian.

Let us split the field momentum $\piTT ij$ in the following way
[cf.\ Eq.\ \eqref{piTTbyhTT}]:
\begin{subequations}
\begin{align}
\label{piTTsplitting}
{\piTT ij} &= \frac{1}{2} {\hTTdot ij} + \frac{d-2}{d-1} \big(\fii{\pitiii ij}\big)^\mathrm{TT} + {\dpiTT ij},
\\[1ex]
{\dpiTT ij} &= {\ore 7}, \quad {\dpiTT ij} = {\ora {1-d}} \quad\text{for $r\to\infty$}.
\end{align}
\end{subequations}
We will show that the density of the 4PN-accurate Routhian \eqref{Routhian2} does not depend on ${\dpiTT ij}$.
The only part of the density which could depend on ${\dpiTT ij}$
[see Eq.\ \eqref{hred-h10}] consists of three terms,
\be
\label{dr3pn1}
\delta\mathfrak{r} := ({\piTT ij})^2 - \frac{2(d-2)}{d-1}\fii{\pitiii ij}{\piTT ij}
- {\hTTdot ij}{\piTT ij}.
\ee
Making use of the splitting \eqref{piTTsplitting} and the relation \eqref{fi2pi3TT}
we rewrite the quantity $\delta\mathfrak{r}$ in the form
\begin{align}
\label{dr3pn2}
\delta\mathfrak{r} &= -\frac{1}{4}\big({\hTTdot ij}\big)^2
- \frac{d-2}{d-1}\fii{\pitiii ij}{\hTTdot ij}
\nonumber\\[1ex]&
+ \big({\dpiTT ij}\big)^2
+ 2(d-2){\pitv ij}{\dpiTT ij}
\nonumber\\[1ex]&
+ \left(\frac{d-2}{d-1}\right)^2 \big(\fii{\pitiii ij}\big)^\mathrm{TT} \Big((d-1){\pitv ij}-\fii{\pitiii ij}\Big).
\end{align}
By virtue of the representation \eqref{pitbyv} we rewrite Eq.\ \eqref{dr3pn2} as
\begin{align}
\label{dr3pn3}
\delta\mathfrak{r} &= -\frac{1}{4} \big({\hTTdot ij}\big)^2
- \frac{d-2}{d-1}\fii{\pitiii ij}{\hTTdot ij}
\nonumber\\[1ex]&\qquad
- \left(\frac{d-2}{d-1}\right)^2\fii{\pitiii ij}\big(\fii{\pitiii ij}\big)^\mathrm{TT}
\nonumber\\[1ex]&\qquad
+ \big({\dpiTT ij}\big)^2
+ \partial_k \mathrm{ED}_{3k},
\end{align}
with
\be
\mathrm{ED}_{3k} := {\Vv i} \Big( 4(d-2){\dpiTT ik} + 2\frac{(d-2)^2}{d-1}\big(\fii{\pitiii ik}\big)^\mathrm{TT} \Big).
\ee
Because $\mathrm{ED}_{3k}$ decays as $1/r^{2d-3}$ for $r\to\infty$
[the quantity ${\Vv i}={\ora {2-d}}$ when $r\to\infty$], it does not contribute to the Routhian,
and because $\big({\dpiTT ij}\big)^2$ is of the order of $\epsilon^{14}$,
only the first three terms on the right-hand side of \eqref{dr3pn3} contribute to the 4PN-accurate Routhian.

By virtue of the above result, to get the 4PN-accurate Routhian
it is enough to eliminate from the 4PN-accurate reduced Hamiltonian \eqref{hrede3}
the TT part ${\piTT ij}$ of the field momentum by means of the relation \eqref{piTTbyhTT}.
Making use of Eqs.\ \eqref{hred-h4-h6}--\eqref{hred-h12} and Eq.\ \eqref{piTTbyhTT},
the 4PN-accurate Routhian reads
\begin{align}
\label{Routhian3}
R_\mathrm{\le 4PN}\big[\xa,\pa,{\hTT ij},{\hTTdot ij}\big]
\nonumber\\[1ex]&\kern-15ex
= \int\md^dx\,\mathfrak{r}_\mathrm{\le 4PN}\big[\x;\xa,\pa,{\hTT ij},{\hTTdot ij}\big],
\end{align}
where
\begin{align}
\mathfrak{r}_\mathrm{\le 4PN}\big[\x;\xa,\pa,{\hTT ij},{\hTTdot ij}\big]
&= \sum_a m_a\da + h_{(4)}\big(\x;\xa,\pa\big)
\nonumber\\&\kern-15ex
+ h_{(6)}\big(\x;\xa,\pa\big)
+ h_{(8)}\big(\x;\xa,\pa,{\hTT ij}\big)
\nonumber\\[1ex]&\kern-15ex
+ \mathfrak{r}_{(10)}\big(\x;\xa,\pa,{\hTT ij},{\hTTdot ij}\big)
\nonumber\\[1ex]&\kern-15ex
+ \mathfrak{r}_{(12)}\big[\x;\xa,\pa,{\hTT ij},{\hTTdot ij}\big].
\end{align}
The Routhian densities $h_{(4)}$, $h_{(6)}$, and $h_{(8)}$ are identical with the corresponding densities
of the reduced Hamiltonian \eqref{hrede3} and they are given in Eqs.\ \eqref{hred-h4-h6} and \eqref{hred-h8}.
The 3PN density $\mathfrak{r}_{(10)}$, after some more integrations by parts, can be written as
\begin{align}
\label{barr10}
\mathfrak{r}_{(10)}\big(\x;\xa,\pa,{\hTT ij},{\hTTdot ij}\big) &= \varkappa_{(10)}\big(\x;\xa,\pa\big)
\nonumber\\[1ex]&\kern-20ex
+ 2{\gradfii i}{\Viii j}\big(\fii{\pitiii ij}\big)^\mathrm{TT}
+ {\Bvi ij}(\x;\xa,\pa){\hTT ij}
\nonumber\\[1ex]&\kern-20ex
+ \frac{1}{2(d-1)}\fii{\hTT ij}\Deltad{\hTT ij}
+ \frac{2(d-2)}{d-1}{\gradfii i}{\Viii j}{\hTTdot ij}
\nonumber\\[1ex]&\kern-20ex
-\frac{1}{4} \big({\hTTdot ij}\big)^2.
\end{align}
The 4PN-level Routhian density $\mathfrak{r}_{(12)}$ can be easily obtained by replacing in the 4PN-level Hamiltonian density $h_{(12)}$
[given in Eq.\ \eqref{hred-h12}] the field momentum ${\piTT ij}$ by the two first terms from the right-hand side of Eq.\ \eqref{piTTsplitting},
\begin{align}
\label{barr12}
\mathfrak{r}_{(12)}\big[\x;\xa,\pa,{\hTT ij},{\hTTdot ij}\big]
\nonumber\\[1ex]&\kern-23ex
= h_{(12)}\Big[\x;\xa,\pa,{\hTT ij},\frac{1}{2} {\hTTdot ij} + \frac{d-2}{d-1} \big(\fii{\pitiii ij}\big)^\mathrm{TT}\Big].
\end{align}

\section{4PN-accurate conservative matter Hamiltonian}

In this section we consider the 4PN-accurate \emph{conservative matter} Hamiltonian,
which depends only on the particle variables $\xa$, $\pa$.
To obtain the conservative Hamiltonian,
from now on we use only the time-symmetric part of the field function $\hTT ij$.
To get the matter Hamiltonian we eliminate $\hTT ij$ (and $\hTTdot ij$)
by replacing it by (time-symmetric) solution of the field equation \eqref{FEhTT}.
We thus treat $\hTT ij$ (and its time derivatives) as functions of matter variables $\xa$, $\pa$.
All time derivatives of $\xa$, $\pa$ involved in $\hTT ij$ (and $\hTTdot ij$)
are eliminated by means of the lower-order equations of motion.

We thus start from plugging into the 4PN-accurate Routhian \eqref{Routhian3}
the time-symmetric solution of the field equation for the function ${\hTT ij}$.
This way we obtain the 4PN-accurate conservative matter Hamiltonian
\begin{align}
\label{hm0}
H_\mathrm{\le 4PN}^\textrm{matter}(\xa,\pa)
\nonumber\\[1ex]&\kern-10ex
:= R_\mathrm{\le 4PN}\big[\xa,\pa,{\hTT ij}(\x;\xa,\pa),{\hTTdot ij}(\x;\xa,\pa)\big],
\end{align}
where we have assumed that all time derivatives of $\xa$ and $\pa$ present in $\hTT ij$ and $\hTTdot ij$
were eliminated by means of lower-order equations of motion
(to get the 4PN-accurate results we only need to use Newtonian and 1PN equations of motion).
We reorganize now density of the Hamiltonian \eqref{hm0}
by employing the field equation \eqref{FEhTT} for the function ${\hTT ij}$.

We first split ${\hTT ij}$ into the leading-order term ${\hTTiv ij}$ and the rest ${\dhTT ij}$:
\begin{align}
\label{hTTsplitting}
{\hTT ij} &= {\hTTiv ij} + {\dhTT ij},
\nonumber\\[1ex]
{\dhTT ij} &= {\ore 6}, \quad {\dhTT ij} = {\ora {2-d}} \quad\text{for $r\to\infty$}.
\end{align}
The part of the 2PN density $h_{(8)}$ which depends on $\hTT ij$ reads
[see Eq.\ \eqref{hred-h8}] 
\begin{align}
\label{hd2pn1}
{\delta h}_1 := \frac{1}{2}{\Siv ij}{\hTT ij}
+ \frac{1}{4}({\gradhTT ijk})^2,
\end{align}
where ${\Siv ij}$ is the source term for ${\hTTiv ij}$ [see Eq.\ \eqref{feqhTT4}].
Making use of the splitting \eqref{hTTsplitting},
Eq.\ \eqref{hd2pn1} can be rewritten as
\begin{align}
\label{hd2pn2}
{\delta h}_1 &= \frac{1}{2} {\Siv ij}{\hTTiv ij}
+ \frac{1}{4} ({\gradhTTiv ijk})^2
+ \frac{1}{2}{\gradhTTiv ijk}{\graddhTT ijk}
\nonumber\\[1ex]&\quad
+ \frac{1}{4}({\graddhTT ijk})^2
+ \frac{1}{2}{\Siv ij}{\dhTT ij}.
\end{align}
We still rewrite Eq.\ \eqref{hd2pn2} in the following way:
\begin{align}
\label{hd2pn3}
{\delta h}_1 &= \frac{1}{2} {\Siv ij}{\hTTiv ij}
+ \frac{1}{4} ({\gradhTTiv ijk})^2
+ \frac{1}{2} (\Deltad{\hTTiv ij}) {\dhTT ij}
\nonumber\\[1ex]&\quad
+ \frac{1}{2} {\gradhTTiv ijk}{\graddhTT ijk}
+ \frac{1}{4} ({\graddhTT ijk})^2
\nonumber\\[1ex]&\quad
+ \frac{1}{2}\big({\Siv ij} - {\STTiv ij}\big) {\dhTT ij},
\end{align}
where we have employed the field equation \eqref{feqhTT4} for the function ${\hTTiv ij}$.
Finally, making use of the identity
\be
(\Deltad{\hTTiv ij}) {\dhTT ij} = \partial_k \big({\gradhTTiv ijk}{\dhTT ij}\big)
- {\gradhTTiv ijk}{\graddhTT ijk},
\ee
equation \eqref{hd2pn3} takes the form
\begin{align}
\label{hd2pn4}
{\delta h}_1 = \frac{1}{2} {\Siv ij}{\hTTiv ij} + \frac{1}{4} ({\gradhTTiv ijk})^2
+ \frac{1}{4} ({\graddhTT ijk})^2
+ \partial_k \mathrm{ED}_{1k},
\end{align}
where the exact divergence $\mathrm{ED}_{1k}$ is defined as
[we have used here the explicit form \eqref{defTT} of the TT projection operator]
\begin{align}
\label{ED1}
\mathrm{ED}_{1k} &:= \frac{1}{2} {\gradhTTiv ijk}{\dhTT ij}
+ {\dhTT ik} \Big( \Deltad^{-1}{\gradSiv ijj}
\nonumber\\[1ex]&\quad
- \frac{1}{2(d-1)}\Deltad^{-1}{\gradSiv jji}
- \frac{d-2}{2(d-1)}\Deltad^{-2}{\gradiiiSiv jljli} \Big).
\end{align}
Because $\mathrm{ED}_{1k}$ decays as $1/r^{2d-3}$ for $r\to\infty$, it does not contribute to the Hamiltonian.
Therefore the Hamiltonian density ${\delta h}_1$
can be replaced by ${\delta h}'_1$, where
\begin{align}
\label{hd2pn5}
{\delta h}'_1 &:= \frac{1}{2} {\Siv ij}{\hTTiv ij} + \frac{1}{4} ({\gradhTTiv ijk})^2 + \frac{1}{4} ({\graddhTT ijk})^2,
\nonumber\\[1ex]
({\graddhTT ijk})^2 &\phantom{:}= {\ore {12}}.
\end{align}
Because
\begin{align}
({\graddhTT ijk})^2 &= -{\dhTT ij}(\Deltad{\dhTT ij}) + \partial_k\mathrm{ED}_{2k},
\nonumber\\[1ex]
\mathrm{ED}_{2k} &:= {\dhTT ij}{\graddhTT ijk},
\end{align}
and $\mathrm{ED}_{2k}={\ora {3-2d}}$ for $r\to\infty$,
instead of the density \eqref{hd2pn5} one can also use
\be
\label{deltar2bis}
{\delta h}''_1 := \frac{1}{2} {\Siv ij}{\hTTiv ij} + \frac{1}{4} ({\gradhTTiv ijk})^2
- \frac{1}{4} {\dhTT ij}(\Deltad{\dhTT ij}).
\ee

The part of the 3PN Routhian density $\mathfrak{r}_{10}$ [see Eq.\ \eqref{barr10}]
which depends on $\hTT ij$ or $\hTTdot ij$ reads
\begin{align}
&{\Bvi ij}{\hTT ij}
+ \frac{1}{2(d-1)}\fii{\hTT ij}\Deltad{\hTT ij}
\nonumber\\[1ex]
&+ \frac{2(d-2)}{d-1}{\gradfii i}{\Viii j}{\hTTdot ij}
-\frac{1}{4} \big({\hTTdot ij}\big)^2.
\end{align}
Let us denote the sum of these terms and of the expression \eqref{deltar2bis} by ${\delta h}_2$,
\begin{align}
\label{r23pn1}
{\delta h}_2 &:= \frac{1}{2} {\Siv ij}{\hTTiv ij}
+ \frac{1}{4} ({\gradhTTiv ijk})^2
+ B_{(6)ij}{\hTT ij}
\nonumber\\[1ex]&\qquad
+ \frac{1}{2(d-1)}\fii{\hTT ij}\Deltad{\hTT ij}
- \frac{1}{4} {\dhTT ij}(\Deltad{\dhTT ij})
\nonumber\\[1ex]&\qquad
+ \frac{2(d-2)}{d-1}{\gradfii i}{\Viii j}{\hTTdot ij}
- \frac{1}{4} \big({\hTTdot ij}\big)^2.
\end{align}
In the field equation \eqref{FEhTT} we split the source term ${\STT ij}$
into the leading-order part ${\STTiv ij}$ and the rest ${\dSTT ij}$,
\be
\label{feqhTT1}
\Box_{d+1}{\hTT ij} = {\STTiv ij} + {\dSTT ij},
\qquad {\dSTT ij} = {\ore 6}.
\ee
After substituting the splitting \eqref{hTTsplitting} into \eqref{feqhTT1} we obtain
(remembering that $\Box_{d+1}=-\partial^2_t+\Deltad$)
\be
\label{feqhTT2}
\Deltad{\hTT ij} = {\hTTddot ij} + {\STTiv ij} + {\dSTT ij}.
\ee
By virtue of Eqs.\ \eqref{hTTsplitting} and \eqref{feqhTT4},
Eq.\ \eqref{feqhTT2} finally leads to
\be
\label{feqhTT3}
\Deltad{\dhTT ij} = {\hTTddot ij} + {\dSTT ij}.
\ee
Making use of Eqs.\ \eqref{hTTsplitting} and \eqref{feqhTT3},
we can prove the following relation:
\begin{align}
\label{ttd}
- \frac{1}{4} \big({\hTTdot ij}\big)^2
&- \frac{1}{4} {\dhTT ij}(\Deltad{\dhTT ij})
\nonumber\\[1ex]
= &- \frac{\md}{\md t} \bigg( {\dhTT ij} \Big(\frac{1}{4}{\dhTTdot ij}
+ \frac{1}{2}{\hTTivdot ij}\Big) \bigg)
\nonumber\\[1ex]
&- \frac{1}{4} ({\hTTivdot ij})^2
+ \frac{1}{4} {\hTTivddot ij}{\dhTT ij}
- \frac{1}{4} {\dhTT ij}{\dSTT ij}.
\end{align}
With the aid of the above relation, after dropping the total time derivative
on the right-hand side of Eq.\ \eqref{ttd},
the density ${\delta h}_2$ from \eqref{r23pn1} can be replaced by
\begin{align}
\label{r23pn2}
{\delta h}'_2 &:= \frac{1}{2} {\Siv ij}{\hTTiv ij} + \frac{1}{4} ({\gradhTTiv ijk})^2
+ B_{(6)ij}{\hTT ij}
\nonumber\\[1ex]&\quad
+ \frac{1}{2(d-1)}\fii{\hTT ij}\Deltad{\hTT ij}
\nonumber\\[1ex]&\quad
+ \frac{2(d-2)}{d-1}{\gradfii i}{\Viii j}{\hTTdot ij}
- \frac{1}{4} ({\hTTivdot ij})^2
\nonumber\\[1ex]&\quad
+ \frac{1}{4} {\hTTivddot ij}{\dhTT ij}
- \frac{1}{4} {\dhTT ij}{\dSTT ij}.
\end{align}

The expression \eqref{r23pn2} contains all terms related with the 2PN-level and 3PN-level parts of the Routhian \eqref{Routhian3}
which depend on $\hTT ij$ or $\hTTdot ij$. Taking into account this expression and all other terms entering
the Routhian \eqref{Routhian3}, one can write the 4PN-accurate conservative matter Hamiltonian \eqref{hm0} in the form
\begin{align}
\label{hm1}
H_\mathrm{\le 4PN}^\textrm{matter}(\xa,\pa)
&= \int\md^dx\,h_\mathrm{\le 4PN}^\textrm{matter}(\x;\xa,\pa),
\end{align}
where
\begin{align}
h_\mathrm{\le 4PN}^\textrm{matter}(\x;\xa,\pa) &= \sum_a m_a\da + h_{(4)}(\x;\xa,\pa)
\nonumber\\&\kern-10ex
+ h_{(6)}(\x;\xa,\pa) + h_{(8)}^\textrm{matter}(\x;\xa,\pa)
\nonumber\\[1ex]&\kern-10ex
+ h_{(10)}^\textrm{matter}(\x;\xa,\pa) + h_{(12)}^\textrm{matter}(\x;\xa,\pa).
\end{align}
Here the densities $h_{(4)}$ (at the Newtonian order) and $h_{(6)}$ (at the 1PN order)
can be found in Eqs.\ \eqref{hred-h4-h6} and the densities at the 2PN, 3PN, and 4PN orders are as follows:
\begin{subequations}
\label{hm1e}
\begin{align}
h_{(8)}^\textrm{matter}(\x;\xa,\pa) &= \varkappa_{(8)}(\x;\xa,\pa) + \frac{1}{2}{\Siv ij}{\hTTiv ij}
\nonumber\\[1ex]&\kern-10ex
+ \frac{1}{4}({\gradhTTiv ijk})^2,
\\[2ex]
h_{(10)}^\textrm{matter}(\x;\xa,\pa) &= \varkappa_{(10)}\big(\x;\xa,\pa\big)
\nonumber\\[1ex]&\kern-10ex
+ 2{\gradfii i}{\Viii j}\big(\fii{\pitiii ij}\big)^\mathrm{TT}
+ B_{(6)ij}{\hTT ij}
\nonumber\\[1ex]&\kern-10ex
+ \frac{1}{2(d-1)}\fii{\hTT ij}\Deltad{\hTT ij}
\nonumber\\[1ex]&\kern-10ex
+ \frac{2(d-2)}{d-1}{\gradfii i}{\Viii j}{\hTTdot ij}
- \frac{1}{4} ({\hTTivdot ij})^2,
\\[2ex]
h_{(12)}^\textrm{matter}(\x;\xa,\pa)&
\nonumber\\[1ex]&\kern-15ex
= h_{(12)}\Big[\x;\xa,\pa,{\hTT ij},\frac{1}{2} {\hTTdot ij} + \frac{d-2}{d-1} \big(\fii{\pitiii ij}\big)^\mathrm{TT}\Big]
\nonumber\\[1ex]&\kern-10ex
+ \frac{1}{4} {\hTTivddot ij}{\dhTT ij}
- \frac{1}{4} {\dhTT ij}{\dSTT ij}.
\end{align}
\end{subequations}

\section{4PN-accurate near-zone conservative matter Hamiltonian}

We employ now the crucial near-zone (time-symmetric) PN expansion \eqref{hTTe} of the field variable $\hTT ij$,
i.e., we use ${\hTT ij}={\hTTiv ij}+{\dhTT ij}$  with ${\dhTT ij}={\hTTvi ij} + {\ore 8}$.
We also employ [see Eq.\ \eqref{feqhTT1}]
\be
\label{dSTTe}
{\dSTT ij} = {\STTvi ij} + {\ore 8},
\ee
where the next-to-leading-order source function $\Svi ij$ is defined in Eq.\ \eqref{STT6}.
After plugging \eqref{hTTe} and \eqref{dSTTe} into the 4PN-accurate conservative matter Hamiltonian of Eq.\ \eqref{hm1},
we obtain the 4PN-accurate \textit{near-zone} conservative matter Hamiltonian,
\begin{align}
\label{hm2}
H_\mathrm{\le 4PN}^\textrm{near-zone}(\xa,\pa)
&= \int\md^dx\,h_\mathrm{\le 4PN}^\textrm{near-zone}(\x;\xa,\pa),
\end{align}
where
\begin{align}
h_\mathrm{\le 4PN}^\textrm{near-zone}(\x;\xa,\pa) = \sum_a m_a\da & + h_{(4)}(\x;\xa,\pa)
\nonumber\\[1ex]&\kern-20ex
+ h_{(6)}(\x;\xa,\pa) + h_{(8)}^\textrm{near-zone}(\x;\xa,\pa)
\nonumber\\[1ex]&\kern-20ex
+ h_{(10)}^\textrm{near-zone}(\x;\xa,\pa) + h_{(12)}^\textrm{near-zone}(\x;\xa,\pa).
\end{align}
Here again the densities $h_{(4)}$ and $h_{(6)}$ can be found in Eqs.\ \eqref{hred-h4-h6}, the 2PN density
$h_{(8)}^\textrm{near-zone}(\x;\xa,\pa)=h_{(8)}^\textrm{matter}(\x;\xa,\pa)$, where $h_{(8)}^\textrm{matter}$ is given in Eq.\ \eqref{hm1e},
and the 3PN/4PN densities $h_{(10)}^\textrm{near-zone}$ and $h_{(12)}^\textrm{near-zone}$
follow from the sum of densities $h_{(10)}^\textrm{matter}+h_{(12)}^\textrm{matter}$
after plugging the expansions \eqref{hTTe} and \eqref{dSTTe} into the latter.
The 3PN-level density reads
\begin{align}
\label{nzh10}
h_{(10)}^\textrm{near-zone}(\x;\xa,\pa) &= \varkappa_{(10)}\big(\x;\xa,\pa\big)
\nonumber\\[1ex]&\kern-15ex
+ 2{\gradfii i}{\Viii j}\big(\fii{\pitiii ij}\big)^\mathrm{TT}
+ {\Bvi ij}{\hTTiv ij}
\nonumber\\[1ex]&\kern-15ex
+ \frac{1}{2(d-1)}\fii{\hTTiv ij}\Deltad{\hTTiv ij}
\nonumber\\[1ex]&\kern-15ex
+ \frac{2(d-2)}{d-1}{\gradfii i}{\Viii j} {\eldt {\hTTivdot ij}0}
- \frac{1}{4}({\eldt {\hTTivdot ij}0})^2,
\end{align}
where the notation ${\eldt {\hTTivdot ij}0}$ means elimination of time derivatives
$\dot{\x}_a$, $\dot{\p}_a$ in ${\hTTivdot ij}$ by means of Newtonian equations of motion.

The 4PN-order density we split into two parts,
\be
h_{(12)}^\textrm{near-zone}(\x;\xa,\pa) = h_{(12)}^1(\x;\xa,\pa) + h_{(12)}^2(\x;\xa,\pa).
\ee
The first part $h_{(12)}^1$ reads
\begin{widetext}
\begin{align}
\label{h12-1}
h_{(12)}^1(\x;\xa,\pa) &= \varkappa_{(12)}^1(\x;\xa,\pa) + \varkappa_{(12)}^2(\x;\xa,\pa,{\hTTiv ij}) + \varkappa_{(12)}^3[\x;\xa,\pa,{\hTTiv ij}]
\nonumber\\[1ex]&\kern-10ex
- \fii \Big(\big(\fii{\pitiii ij}\big)^\mathrm{TT}\Big)^2
+ \frac{1}{d-1} \left( \frac{(3d-2)(3d-4)}{8(d-1)}\fii^2
+ d\SivI - \frac{(3d-4)(d-2)}{4(d-1)}\SivII \right) {\pitiii ij}\big(\fii{\pitiii ij}\big)^\mathrm{TT}
\nonumber\\[1ex]&\kern-10ex
+ \bigg\{ \bigg(\frac{3(d-2)^2}{4(d-1)^2}{\gradSivII j}-{\gradSivI j}
+ \frac{7(d-2)^2}{4(d-1)^2}\fii{\gradfii j}\bigg){\Viii i}
+ 2(d-2){\gradfii j}{\Vv i} - \frac{4}{d}{\gradViii kk}{\hTTiv ij}
\nonumber\\[1ex]&\kern-10ex
+ 2{\gradViii ik}{\hTTiv jk} - {\Viii k}{\gradhTTiv ijk}
\bigg\}{\eldt {\hTTivdot ij}0} - \frac{d-3}{4(d-1)}\fii\big({\eldt {\hTTivdot ij}0}\big)^2
+ \frac{2(d-2)}{d-1}{\gradfii i}{\Viii j} {\eldt {\hTTivdot ij}1}
\nonumber\\[1ex]&\kern-10ex
- \frac{1}{2}{\eldt {\hTTivdot ij}0}{\eldt {\hTTivdot ij}1}.
\end{align}
\end{widetext}
The density $h_{(12)}^1$ depends on the following functions:
$\fii$, $\SivI$ and $\SivII$ [they determine, via Eq.\ \eqref{phi4byS}, the function $\fiv$],
${\Viii i}$ and ${\Vv i}$ [they define, through Eq.\ \eqref{pitbyv}, the functions ${\pitiii ij}$ and ${\pitv ij}$],
${\hTTiv ij}$, and $\fvinoTT$, $\fviTT$. The function $\fvinoTT$ is the solution of Eq.\ \eqref{lapphi61};
the equation fulfilled by the function $\fviTT$ one obtains from Eq.\ \eqref{lapphi62b} for the function $\fvibTT$ after replacing
in the latter ${\hTT ij}$ by the leading-order ${\hTTiv ij}$,
\be
\label{lapphi62}
\Deltad\fviTT = \frac{d-2}{d-1}{\ggradfii ij}{\hTTiv ij}.
\ee
Formulas needed to express in an explicit way all these functions
in terms of $\xa$, $\pa$, and $\x$ are given in Appendix \ref{functions}.
The last two terms in Eq.\ \eqref{h12-1} are related with the last two terms in Eq.\ \eqref{nzh10};
${\eldt {\hTTivdot ij}1}$ means here elimination of the time derivatives $\dot{\x}_a$, $\dot{\p}_a$ by means of 1PN equations of motion.
In the rest of this section and in Sec.\ \ref{computation} all time derivatives of $\xa$, $\pa$ are eliminated by means
of Newtonian equations of motion; therefore from now on we drop the notation ${\eldt \cdot0}$.

The second part $h_{(12)}^2$ of the 4PN density $h_{(12)}^\textrm{near-zone}$
is the 4PN-level contribution coming from the density \eqref{r23pn2}; it equals
\begin{align}
\label{h12-2}
h_{(12)}^2(\x;\xa,\pa) &= -\frac{d-2}{d-1} \fii{\pitiii ij}{\hTTvidot ij}
\nonumber\\[1ex]&\kern-5ex
+ \frac{1}{2(d-1)} \fii \Big( {\hTTiv ij}\,\Deltad{\hTTvi ij} + {\hTTvi ij}\,\Deltad{\hTTiv ij} \Big)
\nonumber\\[1ex]&\kern-5ex
+ \Big(B_{(6)ij} + \frac{1}{4} {\hTTivddot ij} - \frac{1}{4}{\STTvi ij}\Big) {\hTTvi ij}.
\end{align}
The function ${\hTTvi ij}$, according to Eq.\ \eqref{feqhTT6}, is the sum of the two terms,
\begin{align}
\label{hTT6}
{\hTTvi ij} &= {\CTTvi ij} + {\ilhTTivddot ij},
\end{align}
where
\be
\label{CTT6}
{\CTTvi ij} = \Deltad^{-1}{\STTvi ij}.
\ee
After plugging Eq.\ \eqref{hTT6} into Eq.\ \eqref{h12-2}
and shifting some time derivatives,\footnote{
Shifting time derivatives means replacing $\dot{A}B$ by $-A\dot{B}$. This implies adding a total time derivative to the Hamiltonian density
and on the level of Hamiltonian is equivalent to performing a canonical transformation.} 
the density $h_{(12)}^2$ can be rewritten as
\begin{align}
\label{h12splitting}
h_{(12)}^2 &= h_{(12)}^{2,1} + h_{(12)}^{2,2} + h_{(12)}^{2,3},
\end{align}
where\footnote{
The density $h_{(12)}^{2,2}$, after ignoring its last term, is identical to the density $r^2_{\textrm{4PN}}$
introduced in Eq.\ (3.4) of Ref.\ \cite{DJS14} (the reason of omission of this term in \cite{DJS14} is explained
at the end of Sec.\ \ref{integral-of-h12-2,2}).}
\begin{widetext}
\begin{subequations}
\begin{align}
\label{h12-2-1}
h_{(12)}^{2,1} &=
\frac{1}{2(d-1)}\fii{\hTTiv ij}{\STTvi ij}
+ \bigg({\Bvi ij} + \frac{1}{2(d-1)}\fii(\Deltad{\hTTiv ij}) - \frac{1}{4}{\STTvi ij} \bigg){\CTTvi ij}
- \frac{d-2}{d-1}\fii{\pitiii ij}{\CTTdotvi ij},
\\[1ex]
\label{h12-2-2}
h_{(12)}^{2,2} &=
\frac{1}{2(d-1)}\fii{\hTTiv ij}{\hTTivddot ij}
- \frac{1}{4}{\hTTivdot ij} {\ilhTTivdddot ij}
+ \bigg( \frac{1}{2(d-1)}\fii(\Deltad{\hTTiv ij}) + \frac{d-2}{d-1}\,\partial_t\big(\fii{\pitiii ij}\big)
\bigg) {\ilhTTivddot ij}
\nonumber\\[1ex]&\quad
+ {\Bvi ij}{\ilhTTivddot ij},
\\[1ex]
\label{h12-2-3}
h_{(12)}^{2,3} &= \frac{1}{4}\Big({\CTTvi ij}{\hTTivddot ij}-{\STTvi ij}({\ilhTTivddot ij})\Big).
\end{align}
\end{subequations}
\end{widetext}

\section{Computation of the 4PN-accurate near-zone conservative matter Hamiltonian}
\label{computation}

The bulk of computations we did to derive the explicit form of the 4PN-accurate near-zone conservative matter Hamiltonian,
i.e.\ to perform integration in Eq.\ \eqref{hm2},
was performed in $d=3$ dimensions, where our working horse was Riesz-implemented Hadamard regularization
supplemented by a Hadamard ``partie finie'' concept of a function at its singular point.
The results of global (i.e., extending to the whole ${\mathbb R}^3$ space)
three-dimensional integrations were then corrected in two different ways:
(i) the UV divergences were locally (i.e., within small balls surrounding particles' positions) recomputed by using dimensional regularization;
(ii) the IR divergences were also ``locally'' (i.e., outside a large ball enclosing particles, or in a neighborhood of $r=\infty$)
recomputed by introducing an additional regularization factor $(r/s)^B$ (with a new IR length scale $s$),
which modifies the behavior of the part of the field function ${\hTTvi ij}$ which diverges at $r\to\infty$ [see Eq.\ \eqref{infdi3}].
The details of regularization procedure are explained in Appendix \ref{Regularization}.

Dimensional regularization introduces a natural length scale $\ell_0$,
which relates gravitational constants in $d$ and 3 dimensions [see Eq.\ \eqref{ell0}].
As explained below correction of the UV divergences by means of dimensional regularization
produces some poles proportional to $1/(d-3)$ together with $\ell_0$-dependent logarithms $\ln(r_{12}/\ell_0)$.
All these poles and logarithms were removed
by adding to the Hamiltonian a total time derivative.
The terms which contribute to IR divergences depend, after regularization, on the IR length scale $s$.
Moreover, as we also explain below, different ways of regularizing IR divergences
lead to different results, so the final result of IR regularization is ambiguous.
We were able to express this ambiguity in terms of only one dimensionless constant,
which we denote by $C$.

The UV divergences is not the only problem caused by usage of distributional sources. 
General relativity uses standard algebraic and differential calculus of ordinary functions,
which includes e.g.\ the Leibniz rule.
By introducing distributional sources we violate this framework:
in some places we have to use distributional differentiation which does not fulfill the Leibniz rule
[see Appendix \ref{gendiff}].
The field equations \eqref{delta-fii-fiv}, \eqref{deltaSivI/II}, and \eqref{lapphi6b}--\eqref{grad-pitiii-ixb}
tell us that for the second (space and time) derivatives of $\phi_{(n)}$
and for the first (space and time) derivatives of ${\tilde \pi}^{ij}_{(n)}$ the distributional derivative has to be applied.
The same also holds for $h^{\rm TT}_{(n)ij}$ and $\pi^{ij}_{(n)\rm TT}$, respectively,
taking into account the Eqs.\ \eqref{feqhTT4}--\eqref{feqhTT6} and \eqref{piTTbyhTT}.
The differentiation of any product of these functions goes with the Leibniz rule.

The 4PN-accurate near-zone Hamiltonian is the sum of Hamiltonians at different PN orders,
\begin{align}
H_\mathrm{\le4PN}^\textrm{near-zone ($s$)}(&\xa,\pa;C) = \sum_a m_a
+ \hn(\xa,\pa)
\nonumber\\
+ \hi(&\xa,\pa) + \hii(\xa,\pa)
\nonumber\\[1ex]
+ \hiii(&\xa,\pa) + H_\mathrm{4PN}^\textrm{near-zone ($s$)}(\xa,\pa;C),
\end{align}
where we have introduced notation which indicates that the 4PN near-zone Hamiltonian
depends on the IR regularization scale $s$ and on one dimensionless constant $C$
parametrizing ambiguity in the regularization of IR divergences.
Hamiltonians $\hn$ through $\hiii$ were uniquely recomputed by us
using three-dimensional Riesz-implemented Hadamard regularization
supplemented by dimensional regularization.
These Hamiltonians do not develop IR divergences.

\subsection{Computation of the 3PN-accurate Hamiltonian}

The three-dimensional 3PN-accurate Hamiltonian,
\begin{align}
H_\mathrm{\le3PN}(\xa,\pa) &= \sum_a m_a
+ \hn(\xa,\pa)
\nonumber\\&\quad
+ \hi(\xa,\pa) + \hii(\xa,\pa)
\nonumber\\[1ex]&\quad
+ \hiii(\xa,\pa),
\end{align}
was computed for the first time in the series of papers \cite{JS98,JS99,JS00a,DJS00,DJS2001}.
We have recomputed it here and got the result identical with the previously obtained.
The explicit formulas for the three-dimensional Hamiltonians $\hn$ through $\hiii$ written in general reference frame
are given at the end of the current section.


In the computation of the 3PN/4PN Hamiltonians
we eliminate time derivatives of $\xa$ and $\pa$ [present in Eqs.\ \eqref{nzh10} and \eqref{h12-1}]
by means of lower-order equations of motion: we use Newtonian [in both \eqref{nzh10} and \eqref{h12-1}]
or 1PN [in \eqref{h12-1}] equations of motion.
To perform dimensional regularization of UV divergences
we have to use the $d$-dimensional version of these equations, which follows from $d$-dimensional Hamiltonians.
We have explicitly computed the Newtonian and 1PN Hamiltonians in $d$ dimensions. They read
[notation ``$+(1\leftrightarrow2)$'' used here means adding to each term
another term obtained by the exchange of the bodies' labels]
\begin{subequations}
\begin{align}
\hn(\xa,\pa) &= \frac{\pipi}{2m_1} - \frac{\kappa(d-2)}{4(d-1)}\frac{m_1m_2}{r_{12}^{d-2}}
+ (1\leftrightarrow2),
\\
\hi(\xa,\pa) &= -\frac{\pipip^2}{8m_1^3}
+ \frac{\kappa}{4(d-1)}\bigg(\frac{1}{2}(3d-2)\pipii
\nonumber\\[1ex]
&\kern-8ex - d\,\frac{m_2}{m_1}\pipi + \frac{1}{2}(d-2)^2\npi\npii\bigg)\frac{1}{r_{12}^{d-2}}
\nonumber\\[1ex]
&\kern-8ex + \frac{\kappa^2(d-2)^2}{8(d-1)^2} \frac{m_1^2 m_2}{r_{12}^{2d-4}}
+ (1\leftrightarrow2),
\end{align}
\end{subequations}
where we have introduced
\be
\label{kappa}
\kappa := \frac{\Gamma(d/2-1)}{4\pi^{d/2}}.
\ee
Let us note that $\kappa=1/(4\pi)$ in $d=3$ dimensions.

\subsection{Integral of $h_{(12)}^1$}

The part $h_{(12)}^1$ of the 4PN density given in Eq.\ \eqref{h12-1} is UV divergent but it does not develop IR poles
with the exception of the term $\propto\fii({\hTTivdot ij})^2$.
This term contains however the multiplication factor $d-3$
which causes the IR divergence of $\fii({\hTTivdot ij})^2$ to not contribute to the Hamiltonian
(but the UV divergence of this term produces some nonzero contribution).
After employing the regularization procedures described in Appendix \ref{Regularization} we have obtained
\begin{align}
H_{\textrm{4PN}}^{\textrm{reg 1}}(\xa,\pa;d) &= \int\md^dx\,h_{(12)}^1
\nonumber\\[1ex]
&= \chi_{(12)}^{1,1}(\xa,\pa)
+ \chi_{(12)}^{1,2}(\xa,\pa)\ln\frac{r_{12}}{\bar{\ell}_0}
\nonumber\\[1ex]&\quad
+ \frac{\chi_{(12)}^{1,3}(\xa,\pa)}{d-3}
+ \mathcal{O}(d-3),
\end{align}
where for convenience we have introduced a new dimensional-regularization (DR) scale $\bar{\ell}_0$
related with the original scale $\ell_0$ defined in Eq.\ \eqref{ell0} by the relation
\be
\bar{\ell}_0 = \frac{\ell_0}{2\sqrt{\pi}}e^{-\gamma_{\text{E}}/2}.
\ee
Here $\gamma_{\text{E}}$ denotes the Euler-Mascheroni constant.

\subsection{Integral of $h_{(12)}^2$}

To compute the integral of $h_{(12)}^2$ we have to study asymptotics of the function ${\hTTvi ij}$ for $r\to\infty$.
The function ${\hTTvi ij}$ is the sum of two parts, ${\CTTvi ij}$ and ${\ilhTTivddot ij}$ [see Eq.\ \eqref{hTT6}].
By virtue of Eq.\ \eqref{STT6} the function ${\CTTvi ij}$ [defined in Eq.\ \eqref{CTT6}]
can be written more explicitly as
\begin{align}
\label{CTT6-1}
{\CTTvi ij} &= \bigg( \frac{1}{d-1}\fii{\hTTiv ij} 
+ \frac{2(d-2)}{d-1}\,\partial_t\,\Deltad^{-1}\big(\fii{\pitiii ij}\big)
\nonumber\\[1ex]&\qquad
+ 2 \Deltad^{-1}{\Bvi ij}
+ \frac{1}{d-1}\Deltad^{-1}\big(\fii\Deltad{\hTTiv ij}\big)
\bigg)^\mathrm{TT}.
\end{align}
Analysis of this equation shows that ${\CTTvi ij}={\ora {2-d}}$ for $r\to\infty$.
It is also not difficult to check that the inverse Laplacian ${\ilhTTivddot ij}={\ora {4-d}}$ when $r\to\infty$
(i.e., it diverges linearly in $d=3$ dimensions).
This behavior of ${\CTTvi ij}$ and ${\ilhTTivddot ij}$ for $r\to\infty$ implies that the term $h_{(12)}^{2,1}$ is convergent at infinity,
whereas the terms $h_{(12)}^{2,2}$ and $h_{(12)}^{2,3}$ develop IR divergences.

\subsubsection{Integral of $h_{(12)}^{2,1}$}

The integral of $h_{(12)}^{2,1}$ is convergent at spatial infinity
but it is difficult to express its integrand as an explicit function of $\x$ (and of $\xa$, $\pa$).
To do this we further transform the density $h_{(12)}^{2,1}$, after employing Eq.\ \eqref{CTT6-1},  
by making some additional integrations by parts in space and shifting time derivatives.
We also replace $\Deltad(\fii{\hTTiv ij})^\mathrm{TT}$ [which comes from the first term of Eq.\ \eqref{h12-2-1}]
by more explicit expression using the definition \eqref{defTT} of the TT projection operator.
We finally obtain
\begin{widetext}
\begin{align}
\label{h12-21}
h_{(12)}^{2,1} &= -\frac{d-2}{(d-1)^2}\,\partial_t\big(\fii{\hTTiv ij}\big)\,\big(\fii{\pitiii ij}\big)^\mathrm{TT}
+ \frac{1}{4(d-1)^2}\,\Big(
2{\gradfii i}{\gradfii j}{\hTTiv ik}{\hTTiv jk}
- ({\gradfii k})^2({\hTTiv ij})^2
\nonumber\\[1ex]&\quad
+ \fii^2 {\hTTiv ij}\Deltad{\hTTiv ij}
+ \frac{d-2}{d-1}\,{\ggradfii ij}\,{\hTTiv ij}\,\Deltad^{-1}\big({\ggradfii kl}{\hTTiv kl}\big) \Big)
- \frac{(d-2)^2}{(d-1)^2}\,\fii{\pitiii ij}\,
\partial_t^2\Big(\Deltad^{-1}\big(\fii{\pitiii ij}\big)\Big)^\mathrm{TT}
\nonumber\\[1ex]&\quad
+ \frac{2(d-2)}{d-1} \Big({\Bvi ij} + \frac{1}{2(d-1)}\,\fii\Deltad{\hTTiv ij}\Big)\,
\partial_t\Big(\Deltad^{-1}\big(\fii{\pitiii ij}\big)\Big)^\mathrm{\!\!TT}
+ h_{(12)}^{2,1,1},
\end{align}
where the last term equals
\begin{align}
\label{h12-211}
h_{(12)}^{2,1,1} &= \frac{1}{d-1}\,\fii{\hTTiv ij} \Big({\Bvi ij} + \frac{1}{2(d-1)}\,\fii\,\Deltad{\hTTiv ij}\Big)^\mathrm{TT}
\nonumber\\[1ex]&\quad
+ \Big( {\Bvi ij} + \frac{1}{2(d-1)}\,\fii\,\Deltad{\hTTiv ij} \Big)
\Biglb( \Deltad^{-1}\Big( {\Bvi ij} + \frac{1}{2(d-1)}\,\fii\,\Deltad{\hTTiv ij} \Big) \Bigrb)^\mathrm{TT}.
\end{align}
\end{widetext}
It is crucial to single out the last term in Eq.\ \eqref{h12-21} and to put it exactly in the form of Eq.\ \eqref{h12-211}.
Due to this it is possible to compute fully explicitly (in $d=3$ dimensions) all inverse Laplacians involved in Eq.\ \eqref{h12-211}.
Formulas needed to calculate in $d=3$ all inverse Laplacians and perform TT projections involved in Eqs.\ \eqref{h12-21} and \eqref{h12-211}
are given in Appendix \ref{functions}.

After performing regularization by means of the procedures
described in Appendix \ref{Regularization} we have
\begin{align}
H_{\textrm{4PN}}^{\textrm{reg 2,1}}(\xa,\pa;d) &= \int\md^dx\,h_{(12)}^{2,1}
\nonumber\\[1ex]
&= \chi_{(12)}^{2,1,1}(\xa,\pa)
+ \chi_{(12)}^{2,1,2}(\xa,\pa)\ln\frac{r_{12}}{\bar{\ell}_0}
\nonumber\\[1ex]&\quad
+ \frac{\chi_{(12)}^{2,1,3}(\xa,\pa)}{d-3}
+ \mathcal{O}(d-3).
\end{align}

\subsubsection{Integral of $h_{(12)}^{2,2}$}
\label{integral-of-h12-2,2}

The integral of the density $h_{(12)}^{2,2}$ is not convergent at spatial infinity and it also develops UV poles.
After making use of the procedures described in Appendix \ref{Regularization} one gets
\begin{align}
\label{h12-2-2-a}
\int\md^dx\,h_{(12)}^{2,2} &= \bar{\chi}_{(12)\textrm{amb}}^{2,2,1}(\xa,\pa;C)
\nonumber\\[1ex]&\quad
+ \chi_{(12)}^{2,2,2}(\xa,\pa)\ln\frac{r_{12}}{\bar{\ell}_0} + \frac{\chi_{(12)}^{2,2,3}(\xa,\pa)}{d-3}
\nonumber\\[1ex]&\quad
+ \chi_{(12)}^{2,2,4}(\xa,\pa)\ln\frac{r_{12}}{s}
+ \mathcal{O}(d-3).
\end{align}
Because of the ambiguity of the results of IR regularization discussed in the Appendix \ref{IRcorr},
the term $\bar{\chi}_{(12)\textrm{amb}}^{2,2,1}(\xa,\pa;C)$ depends on some undetermined dimensionless constant $C$ which parametrizes this ambiguity.
We have found that the coefficient $\chi_{(12)}^{2,2,4}$ of the logarithm $\ln(r_{12}/{s})$ can be written,
after adding a total time derivative, in a very specific form.
Thus we have shown that there exists a \textit{unique} term $q(\xa,\pa)$,
\begin{align}
q(\xa,\pa) &= \frac{21}{20} \frac{(m_1 + m_2)m_1 m_2^2}{(16\pi)^4 r_{12}^3} \npi
\nonumber\\[1ex]&\quad
+  \frac{1}{(16\pi)^3 r_{12}^2} \bigg[ \bigg(\frac{97 m_2}{20 m_1} - \frac{21}{16}\bigg) m_2 \npi^3
\nonumber\\[1ex]&\quad
+ \bigg(\frac{21 m_1}{16} - \frac{687 m_2}{40}\bigg) \npi^2 \npii
\nonumber\\[1ex]&\quad
+ \bigg(\frac{77}{48} - \frac{237 m_2}{40 m_1}\bigg) m_2 \npi \pipi
\nonumber\\[1ex]&\quad
+ \bigg(\frac{7 m_1}{48} + \frac{307 m_2}{40}\bigg) \npii \pipi
\nonumber\\[1ex]&\quad
+ \bigg(\frac{1597 m_2}{120} - \frac{7 m_1}{4}\bigg) \npi \pipii \bigg]
\nonumber\\[1ex]&\quad
+ (1\leftrightarrow2),
\end{align}
such that
\be
\label{h12-2-2-b}
\chi_{(12)}^{2,2,4}(\xa,\pa) + \frac{\md}{\md t}q(\xa,\pa) = F(\xa,\pa),
\ee
where
\be
\label{defF}
F(\xa,\pa) := \frac{2M}{5(16\pi)^2} (\dddot{I}_{ij})^2.
\ee
Here $I_{ij}$ is the Newtonian quadrupole moment of the binary system,
\be
\label{defI}
I_{ij} := \sum_a m_a \Big( x_a^i x_a^j - \frac{1}{3}\delta^{ij}\xa^2 \Big).
\ee
All time derivatives on the right-hand side of the formula \eqref{defF}
were eliminated by means of Newtonian equations of motion.
By virtue of Eq.\ \eqref{h12-2-2-b}
the result of Eq.\ \eqref{h12-2-2-a} we can rewrite in the following form:
\begin{align}
\label{h12-2-2-c}
\int\md^dx\,h_{(12)}^{2,2} &+ \frac{\md}{\md t}\Big(q(\xa,\pa)\ln\frac{r_{12}}{s}\Big) = \chi_{(12)\textrm{amb}}^{2,2,1}(\xa,\pa;C)
\nonumber\\[1ex]
&+ \chi_{(12)}^{2,2,2}(\xa,\pa)\ln\frac{r_{12}}{\bar{\ell}_0} + \frac{\chi_{(12)}^{2,2,3}(\xa,\pa)}{d-3}
\nonumber\\[1ex]
&+ F(\xa,\pa)\ln\frac{r_{12}}{s}
+ \mathcal{O}(d-3),
\end{align}
where
\begin{align}
\label{h12-2-2-d}
\chi_{(12)\textrm{amb}}^{2,2,1}(\xa,\pa;C) &:= \bar{\chi}_{(12)\textrm{amb}}^{2,2,1}(\xa,\pa;C)
\nonumber\\[1ex]&\quad
+ q(\xa,\pa)\frac{\md}{\md t}\Big(\ln\frac{r_{12}}{s}\Big).
\end{align}

We have found (see Appendix \ref{IRcorr}) that the ambiguity of the term $\chi_{(12)\textrm{amb}}^{2,2,1}(\xa,\pa;C)$ can be expressed,
up to adding a total time derivative, as a \emph{multiple of the term $F$} introduced in Eq.\ \eqref{defF},
\begin{align}
\label{h12-2-2-e}
\chi_{(12)\textrm{amb}}^{2,2,1}(\xa,\pa;C) &= \chi_{(12)}^{2,2,1}(\xa,\pa) + C\,F(\xa,\pa)
\nonumber\\[1ex]&\quad
+ \textrm{(total time derivative)}.
\end{align}
Therefore as the contribution of the density $h_{(12)}^{2,2}$ to the 4PN Hamiltonian we take
\begin{align}
\label{h12-2-2-f}
H_{\textrm{4PN}}^{\textrm{reg 2,2}}(\xa,\pa;d;s,C) &=  \chi_{(12)}^{2,2,1}(\xa,\pa)
\nonumber\\[1ex]&\kern-8ex
+ \chi_{(12)}^{2,2,2}(\xa,\pa)\ln\frac{r_{12}}{\bar{\ell}_0}
+ \frac{\chi_{(12)}^{2,2,3}(\xa,\pa)}{d-3}
\nonumber\\[1ex]&\kern-8ex
+ F(\xa,\pa)\Big(\ln\frac{r_{12}}{s} + C\Big)
+ \mathcal{O}(d-3).
\end{align}
Let us finally mention that the integral of the last term of the density $h_{(12)}^{2,2}$, ${\Bvi ij}{\ilhTTivddot ij}$,
though IR divergent, can be in fact regularized uniquely. This is so because the difference between results of different methods
of regularizing this term is a total time derivative.

\subsubsection{Integral of $h_{(12)}^{2,3}$}

The density $h_{(12)}^{2,3}$ is an exact divergence, it can be written in the form
\begin{align}
\frac{1}{4}\Big({\CTTvi ij}{\hTTivddot ij}-{\STTvi ij}({\ilhTTivddot ij})\Big)
= \frac{1}{4}\partial_k E_k,
\end{align}
where
\begin{align}
E_k := {\CTTvi ij}({\gradilhTTivddot ijk})-{\gradCTTvi ijk}({\ilhTTivddot ij}).
\end{align}
The surface integral associated with $E_k$ is not convergent at spatial infinity in $d=3$ dimensions.
However in $d$ dimensions $E_k\sim r^{5-2d}$ for $r\to\infty$, so the surface integral behaves like $r^{4-d}$
and it vanishes in the limit $r\to\infty$ when $d$ is large enough.
Therefore this term does not contribute to the Hamiltonian.

\subsection{Removing UV poles}
\label{removeUVpoles}

As the result of UV and IR regularizations described in the previous subsections,
we have obtained the 4PN near-zone Hamiltonian of the following form:
\begin{align}
H_\mathrm{4PN}^\textrm{reg}(\xa,\pa;d;s,C)
&= H_{\textrm{4PN}}^{\textrm{reg 1}}(\xa,\pa;d)
\nonumber\\[1ex]&\kern-14ex
+ H_{\textrm{4PN}}^{\textrm{reg 2,1}}(\xa,\pa;d) + H_{\textrm{4PN}}^{\textrm{reg 2,2}}(\xa,\pa;d;s,C)
\nonumber\\[1ex]&\kern-16ex
= \bar{\chi}_1(\xa,\pa) + \bar{\chi}_2(\xa,\pa)\ln\frac{r_{12}}{\bar{\ell}_0}
+ \frac{\bar{\chi}_3(\xa,\pa)}{d-3}
\nonumber\\[1ex]&\kern-14ex
+  F(\xa,\pa)\Big(\ln\frac{r_{12}}{s} + C\Big)
+ \mathcal{O}(d-3),
\end{align}
where
\begin{align}
\bar{\chi}_k(\xa,\pa) &:= \chi_{(12)}^{1,k}(\xa,\pa) + \chi_{(12)}^{2,1,k}(\xa,\pa)
\nonumber\\[1ex]&\qquad
+ \chi_{(12)}^{2,2,k}(\xa,\pa),
\quad k=1,2,3.
\end{align}
In the next step we remove both the DR-related scale $\bar{\ell}_0$
and the pole terms proportional to $1/(d-3)$ by adding a total time derivative.
To do this we have found a \textit{unique} function $D(\xa,\pa;d)$,
\begin{widetext}
\begin{align}
\label{ttdD}
D(\xa,\pa;d) &= \frac{m_1}{(16\pi)^3 r_{12}^2} \Bigg(
\bigg( \frac{39}{80}\big\{\big[\npi^2-\pipi\big]\npii + 2 \npi \pipii\big\} + \frac{2}{3}\npii\piipii \bigg) \frac{1}{d - 3}
\nonumber\\[1ex]&\quad
+ \bigg(\frac{117}{80} \big\{\big[\pipi-\npi^2\big]\npii - 2 \npi \pipii\big\} - 2 \npii \piipii\bigg) \ln\frac{r_{12}}{\bar{\ell}_0} \Bigg)
+ (1\leftrightarrow2),
\end{align}
\end{widetext}
such that the sum
$$
H_\mathrm{4PN}^\textrm{reg}(\xa,\pa;d;s,C) + \frac{\md}{\md t}D(\xa,\pa;d)
$$
has the finite limit as $d\to3$.
This limit we take as the regularized value of the 4PN near-zone Hamiltonian:
\begin{align}
H_\mathrm{4PN}^\textrm{near-zone ($s$)}(\xa,\pa;C) &:= \lim_{d\to3}
\Big(H_\mathrm{4PN}^\textrm{reg}(\xa,\pa;d;s,C)
\nonumber\\[1ex]&\qquad
+ \frac{\md}{\md t}D(\xa,\pa;d)\Big).
\end{align}
This Hamiltonian depends on the IR regularization scale $s$
and on the dimensionless constant $C$ and can be rewritten as
\begin{align}
\label{nzh4}
H_\mathrm{4PN}^\textrm{near-zone ($s$)}(\xa,\pa;C) &= H_\mathrm{4PN}^\textrm{local 0}(\xa,\pa)
\nonumber\\[1ex]&\qquad
+ F(\xa,\pa)\Big(\ln\frac{r_{12}}{s} + C\Big),
\end{align}
where the uniquely computed part $H_\mathrm{4PN}^\textrm{local 0}$
of the near-zone Hamiltonian reads
\begin{align}
H_\mathrm{4PN}^\textrm{local 0}(\xa,\pa) &:= \bar{\chi}_1(\xa,\pa)
+ \lim_{d\to3}\bigg( \bar{\chi}_2(\xa,\pa)\ln\frac{r_{12}}{\bar{\ell}_0}
\nonumber\\[1ex]&\kern-5ex
+ \frac{\bar{\chi}_3(\xa,\pa)}{d-3} + \frac{\md}{\md t}D(\xa,\pa;d) \bigg).
\end{align}
Let us stress again that $H_\mathrm{4PN}^\textrm{local 0}$
is without DR-related scale $\bar{\ell}_0$ and poles in $1/(d-3)$.

\subsection{Total 4PN-accurate conservative matter Hamiltonian}

Reference \cite{DJS14} showed that the total 4PN conservative matter Hamiltonian
is the sum of the local-in-time near-zone Hamiltonian \eqref{nzh4}
and the time-symmetric but nonlocal-in-time \emph{tail Hamiltonian} $H_\mathrm{4PN}^\textrm{tail sym ($s$)}$,
\begin{align}
\label{tH1}
H_\mathrm{4PN}[\xa,\pa] &= H_\mathrm{4PN}^\textrm{near-zone ($s$)}(\xa,\pa;C)
\nonumber\\[1ex]&\quad
+ H_\mathrm{4PN}^\textrm{tail sym ($s$)}[\xa,\pa],
\end{align}
where we have used brackets $[\cdot,\cdot]$ to emphasize that the tail Hamiltonian
$H_\mathrm{4PN}^\textrm{tail sym ($s$)}$ is a \emph{functional} of phase-space trajectories $\xa(t)$, $\pa(t)$.
Reference \cite{DJS14} also computed the value of the constant $C$,
\be
\label{C}
C = -\frac{1681}{1536},
\ee
and showed that the total 4PN Hamiltonian \eqref{tH1} does not depend on the scale $s$
and can be written as
\begin{align}
\label{tH2}
H_\mathrm{4PN}[\xa,\pa] = H_\mathrm{4PN}^\textrm{local}(\xa,\pa)
+ H_\mathrm{4PN}^\textrm{nonlocal}[\xa,\pa],
\end{align}
where the local piece of the 4PN Hamiltonian reads
\be
\label{locH4}
H^\textrm{local}_\textrm{4PN}(\xa,\pa) =  H_\mathrm{4PN}^\textrm{local 0}(\xa,\pa) +  C F(\xa,\pa),
\ee
and the nonlocal-in-time piece can be written as
(from now on we restore the constants $c$ and $G$)
\begin{multline}
\label{nonlocH4}
H_\mathrm{4PN}^\textrm{nonlocal}[\xa,\pa] = -\frac{1}{5}\frac{G^2M}{c^8}\dddot{I}_{ij}
\\[1ex]
\times \mathrm{Pf}_{2r_{12}/c} \int_{-\infty}^{+\infty}\dddot{I}_{ij}(t+v)\frac{\md v}{|v|}.
\end{multline}
Here $\dddot{I}_{ij}$ denotes a third time derivative of the Newtonian quadrupole moment of the binary
[defined in Eq.\ \eqref{defI}]
and $\mathrm{Pf}_{2r_{12}/c}$ is a Hadamard partie finie with time scale $2r_{12}/c$
[see Eq.\ (4.2) in \cite{DJS14} for the definition of the $\mathrm{Pf}$ operation].


The total 4PN-accurate conservative matter Hamiltonian is the sum
\begin{align}
H_\mathrm{\le4PN}[\xa,\pa] &= \sum_a m_a c^2
+ \hn(\xa,\pa)
\nonumber\\&
+ \hi(\xa,\pa) + \hii(\xa,\pa)
\nonumber\\[1ex]&
+ \hiii(\xa,\pa) + H_\mathrm{4PN}[\xa,\pa].
\end{align}
The explicit formulas for the local-in-time Hamiltonians 
from the Newtonian up to the 4PN level
[the local piece \eqref{locH4} of the 4PN Hamiltonian given below
incorporates the value \eqref{C} of the constant $C$],
valid in the generic, i.e., noncenter-of-mass, reference frame, are as follows:
\begin{widetext}
\begin{subequations}
\begin{align}
H_{\text{N}}(\xa,\pa) &= \frac{{\bf p}_1^2}{2\,m_1}
- \frac{1}{2}\frac{G \, m_1 \, m_2}{r_{12}}
+ \big(1\leftrightarrow 2\big),
\\[2ex]
c^2\,H_{\text{1PN}}(\xa,\pa) &=
- \frac{1}{8}\frac{\pipip^2}{m_1^3}
+ \frac{1}{8}\frac{Gm_1m_2}{r_{12}} \left(
- 12\,\frac{\pipi}{m_1^2}
+ 14\,\frac{\pipii}{m_1m_2}
+ 2\,\frac{\npi\npii}{m_1m_2} \right)
\nonumber\\[1ex]&\kern-10ex
+ \frac{1}{4}\frac{Gm_1m_2}{r_{12}}\frac{G(m_1+m_2)}{r_{12}}
+ \big(1\leftrightarrow 2\big),
\\[2ex]
c^4\,H_{\text{2PN}}(\xa,\pa) &= \frac{1}{16}\frac{\pipip^3}{m_1^5}
+ \frac{1}{8} \frac{Gm_1m_2}{r_{12}} \Bigg(
5\,\frac{\pipip^2}{m_1^4}
- \frac{11}{2}\frac{\pipi\,\piipii}{m_1^2m_2^2}
- \frac{\pipii^2}{m_1^2m_2^2}
+ 5\,\frac{\pipi\,\npii^2}{m_1^2m_2^2}
\nonumber\\[1ex]&\kern-10ex
- 6\,\frac{\pipii\,\npi\npii}{m_1^2m_2^2}
- \frac{3}{2}\frac{\npi^2\npii^2}{m_1^2m_2^2} \Bigg)
\nonumber\\[1ex]&\kern-10ex
+ \frac{1}{4}\frac{G^2m_1m_2}{r_{12}^2} \Bigglb(
m_2\left(10\frac{\pipi}{m_1^2}+19\frac{\piipii}{m_2^2}\right)
- \frac{1}{2}(m_1+m_2)\frac{27\,\pipii+6\,\npi\npii}{m_1m_2} \Biggrb)
\nonumber\\[1ex]&\kern-10ex
- \frac{1}{8} \frac{Gm_1m_2}{r_{12}}\frac{G^2 (m_1^2+5m_1m_2+m_2^2)}{r_{12}^2}
+ \big(1\leftrightarrow 2\big),
\\[2ex]
c^6\,H_{\text{3PN}}(\xa,\pa) &=
-\frac{5}{128}\frac{\pipip^4}{m_1^7}
+ \frac{1}{32} \frac{Gm_1m_2}{r_{12}} \Bigg(
- 14\,\frac{\pipip^3}{m_1^6}
+ 4\,\frac{\biglb(\pipii^2+4\,\pipi\,\piipii\bigrb)\pipi}{m_1^4m_2^2}
\nonumber\\[2ex]&\kern-10ex
+ 6\,\frac{\pipi\,\npi^2\npii^2}{m_1^4m_2^2}
- 10\,\frac{\biglb(\pipi\,\npii^2+\piipii\,\npi^2\bigrb)\pipi}{m_1^4m_2^2}
+ 24\,\frac{\pipi\,\pipii\npi\npii}{m_1^4m_2^2}
\nonumber\\[2ex]&\kern-10ex
+ 2\,\frac{\pipi\,\pipii\npii^2}{m_1^3m_2^3}
+ \frac{\biglb(7\,\pipi\,\piipii-10\,\pipii^2\bigrb)\npi\npii}{m_1^3m_2^3}
+ \frac{\biglb(\pipi\,\piipii-2\,\pipii^2\bigrb)\pipii}{m_1^3m_2^3}
\nonumber\\[2ex]&\kern-10ex
+ 15\,\frac{\pipii\npi^2\npii^2}{m_1^3m_2^3}
- 18\,\frac{\pipi\,\npi\npii^3}{m_1^3m_2^3}
+ 5\,\frac{\npi^3\npii^3}{m_1^3m_2^3} \Bigg)
\nonumber\\[2ex]&\kern-10ex
+ \frac{G^2m_1m_2}{r_{12}^2} \Bigg(
\frac{1}{16}(m_1-27m_2)\frac{\pipip^2}{m_1^4}
- \frac{115}{16}m_1\frac{\pipi\,\pipii}{m_1^3m_2}
+ \frac{1}{48}m_2\frac{25\,\pipii^2+371\,\pipi\,\piipii}{m_1^2 m_2^2}
\nonumber\\[2ex]&\kern-10ex
+ \frac{17}{16}\frac{\pipi\npi^2}{m_1^3}
+ \frac{5}{12}\frac{\npi^4}{m_1^3}
- \frac{1}{8}m_1 \frac{\biglb(15\,\pipi\,\npii+11\,\pipii\,\npi\bigrb)\npi}{m_1^3 m_2}
\nonumber\\[2ex]&\kern-10ex
- \frac{3}{2}m_1\frac{\npi^3\npii}{m_1^3m_2}
+ \frac{125}{12}m_2\frac{\pipii\,\npi\npii}{m_1^2m_2^2}
+ \frac{10}{3}m_2\frac{\npi^2\npii^2}{m_1^2m_2^2}
\nonumber\\[1ex]&\kern-10ex
- \frac{1}{48} (220 m_1 + 193 m_2) \frac{\pipi \npii^2}{m_1^2 m_2^2} \Bigg)
+ \frac{G^3m_1m_2}{r_{12}^3} \Bigg(
-\frac{1}{48}
\bigglb(425\,m_1^2+\Big(473-\frac{3}{4}\pi^2\Big)m_1m_2+150\,m_2^2\biggrb)
\frac{\pipi}{m_1^2}
\nonumber\\[2ex]&\kern-10ex
+ \frac{1}{16}
\bigglb(77(m_1^2+m_2^2)+\Big(143-\frac{1}{4}\pi^2\Big)m_1m_2\biggrb)
\frac{\pipii}{m_1m_2}
+ \frac{1}{16}
\bigglb(20\,m_1^2-\Big(43+\frac{3}{4}\pi^2\Big)m_1m_2\biggrb)
\frac{\npi^2}{m_1^2}
\nonumber\\[2ex]&\kern-10ex
+ \frac{1}{16}
\bigglb(21(m_1^2+m_2^2)+\Big(119+\frac{3}{4}\pi^2\Big)m_1m_2\biggrb)
\frac{\npi\npii}{m_1m_2} \Bigg)
+ \frac{1}{8} \frac{G^4m_1m_2^3}{r_{12}^4}
\Bigg( \bigg(\frac{227}{3}-\frac{21}{4}\pi^2\bigg)m_1+m_2 \Bigg)
\nonumber\\[2ex]&\kern-10ex
+ \big(1\leftrightarrow 2\big),
\\[2ex]
c^8\,H_\textrm{4PN}^\textrm{local}(\xa,\pa) &=
\frac{7 \pipip^5}{256 m_1^9}
+ \frac{G m_1 m_2}{r_{12}} H_{48}(\xa,\pa)
+ \frac{G^2 m_1 m_2}{r_{12}^2} m_1\,H_{46}(\xa,\pa)
\nonumber\\[1ex]&\kern-10ex
+ \frac{G^3 m_1 m_2}{r_{12}^3} \Big(m_1^2\,H_{441}(\xa,\pa) + m_1 m_2\,H_{442}(\xa,\pa) \Big)
\nonumber\\[1ex]&\kern-10ex
+ \frac{G^4 m_1 m_2}{r_{12}^4} \Big(m_1^3\,H_{421}(\xa,\pa) + m_1^2 m_2\,H_{422}(\xa,\pa)\Big)
+ \frac{G^5 m_1 m_2}{r_{12}^5} H_{40}(\xa,\pa)
+ \big(1\leftrightarrow 2\big),
\\[2ex]
H_{48}(\xa,\pa) &=\frac{45 \pipip^4}{128 m_1^8}
-\frac{9 \npi^2 \npii^2 \pipip^2}{64 m_1^6 m_2^2}
+\frac{15 \npii^2 \pipip^3}{64m_1^6 m_2^2}
-\frac{9 \npi \npii \pipip^2 \pipii}{16 m_1^6 m_2^2}
\nonumber\\[1ex]&\kern-10ex
-\frac{3 \pipip^2 \pipii^2}{32m_1^6 m_2^2}
+\frac{15 \npi^2 \pipip^2 \piipii}{64 m_1^6 m_2^2}
-\frac{21 \pipip^3 \piipii}{64 m_1^6m_2^2}
-\frac{35 \npi^5 \npii^3}{256 m_1^5 m_2^3}
\nonumber\\[1ex]&\kern-10ex
+\frac{25 \npi^3 \npii^3 \pipi}{128 m_1^5m_2^3}
+\frac{33 \npi \npii^3 \pipip^2}{256 m_1^5 m_2^3}
-\frac{85 \npi^4 \npii^2 \pipii}{256 m_1^5m_2^3}
\nonumber\\[1ex]&\kern-10ex
-\frac{45 \npi^2 \npii^2 \pipi \pipii}{128 m_1^5 m_2^3}
-\frac{\npii^2 \pipip^2 \pipii}{256m_1^5 m_2^3}
+\frac{25 \npi^3 \npii \pipii^2}{64 m_1^5 m_2^3}
\nonumber\\[1ex]&\kern-10ex
+\frac{7 \npi \npii \pipi\pipii^2}{64 m_1^5 m_2^3}
-\frac{3 \npi^2 \pipii^3}{64 m_1^5 m_2^3}
+\frac{3 \pipi \pipii^3}{64 m_1^5m_2^3}
\nonumber\\[1ex]&\kern-10ex
+\frac{55 \npi^5 \npii \piipii}{256 m_1^5 m_2^3}
-\frac{7 \npi^3 \npii \pipi \piipii}{128m_1^5 m_2^3}
-\frac{25 \npi \npii \pipip^2 \piipii}{256 m_1^5 m_2^3}
\nonumber\\[1ex]&\kern-10ex
-\frac{23 \npi^4 \pipii\piipii}{256 m_1^5 m_2^3}
+\frac{7 \npi^2 \pipi \pipii \piipii}{128 m_1^5 m_2^3}
-\frac{7 \pipip^2\pipii \piipii}{256 m_1^5 m_2^3}
\nonumber\\[1ex]&\kern-10ex
-\frac{5 \npi^2 \npii^4 \pipi}{64 m_1^4 m_2^4}
+\frac{7 \npii^4\pipip^2}{64 m_1^4 m_2^4}
-\frac{\npi \npii^3 \pipi \pipii}{4 m_1^4 m_2^4}
\nonumber\\[1ex]&\kern-10ex
+\frac{\npii^2 \pipi\pipii^2}{16 m_1^4 m_2^4}
-\frac{5 \npi^4 \npii^2 \piipii}{64 m_1^4 m_2^4}
+\frac{21 \npi^2 \npii^2\pipi \piipii}{64 m_1^4 m_2^4}
\nonumber\\[1ex]&\kern-10ex
-\frac{3 \npii^2 \pipip^2 \piipii}{32 m_1^4 m_2^4}
-\frac{\npi^3\npii \pipii \piipii}{4 m_1^4 m_2^4}
+\frac{\npi \npii \pipi \pipii \piipii}{16 m_1^4m_2^4}
\nonumber\\[1ex]&\kern-10ex
+\frac{\npi^2 \pipii^2 \piipii}{16 m_1^4 m_2^4}
-\frac{\pipi \pipii^2 \piipii}{32 m_1^4m_2^4}
+\frac{7 \npi^4 \piipiip^2}{64 m_1^4 m_2^4}
\nonumber\\[1ex]&\kern-10ex
-\frac{3 \npi^2 \pipi \piipiip^2}{32 m_1^4m_2^4}
-\frac{7 \pipip^2 \piipiip^2}{128 m_1^4 m_2^4},
\\[2ex]
H_{46}(\xa,\pa) &=
\frac{369 \npi^6}{160 m_1^6}
-\frac{889 \npi^4 \pipi}{192 m_1^6}
+\frac{49 \npi^2 \pipip^2}{16 m_1^6}
-\frac{63\pipip^3}{64 m_1^6}
\nonumber\\[1ex]&\kern-10ex
-\frac{549 \npi^5 \npii}{128 m_1^5 m_2}
+\frac{67 \npi^3 \npii \pipi}{16 m_1^5m_2}
-\frac{167 \npi \npii \pipip^2}{128 m_1^5 m_2}
\nonumber\\[1ex]&\kern-10ex
+\frac{1547 \npi^4 \pipii}{256 m_1^5m_2}
-\frac{851 \npi^2 \pipi \pipii}{128 m_1^5 m_2}
+\frac{1099 \pipip^2 \pipii}{256 m_1^5 m_2}
\nonumber\\[1ex]&\kern-10ex
+\frac{3263\npi^4 \npii^2}{1280 m_1^4 m_2^2}
+\frac{1067 \npi^2 \npii^2 \pipi}{480 m_1^4 m_2^2}
-\frac{4567\npii^2 \pipip^2}{3840 m_1^4 m_2^2}
\nonumber\\[1ex]&\kern-10ex
-\frac{3571 \npi^3 \npii \pipii}{320 m_1^4 m_2^2}
+\frac{3073\npi \npii \pipi \pipii}{480 m_1^4 m_2^2}
+\frac{4349 \npi^2 \pipii^2}{1280 m_1^4 m_2^2}
\nonumber\\[1ex]&\kern-10ex
-\frac{3461\pipi \pipii^2}{3840 m_1^4 m_2^2}
+\frac{1673 \npi^4 \piipii}{1920 m_1^4 m_2^2}
-\frac{1999 \npi^2 \pipi\piipii}{3840 m_1^4 m_2^2}
\nonumber\\[1ex]&\kern-10ex
+\frac{2081 \pipip^2 \piipii}{3840 m_1^4 m_2^2}
-\frac{13 \npi^3 \npii^3}{8m_1^3 m_2^3}
+\frac{191 \npi \npii^3 \pipi}{192 m_1^3 m_2^3}
\nonumber\\[1ex]&\kern-10ex
-\frac{19 \npi^2 \npii^2 \pipii}{384m_1^3 m_2^3}
-\frac{5 \npii^2 \pipi \pipii}{384 m_1^3 m_2^3}
+\frac{11 \npi \npii \pipii^2}{192m_1^3 m_2^3}
\nonumber\\[1ex]&\kern-10ex
+\frac{77 \pipii^3}{96 m_1^3 m_2^3}
+\frac{233 \npi^3 \npii \piipii}{96 m_1^3m_2^3}
-\frac{47 \npi \npii \pipi \piipii}{32 m_1^3 m_2^3}
\nonumber\\[1ex]&\kern-10ex
+\frac{\npi^2 \pipii \piipii}{384 m_1^3m_2^3}
-\frac{185 \pipi \pipii \piipii}{384 m_1^3 m_2^3}
-\frac{7 \npi^2 \npii^4}{4 m_1^2 m_2^4}
\nonumber\\[1ex]&\kern-10ex
+\frac{7\npii^4 \pipi}{4 m_1^2 m_2^4}
-\frac{7 \npi \npii^3 \pipii}{2 m_1^2 m_2^4}
+\frac{21 \npii^2\pipii^2}{16 m_1^2 m_2^4}
\nonumber\\[1ex]&\kern-10ex
+\frac{7 \npi^2 \npii^2 \piipii}{6 m_1^2 m_2^4}
+\frac{49 \npii^2 \pipi\piipii}{48 m_1^2 m_2^4}
-\frac{133 \npi \npii \pipii \piipii}{24 m_1^2 m_2^4}
\nonumber\\[1ex]&\kern-10ex
-\frac{77 \pipii^2\piipii}{96 m_1^2 m_2^4}
+\frac{197 \npi^2 \piipiip^2}{96 m_1^2 m_2^4}
-\frac{173 \pipi \piipiip^2}{48 m_1^2m_2^4}
+\frac{13 \piipiip^3}{8 m_2^6},
\\[2ex]
H_{441}(\xa,\pa) &=
\frac{5027 \npi^4}{384 m_1^4}
-\frac{22993 \npi^2 \pipi}{960 m_1^4}
-\frac{6695 \pipip^2}{1152 m_1^4}
-\frac{3191\npi^3 \npii}{640 m_1^3 m_2}
\nonumber\\[1ex]&\kern-10ex
+\frac{28561 \npi \npii \pipi}{1920 m_1^3 m_2}
+\frac{8777 \npi^2\pipii}{384 m_1^3 m_2}
+\frac{752969 \pipi \pipii}{28800 m_1^3 m_2}
\nonumber\\[1ex]&\kern-10ex
-\frac{16481 \npi^2 \npii^2}{960m_1^2 m_2^2}
+\frac{94433 \npii^2 \pipi}{4800 m_1^2 m_2^2}
-\frac{103957 \npi \npii \pipii}{2400 m_1^2m_2^2}
\nonumber\\[1ex]&\kern-10ex
+\frac{791 \pipii^2}{400 m_1^2 m_2^2}
+\frac{26627 \npi^2 \piipii}{1600 m_1^2 m_2^2}
-\frac{118261 \pipi\piipii}{4800 m_1^2 m_2^2}
+\frac{105 \piipiip^2}{32 m_2^4},
\\[2ex]
H_{442}(\xa,\pa) &=
\left(\frac{2749 \pi ^2}{8192}-\frac{211189}{19200}\right) \frac{\pipip^2}{m_1^4}
+\left(\frac{63347}{1600}-\frac{1059 \pi ^2}{1024}\right)\frac{\npi^2 \pipi}{m_1^4}
+\left(\frac{375\pi^2}{8192}-\frac{23533}{1280}\right)\frac{\npi^4}{m_1^4}
\nonumber\\[1ex]&\kern-10ex
+\left(\frac{10631 \pi ^2}{8192}-\frac{1918349}{57600}\right) \frac{\pipii^2}{m_1^2 m_2^2}
+\left(\frac{13723 \pi^2}{16384}-\frac{2492417}{57600}\right) \frac{\pipi\piipii}{m_1^2 m_2^2}
\nonumber\\[1ex]&\kern-10ex
+\left(\frac{1411429}{19200}-\frac{1059 \pi ^2}{512}\right)\frac{\npii^2\pipi}{m_1^2 m_2^2}
+\left(\frac{248991}{6400}-\frac{6153 \pi ^2}{2048}\right) \frac{\npi\npii\pipii}{m_1^2 m_2^2}
\nonumber\\[1ex]&\kern-10ex
-\left(\frac{30383}{960}+\frac{36405\pi^2}{16384}\right) \frac{\npi^2 \npii^2}{m_1^2m_2^2}
+\left(\frac{1243717}{14400}-\frac{40483 \pi ^2}{16384}\right) \frac{\pipi\pipii}{m_1^3m_2}
\nonumber\\[1ex]&\kern-10ex
+\left(\frac{2369}{60}+\frac{35655 \pi ^2}{16384}\right) \frac{\npi^3 \npii}{m_1^3 m_2}
+\left(\frac{43101\pi ^2}{16384}-\frac{391711}{6400}\right) \frac{\npi \npii \pipi}{m_1^3 m_2}
\nonumber\\[1ex]&\kern-10ex
+\left(\frac{56955 \pi^2}{16384}-\frac{1646983}{19200}\right) \frac{\npi^2 \pipii}{m_1^3 m_2},
\\[2ex]
H_{421}(\xa,\pa) &=
\frac{64861 \pipi}{4800 m_1^2}-\frac{91 \pipii}{8 m_1 m_2}
+\frac{105 \piipii}{32 m_2^2}
-\frac{9841 \npi^2}{1600m_1^2}-\frac{7 \npi \npii}{2 m_1 m_2},
\\[2ex]
H_{422}(\xa,\pa) &=
\left(\frac{1937033}{57600}-\frac{199177 \pi ^2}{49152}\right) \frac{\pipi}{m_1^2}
+\left(\frac{176033 \pi ^2}{24576}-\frac{2864917}{57600}\right)\frac{\pipii}{m_1 m_2}
+\left(\frac{282361}{19200}-\frac{21837 \pi ^2}{8192}\right)\frac{\piipii}{m_2^2}
\nonumber\\[1ex]&\kern-10ex
+\left(\frac{698723}{19200}+\frac{21745 \pi ^2}{16384}\right) \frac{\npi^2}{m_1^2}
+\left(\frac{63641 \pi^2}{24576}-\frac{2712013}{19200}\right) \frac{\npi \npii}{m_1 m_2}
\nonumber\\[1ex]&\kern-10ex
+\left(\frac{3200179}{57600}-\frac{28691 \pi ^2}{24576}\right)\frac{\npii^2}{m_2^2},
\\[2ex]
H_{40}(\xa,\pa) &=
-\frac{m_1^4}{16}
+\left(\frac{6237 \pi^2}{1024}-\frac{169799}{2400}\right) m_1^3 m_2
+\left(\frac{44825 \pi^2}{6144}-\frac{609427}{7200}\right)m_1^2 m_2^2.
\end{align}
\end{subequations}
\end{widetext}

For completeness let us supplement the generic 4PN-accurate local Hamiltonian given above
by its center-of-mass expression.
The center-of-mass reference frame is defined through the condition
\be
\p_1 + \p_2 = \mathbf{0}.
\ee
Let us introduce the reduced mass $\mu$ of the system
and its symmetric mass ratio $\nu$,
\be
\mu := \frac{m_1 m_2}{M},
\quad
\nu := \frac{\mu}{M} = \frac{m_1 m_2}{(m_1+m_2)^2}.
\ee
It is convenient to introduce the reduced variables
\be
\mathbf{r} := \frac{\mathbf{x}_{12}}{GM},
\quad
\mathbf{p} := \frac{\p_1}{\mu} = -\frac{\p_2}{\mu}
\ee
(with $r:=|\rv|$ and $\n:=\rv/r$).
We also define the reduced 4PN-accurate center-of-mass Hamiltonian
\be
\widehat{H}_\textrm{$\le$4PN}[\rv,\p]
:= \frac{{H}_\textrm{$\le$4PN}[\rv,\p]-M c^2}{\mu},
\ee
which is the sum of different PN contributions,
\begin{align}
\widehat{H}_\textrm{$\le$4PN}[\rv,\p]
&= \widehat{H}_{\rm N}(\mathbf{r},\mathbf{p})+ \widehat{H}_{\rm 1PN}(\mathbf{r},\mathbf{p})
+ \widehat{H}_{\rm 2PN}(\mathbf{r},\mathbf{p})
\nonumber\\[1ex]&\quad
+ \widehat{H}_{\rm 3PN}(\mathbf{r},\mathbf{p}) + \widehat{H}_{\rm 4PN}[\mathbf{r},\mathbf{p}],
\end{align}
where the 4PN Hamiltonian is the sum of local- and nonlocal-in-time parts,
\be
\widehat{H}_\mathrm{4PN}[\rv,\p] = \widehat{H}_\mathrm{4PN}^\textrm{local}(\rv,\p)
+ \widehat{H}_\mathrm{4PN}^\textrm{nonlocal}[\rv,\p].
\ee
The local Hamiltonians from $\widehat{H}_{\rm N}$ to $\widehat{H}_\mathrm{4PN}^\textrm{local}$ are equal to
(let us recall that their coefficients depend on masses $m_1$, $m_2$
only through the symmetric mass ratio $\nu$)
\begin{widetext}
\begin{subequations}
\begin{align}
\widehat{H}_{\rm N}\left({\bf r},{\bf p}\right) &= \frac{\pp}{2} - \frac{1}{r},
\\[2ex]
c^2\,\widehat{H}_{\rm 1PN}\left({\bf r},{\bf p}\right) &=
\frac{1}{8}(3\nu-1) \ppn^2 - \frac{1}{2}\Big\{(3+\nu)\pp + \nu \np^2
\Big\}\frac{1}{r} + \frac{1}{2r^2},
\\[2ex]
c^4\,\widehat{H}_{\rm 2PN}\left({\bf r},{\bf p}\right) &= 
\frac{1}{16}\left(1-5\nu+5\nu^2\right) \ppn^3
+ \frac{1}{8}\Big\{ \left(5-20\nu-3\nu^2\right)\ppn^2 - 2\nu^2 \np^2\pp - 3\nu^2\np^4 \Big\}\frac{1}{r}
\nonumber\\[1ex]&\kern-10ex
+ \frac{1}{2} \Big\{(5+8\nu)\pp+3\nu\np^2\Big\}\frac{1}{r^2}
- \frac{1}{4}(1+3\nu)\frac{1}{r^3},
\\[2ex]
c^6\,\widehat{H}_{\rm 3PN}\left({\bf r},{\bf p}\right)
&= \frac{1}{128}\left(-5+35\nu-70\nu^2+35\nu^3\right)\ppn^4
+ \frac{1}{16}\bigg\{
\left(-7+42\nu-53\nu^2-5\nu^3\right)\ppn^3
\nonumber\\[1ex]&\kern-10ex
+ (2-3\nu)\nu^2\np^2\ppn^2
+ 3(1-\nu)\nu^2\np^4\pp - 5\nu^3\np^6
\bigg\}\frac{1}{r}
\nonumber\\[1ex]&\kern-10ex
+\bigg\{ \frac{1}{16}\left(-27+136\nu+109\nu^2\right)\ppn^2
+ \frac{1}{16}(17+30\nu)\nu\np^2\pp + \frac{1}{12}(5+43\nu)\nu\np^4
\bigg\}\frac{1}{r^2}
\nonumber\\[1ex]&\kern-10ex
+\Bigg\{ \bigglb(
-\frac{25}{8}+\left(\frac{\pi^2}{64}-\frac{335}{48}\right)\nu-\frac{23\nu^2}{8} \biggrb)\pp
+ \left(-\frac{85}{16}-\frac{3\pi^2}{64}-\frac{7\nu}{4}\right)\nu\np^2 
\Bigg\}\frac{1}{r^3}
+ \Bigg\{\frac{1}{8}+\Big(\frac{109}{12}-\frac{21}{32}\pi^2\Big)\nu\Bigg\}\frac{1}{r^4},
\\[2ex]
c^8\,\widehat{H}_{\rm 4PN}^{\rm local}(\mathbf{r},\mathbf{p}) &= 
\bigg( \frac{7}{256} - \frac{63}{256}\nu +\frac{189}{256}\nu^2 - \frac{105}{128}\nu^3 + \frac{63}{256}\nu^4 \bigg) \ppn^5
\nonumber\\[1ex]&\kern-10ex
+ \Bigg\{
\frac{45}{128}\ppn^4 - \frac{45}{16}\ppn^4\,\nu
+\left( \frac{423}{64}\ppn^4 -\frac{3}{32}\np^2\ppn^3 - \frac{9}{64}\np^4\ppn^2 \right)\,\nu^2
\nonumber\\[1ex]&\kern-10ex
+ \left( -\frac{1013}{256}\ppn^4 + \frac{23}{64}\np^2\ppn^3 + \frac{69}{128}\np^4\ppn^2
- \frac{5}{64}\np^6\pp + \frac{35}{256}\np^8 \right)\,\nu^3
\nonumber\\[1ex]&\kern-10ex
+ \left( -\frac{35}{128}\ppn^4 - \frac{5}{32}\np^2\ppn^3 - \frac{9}{64}\np^4\ppn^2 -\frac{5}{32}\np^6\pp
- \frac{35}{128}\np^8 \right)\,\nu^4
\Bigg\}\frac{1}{r}
\nonumber\\[1ex]&\kern-10ex
+ \Bigg\{ \frac{13}{8}\ppn^3
+ \left( -\frac{791}{64}\ppn^3 + \frac{49}{16}\np^2\ppn^2 -\frac{889}{192}\np^4\pp + \frac{369}{160}\np^6 \right)\,\nu
\nonumber\\[1ex]&\kern-10ex
+\left( \frac{4857}{256}\ppn^3 -\frac{545}{64}\np^2\ppn^2 +\frac{9475}{768}\np^4\pp - \frac{1151}{128}\np^6 \right)\,\nu^2
\nonumber\\[1ex]&\kern-10ex
+ \left( \frac{2335}{256}\ppn^3 + \frac{1135}{256}\np^2\ppn^2 - \frac{1649}{768}\np^4\pp + \frac{10353}{1280}\np^6 \right)\,\nu^3
\Bigg\}\frac{1}{r^2}
\nonumber\\[1ex]&\kern-10ex
+ \Bigg\{ \frac{105}{32}\ppn^2
+ \bigglb( \left(\frac{2749\pi^2}{8192}-\frac{589189}{19200}\right)\ppn^2
+ \left(\frac{63347}{1600}-\frac{1059\pi^2}{1024}\right)\np^2\pp
+ \left(\frac{375\pi^2}{8192}-\frac{23533}{1280}\right)\np^4 \biggrb)\,\nu
\nonumber\\[1ex]&\kern-10ex
+ \bigglb( \left(\frac{18491\pi^2}{16384}-\frac{1189789}{28800}\right)\ppn^2
+ \left(-\frac{127}{3}-\frac{4035\pi^2}{2048}\right)\np^2\pp
+ \left(\frac{57563}{1920}-\frac{38655\pi^2}{16384}\right)\np^4 \biggrb)\,\nu^2
\nonumber\\[1ex]&\kern-10ex
+\left( -\frac{553}{128}\ppn^2 -\frac{225}{64}\np^2\pp -\frac{381}{128}\np^4 \right)\,\nu^3
\Bigg\}\frac{1}{r^3}
\nonumber\\[1ex]&\kern-10ex
+ \Bigg\{
\frac{105}{32}\pp
+ \bigglb( \left(\frac{185761}{19200}-\frac{21837\pi^2}{8192}\right) \pp
+ \left(\frac{3401779}{57600}-\frac{28691\pi^2}{24576}\right) \np^2 \biggrb)\,\nu
\nonumber\\[1ex]&\kern-10ex
+ \bigglb( \left(\frac{672811}{19200}-\frac{158177\pi^2}{49152}\right) \pp
+ \left(\frac{110099\pi^2}{49152}-\frac{21827}{3840}\right) \np^2 \biggrb)\,\nu^2 \Bigg\}\frac{1}{r^4}
\nonumber\\[1ex]&\kern-10ex
+ \Bigg\{ -\frac{1}{16} + \left(\frac{6237\pi^2}{1024}-\frac{169199}{2400}\right)\,\nu 
+ \left(\frac{7403\pi^2}{3072}-\frac{1256}{45} \right)\,\nu^2 \Bigg\}\frac{1}{r^5}.
\end{align}
\end{subequations}
\end{widetext}

\section{The Poincar\'e invariance}

There are several possibilities for partial checks of our results:
first the test-body limit and second the linear in $G$ part of the 4PN Hamiltonian
through comparison with the post-Minkowskian results achieved in Ref.\ \cite{LSB08}.
Both these checks were already performed by us in Ref.\ \cite{JS12}.
The most important tool however is the Poincar\'e invariance,
discussed for the first time in the context of the ADM Hamiltonian approach to the two-body problem in Ref.\ \cite{DJS00}.
Poincar\'e symmetry holds because our (isolated and conservative) two-point-mass system
is living in asymptotically flat spacetime with its globally conserved quantities:
energy $H$, linear momentum ${\bf P}$, angular momentum ${\bf J}$,
and Lorentz boost vector ${\bf K} = - {\bf P}t + {\bf G}$,
where ${\bf G}$ denotes the center-of-(mass)energy vector.
All these quantities are realized as functions on the two-body phase space $(\x_1,\x_2,\p_1,\p_2)$,
whose usual Poisson brackets,
\be
\{f(\xa,\pa),g(\xa,\pa)\} := \sum_a \bigg(
\frac{\partial f}{\partial x_a^i}\frac{\partial g}{\partial p_{ai}}
- \frac{\partial f}{\partial p_{ai}}\frac{\partial g}{\partial x_a^i} \bigg),
\ee
satisfy the Poincar\'e algebra relations,
\begin{subequations}
\label{PA}
\begin{align}
\label{PAa}
\left\{P_i\,,P_j\right\} &= 0,
\quad \left\{J_i,\,J_j\right\} = \epsilon_{ijk}J_k,
\\[1ex]
\label{PAb}
\left\{J_i, \, P_j \right\} & =\epsilon_{ijk} P_k,
\\[1ex]
\label{PAc}
\left\{P_i, \, H\right\} &= 0,
\quad \left\{J_i,\, H \right\} = 0,
\\[1ex]
\label{PAd}
\left\{J_i, \, G_j \right\} &= \epsilon_{ijk}G_k,
\\[1ex]
\label{PAe}
\left\{G_i, \, H \right\} &= P_i,
\\[1ex]
\label{PAf}
\left\{G_i,\, P_j\right\} &= c^{-2}H\delta_{ij},
\\[1ex]
\label{PAg}
\left\{G_i, \, G_j \right\} &= -c^{-2}\epsilon_{ijk}J_k.
\end{align}
\end{subequations}
These relations have to be fulfilled through 4PN order.

The total linear and angular momenta have universal forms,
\be
P_i(\xa,\pa) = \sum_a p_{ai}, \quad J_i(\xa,\pa) = \sum_a \epsilon_{ijk} x^j_a p_{ak},
\ee
and they exactly satisfy Eqs.\ \eqref{PAa}--\eqref{PAc}.
We will construct the boost vector $\mathbf{G}$ as a three-vector from $\xa$ and $\pa$ only;
therefore, the relation \eqref{PAd} will also be exactly satisfied.
Consequently Eqs.\ \eqref{PAe}--\eqref{PAg}
are the only nontrivial relations which have to be satisfied by the vector $\mathbf{G}$.
The Hamitonian $H$ entering Poincar\'e algebra \eqref{PA} is the full 4PN-accurate Hamiltonian,
\be
H_{\le\mathrm{4PN}}[\xa,\pa] = H_{\le\mathrm{4PN}}^\textrm{local}(\xa,\pa)
+ H_\mathrm{4PN}^\textrm{nonlocal}[\xa,\pa],
\ee
where the local-in-time part reads
\begin{align}
H_{\le\mathrm{4PN}}^\textrm{local}(\xa,\pa) &= \sum_a m_a c^2
+ \hn(\xa,\pa)
\nonumber\\&\quad
+ \hi(\xa,\pa) + \hii(\xa,\pa)
\nonumber\\[1ex]&\quad
+ \hiii(\xa,\pa) + H_{\mathrm{4PN}}^\textrm{local}(\xa,\pa).
\end{align}
Because the nonlocal-in-time piece $H_\mathrm{4PN}^\textrm{nonlocal}$ is Galileo invariant
[see Eq.\ (5.15) in \cite{DJS14} for the proof],
it is enough to restrict the 4PN-accurate Hamiltonian to its local part $H_{\le\mathrm{4PN}}^\textrm{local}$
when looking for the 4PN-accurate boost vector $\mathbf{G}$.

We have found the 4PN-accurate boost vector $\mathbf{G}$ using
the method of undetermined coefficients employed at the 3PN level in Ref.\ \cite{DJS00}.
The generic form of the three-vector $\mathbf{G}$ reads
\be
\mathbf{G}(\xa,\pa) = \sum_a \Big( M_a(\x_b,\p_b)\,\xa + N_a(\x_b,\p_b)\,\pa \Big),
\ee
where the scalars $M_a$ and $N_a$ possess the following 4PN-accurate expansions
\begin{subequations}
\begin{align}
M_a &= m_a + M_a^{\rm 1PN} + M_a^{\rm 2PN} + M_a^{\rm 3PN} + M_a^{\rm 4PN},
\\[1ex]
N_a &= N_a^{\rm 2PN} + N_a^{\rm 3PN} + N_a^{\rm 4PN}.
\end{align}
\end{subequations}
Let us note that $M_a$ and $N_a$ reduce to $m_a$ and 0, respectively, in the Newtonian approximation.
Next we write the most general expressions for the successive 
PN approximations to the functions $M_a$ and $N_a$
as sums of scalar monomials of the form
\begin{multline}
c_\mathbf{n}\,r_{12}^{-n_0}\,
\bigg(\frac{\pipi}{m_1^2}\bigg)^{n_1}\,\bigg(\frac{\p_1\cdot\p_2}{m_1 m_2}\bigg)^{n_2}\,\bigg(\frac{\piipii}{m_2^2}\bigg)^{n_3}
\\[1ex]
\times\bigg(\frac{\n_{12}\cdot\p_1}{m_1}\bigg)^{n_4}\,\bigg(\frac{\n_{12}\cdot\p_2}{m_2}\bigg)^{n_5}\,m_1^{n_6}\,m_2^{n_7},
\end{multline}
where $n_0,\ldots,n_7$ are non-negative integers.
To constrain the possible values of $n_0,\ldots,n_7$
we employ dimensional analysis, Euclidean covariance (including parity symmetry),
and time reversal symmetry (which imposes that $M_a$ is even and $N_a$ is odd under the operation $\pa\to-\pa$).
We also use the $1 \leftrightarrow 2$ relabeling symmetry.
At the 4PN level the most general pattern for the functions $M_a^{\rm 4PN}$ and  $N_a^{\rm 4PN}$
involves 210 \emph{dimensionless} coefficients $c_\mathbf{n}$.

To find the functions $M_a^{\rm 1PN}$, \ldots, $M_a^{\rm 4PN}$ and $N_a^{\rm 2PN}$, \ldots, $N_a^{\rm 4PN}$
it is enough to use Eq.\ \eqref{PAe} only.
The 3PN-accurate functions were constructed in Ref.\ \cite{DJS00}
(for completeness we give below their explicit expressions).
At the 4PN level the relation \eqref{PAe} yields 525 equations to be satisfied.
We have found a \emph{unique} solution to these equations
[and we have then checked that this solution
satisfies the remaining Poincar\'e algebra relations \eqref{PAf} and \eqref{PAg}].
The explicit forms of the functions $M_a^{\rm 1PN}$, \ldots, $M_a^{\rm 4PN}$
and $N_a^{\rm 2PN}$, \ldots, $N_a^{\rm 4PN}$ read
\begin{widetext}
\begin{subequations}
\begin{align}
c^2\,M_1^{\rm 1PN}(\xa,\pa) &= \frac{1}{2}\frac{\pipi}{m_1}
- \frac{1}{2}\frac{Gm_1m_2}{r_{12}},
\\[2ex]
c^4\,M^{\text{2PN}}_1(\xa,\pa) &= -\frac{1}{8}\frac{\pipip^2}{m_1^3}
+ \frac{1}{4}\frac{Gm_1m_2}{r_{12}} \bigg( -5\,\frac{\pipi}{m_1^2}
- \frac{\piipii}{m_2^2} + 7\,\frac{\pipii}{m_1m_2} + \frac{\npi\npii}{m_1m_2} \bigg)
\nonumber\\[1ex]&\kern-10ex
+ \frac{1}{4}\frac{Gm_1m_2}{r_{12}}\frac{G(m_1+m_2)}{r_{12}},
\\[2ex]
c^6\,M^{\text{3PN}}_1(\xa,\pa) &=
\frac{1}{16}\frac{\pipip^3}{m_1^5}
+ \frac{1}{16} \frac{Gm_1m_2}{r_{12}} \Bigg(
9\,\frac{\pipip^2}{m_1^4}
+ \frac{\piipiip^2}{m_2^4}
- 11\,\frac{\pipi\,\piipii}{m_1^2m_2^2}
- 2\,\frac{\pipii^2}{m_1^2m_2^2}
+ 3\,\frac{\pipi\,\npii^2}{m_1^2m_2^2}
\nonumber\\[1ex]&\kern-10ex
+ 7\,\frac{\piipii\,\npi^2}{m_1^2m_2^2}
- 12\,\frac{\pipii\,\npi\npii}{m_1^2m_2^2}
- 3\,\frac{\npi^2\npii^2}{m_1^2m_2^2} \Bigg)
\nonumber\\[1ex]&\kern-10ex
+ \frac{1}{24}\frac{G^2m_1m_2}{r_{12}^2} \Bigg(
(112m_1+45m_2)\frac{\pipi}{m_1^2}
+ (15m_1+2m_2)\frac{\piipii}{m_2^2}
- \frac{1}{2}(209m_1+115m_2)\frac{\pipii}{m_1m_2}
\nonumber\\[1ex]&\kern-10ex
- (31m_1+5m_2)\frac{\npi\npii}{m_1m_2}
+ \frac{\npi^2}{m_1}
- \frac{\npii^2}{m_2} \Bigg)
- \frac{1}{8} \frac{Gm_1m_2}{r_{12}}\frac{G^2 (m_1^2+5m_1m_2+m_2^2)}{r_{12}^2},
\\[2ex]
c^8\,M_1^{\rm 4PN}(\xa,\pa) &= -\frac{5\pipip^4}{128 m_1^7}
+ \frac{G m_1 m_2}{r_{12}} M_{46}(\xa,\pa)
+ \frac{G^2 m_1 m_2}{r_{12}^2} \Big(m_1\,M_{441}(\xa,\pa) + m_2\,M_{442}(\xa,\pa)\Big)
\nonumber\\[1ex]&\kern-10ex
+ \frac{G^3 m_1 m_2}{r_{12}^3} \Big( m_1^2\,M_{421}(\xa,\pa)
+ m_1 m_2\,M_{422}(\xa,\pa) + m_2^2\,M_{423}(\xa,\pa) \Big)
+ \frac{G^4 m_1 m_2}{r_{12}^4}M_{40}(\xa,\pa),
\\[2ex]
M_{46}(\xa,\pa) &=
-\frac{13 \pipip^3}{32 m_1^6}
-\frac{15 \npi^4 \npii^2}{256 m_1^4m_2^2}
+\frac{45 \npi^2 \npii^2 \pipi}{128 m_1^4m_2^2}
-\frac{91 \npii^2 \pipip^2}{256 m_1^4 m_2^2}
\nonumber\\[1ex]&\kern-10ex
-\frac{5\npi^3 \npii \pipii}{32 m_1^4 m_2^2}
+\frac{25 \npi\npii \pipi \pipii}{32 m_1^4 m_2^2}
+\frac{5 \npi^2\pipii^2}{64 m_1^4 m_2^2}
\nonumber\\[1ex]&\kern-10ex
+\frac{7 \pipi \pipii^2}{64 m_1^4m_2^2}
+\frac{11 \npi^4 \piipii}{256 m_1^4 m_2^2}
-\frac{47\npi^2 \pipi \piipii}{128 m_1^4 m_2^2}
+\frac{91 \pipip^2\piipii}{256 m_1^4 m_2^2}
\nonumber\\[1ex]&\kern-10ex
+\frac{5 \npi^3 \npii^3}{32 m_1^3m_2^3}
-\frac{7 \npi \npii^3 \pipi}{32 m_1^3 m_2^3}
+\frac{15 \npi^2 \npii^2 \pipii}{32 m_1^3 m_2^3}
\nonumber\\[1ex]&\kern-10ex
+\frac{7 \npii^2\pipi \pipii}{32 m_1^3 m_2^3}
-\frac{5 \npi \npii\pipii^2}{16 m_1^3 m_2^3}
-\frac{\pipii^3}{16 m_1^3 m_2^3}
-\frac{11\npi^3 \npii \piipii}{32 m_1^3 m_2^3}
\nonumber\\[1ex]&\kern-10ex
+\frac{7 \npi\npii \pipi \piipii}{32 m_1^3 m_2^3}
-\frac{5 \npi^2 \pipii\piipii}{32 m_1^3 m_2^3}
+\frac{\pipi \pipii \piipii}{32 m_1^3m_2^3}
\nonumber\\[1ex]&\kern-10ex
+\frac{15 \npi^2 \npii^4}{256 m_1^2 m_2^4}
-\frac{11\npii^4 \pipi}{256 m_1^2 m_2^4}
+\frac{5 \npi \npii^3\pipii}{32 m_1^2 m_2^4}
-\frac{5 \npii^2 \pipii^2}{64 m_1^2m_2^4}
\nonumber\\[1ex]&\kern-10ex
-\frac{21 \npi^2 \npii^2 \piipii}{128 m_1^2 m_2^4}
+\frac{7\npii^2 \pipi \piipii}{128 m_1^2 m_2^4}
-\frac{\npi \npii\pipii \piipii}{32 m_1^2 m_2^4}
\nonumber\\[1ex]&\kern-10ex
+\frac{\pipii^2 \piipii}{64 m_1^2m_2^4}
+\frac{11 \npi^2 \piipiip^2}{256 m_1^2 m_2^4}
+\frac{37 \pipi\piipiip^2}{256 m_1^2 m_2^4}
-\frac{\piipiip^3}{32 m_2^6},
\\[2ex]
M_{441}(\xa,\pa) &=
\frac{7711 \npi^4}{3840 m_1^4}
-\frac{2689 \npi^2 \pipi}{3840m_1^4}
+\frac{2683 \pipip^2}{1920 m_1^4}
-\frac{67 \npi^3 \npii}{30m_1^3 m_2}
\nonumber\\[1ex]&\kern-10ex
+\frac{1621 \npi \npii \pipi}{1920 m_1^3m_2}
-\frac{411 \npi^2 \pipii}{1280 m_1^3 m_2}
-\frac{25021 \pipi\pipii}{3840 m_1^3 m_2}
\nonumber\\[1ex]&\kern-10ex
+\frac{289 \npi^2 \npii^2}{128 m_1^2m_2^2}
-\frac{259 \npii^2 \pipi}{128 m_1^2 m_2^2}
+\frac{689\npi \npii \pipii}{192 m_1^2 m_2^2}
+\frac{11 \pipii^2}{48m_1^2 m_2^2}
\nonumber\\[1ex]&\kern-10ex
-\frac{147 \npi^2 \piipii}{64 m_1^2 m_2^2}
+\frac{283\pipi \piipii}{64 m_1^2 m_2^2}
+\frac{7 \npi \npii^3}{12m_1 m_2^3}
+\frac{49 \npii^2 \pipii}{48 m_1 m_2^3}
\nonumber\\[1ex]&\kern-10ex
-\frac{7\npi \npii \piipii}{6 m_1 m_2^3}
-\frac{7 \pipii \piipii}{48m_1 m_2^3}
-\frac{9 \piipiip^2}{32 m_2^4},
\\[2ex]
M_{442}(\xa,\pa) &=
-\frac{45 \pipip^2}{32 m_1^4}
+\frac{7 \pipi \pipii}{48 m_1^3m_2}
+\frac{7 \npi \npii \pipi}{6 m_1^3 m_2}
-\frac{49\npi^2 \pipii}{48 m_1^3 m_2}
\nonumber\\[1ex]&\kern-10ex
-\frac{7 \npi^3 \npii}{12 m_1^3 m_2}
+\frac{7 \pipii^2}{24 m_1^2 m_2^2}
+\frac{635 \pipi\piipii}{192 m_1^2 m_2^2}
-\frac{983 \npi^2 \piipii}{384 m_1^2m_2^2}
\nonumber\\[1ex]&\kern-10ex
+\frac{413 \npi^2 \npii^2}{384 m_1^2 m_2^2}
-\frac{331\npii^2 \pipi}{192 m_1^2 m_2^2}
+\frac{437 \npi \npii\pipii}{64 m_1^2 m_2^2}
\nonumber\\[1ex]&\kern-10ex
+\frac{11 \npi \npii^3}{15 m_1m_2^3}
-\frac{1349 \npii^2 \pipii}{1280 m_1 m_2^3}
-\frac{5221\npi \npii \piipii}{1920 m_1 m_2^3}
\nonumber\\[1ex]&\kern-10ex
-\frac{2579 \pipii\piipii}{3840 m_1 m_2^3}
+\frac{6769 \npii^2 \piipii}{3840m_2^4}
-\frac{2563 \piipiip^2}{1920 m_2^4}
-\frac{2037 \npii^4}{1280 m_2^4},
\\[2ex]
M_{421}(\xa,\pa) &=
-\frac{179843 \pipi}{14400 m_1^2}
+\frac{10223 \pipii}{1200 m_1m_2}
-\frac{15 \piipii}{16 m_2^2}
+\frac{8881 \npi \npii}{2400 m_1m_2}
+\frac{17737 \npi^2}{1600 m_1^2},
\\[2ex]
M_{422}(\xa,\pa) &=
\left(\frac{8225 \pi ^2}{16384}-\frac{12007}{1152}\right)\frac{\pipi}{m_1^2}
+\left(\frac{143}{16}-\frac{\pi ^2}{64}\right)\frac{\pipii}{m_1m_2}
+\left(\frac{655}{1152}-\frac{7969 \pi ^2}{16384}\right)\frac{\piipii}{m_2^2}
\nonumber\\[1ex]&\kern-10ex
+\left(\frac{6963 \pi ^2}{16384}-\frac{40697}{3840}\right)\frac{\npi^2}{m_1^2}
+\left(\frac{119}{16}+\frac{3 \pi ^2}{64}\right) \frac{\npi\npii}{m_1 m_2}
\nonumber\\[1ex]&\kern-10ex
+\left(\frac{30377}{3840}-\frac{7731 \pi ^2}{16384}\right)\frac{\npii^2}{m_2^2},
\\[2ex]
M_{423}(\xa,\pa) &=
- \frac{35\pipi}{16 m_1^2}
+ \frac{1327\pipii}{1200 m_1 m_2}
+ \frac{52343\piipii}{14400 m_2^2}
- \frac{2581\npi\npii}{2400 m_1 m_2}
- \frac{15737\npii^2}{1600 m_2^2},
\\[2ex]
M_{40}(\xa,\pa) &=
\frac{m_1^3}{16}
+ \left(\frac{3371\pi^2}{6144}-\frac{6701}{1440}\right) m_1^2 m_2
+ \left(\frac{20321}{1440}-\frac{7403\pi^2}{6144}\right) m_1 m_2^2
+ \frac{m_2^3}{16},
\\[2ex]
c^4\,N^{\text{2PN}}_1(\xa,\pa) &= -\frac{5}{4}\, G\, \npii,
\\[2ex]
c^6\,N^{\text{3PN}}_1(\xa,\pa) &= \frac{1}{8}\frac{G}{m_1m_2}
\Big( 2\,\pipii\npii - \piipii\,\npi + 3\,\npi\npii^2 \Big)
\nonumber\\[2ex]&\quad
+ \frac{1}{48}\frac{G^2}{r_{12}}
\Big( 19\,m_2\,\npi + \left(130\,m_1+137\,m_2\right)\npii \Big),
\\[2ex]
c^8\,N_1^{\rm 4PN}(\xa,\pa) &= G m_2 N_{45}(\xa,\pa)
+ \frac{G^2 m_2}{r_{12}} \Big(m_1\,N_{431}(\xa,\pa) + m_2\,N_{432}(\xa,\pa)\Big)
\nonumber\\[1ex]&\kern-10ex
+ \frac{G^3 m_2}{r_{12}^2} \Big(m_1^2\,N_{411}(\xa,\pa)
+ m_1 m_2\,N_{412}(\xa,\pa) + m_2^2\,N_{413}(\xa,\pa)\Big),
\\[2ex]
N_{45}(\xa,\pa) &=
-\frac{5 \npi^3 \npii^2}{64 m_1^3 m_2^2}
+\frac{\npi \npii^2\pipi}{64 m_1^3 m_2^2}
+\frac{5 \npi^2 \npii \pipii}{32m_1^3 m_2^2}
\nonumber\\[1ex]&\kern-10ex
-\frac{\npii \pipi \pipii}{32 m_1^3m_2^2}
+\frac{3 \npi \pipii^2}{32 m_1^3 m_2^2}
-\frac{\npi^3\piipii}{64 m_1^3 m_2^2}
-\frac{\npi \pipi \piipii}{64 m_1^3m_2^2}
\nonumber\\[1ex]&\kern-10ex
+\frac{\npi^2 \npii^3}{32 m_1^2 m_2^3}
-\frac{7 \npii^3\pipi}{32 m_1^2 m_2^3}
+\frac{3 \npi \npii^2 \pipii}{16m_1^2 m_2^3}
+\frac{\npii \pipii^2}{16 m_1^2 m_2^3}
\nonumber\\[1ex]&\kern-10ex
-\frac{9\npi^2 \npii \piipii}{32 m_1^2 m_2^3}
+\frac{5 \npii \pipi\piipii}{32 m_1^2 m_2^3}
-\frac{3 \npi \pipii \piipii}{16 m_1^2m_2^3}
-\frac{11 \npi \npii^4}{128 m_1 m_2^4}
\nonumber\\[1ex]&\kern-10ex
+\frac{\npii^3\pipii}{32 m_1 m_2^4}
+\frac{7 \npi \npii^2 \piipii}{64 m_1m_2^4}
+\frac{\npii \pipii \piipii}{32 m_1 m_2^4}
-\frac{3\npi \piipiip^2}{128 m_1 m_2^4},
\\[2ex]
N_{431}(\xa,\pa) &=
-\frac{387 \npi^3}{1280 m_1^3}
+\frac{10429 \npi \pipi}{3840m_1^3}
-\frac{751 \npi^2 \npii}{480 m_1^2 m_2}
+\frac{2209\npii \pipi}{640 m_1^2 m_2}
\nonumber\\[1ex]&\kern-10ex
-\frac{6851 \npi \pipii}{1920m_1^2 m_2}
+\frac{43 \npi \npii^2}{192 m_1 m_2^2}
-\frac{125\npii \pipii}{192 m_1 m_2^2}
+\frac{25 \npi \piipii}{48 m_1m_2^2}
\nonumber\\[1ex]&\kern-10ex
-\frac{7 \npii^3}{8 m_2^3}
+\frac{7 \npii \piipii}{12 m_2^3},
\\[2ex]
N_{432}(\xa,\pa) &=
\frac{7 \npii \pipi}{48 m_1^2 m_2}
+\frac{7 \npi \pipii}{24m_1^2 m_2}
-\frac{49 \npi^2 \npii}{48 m_1^2 m_2}
+\frac{295\npi \npii^2}{384 m_1 m_2^2}
\nonumber\\[1ex]&\kern-10ex
-\frac{5 \npii \pipii}{24m_1 m_2^2}
-\frac{155 \npi \piipii}{384 m_1 m_2^2}
-\frac{5999\npii^3}{3840 m_2^3}
+\frac{11251 \npii \piipii}{3840 m_2^3},
\\[2ex]
N_{411}(\xa,\pa) &=
-\frac{37397 \npi}{7200 m_1}-\frac{12311 \npii}{2400 m_2},
\\[2ex]
N_{412}(\xa,\pa) &=
\left(\frac{5005 \pi ^2}{8192}-\frac{81643}{11520}\right)\frac{\npi}{m_1}
+\left(\frac{773 \pi ^2}{2048}-\frac{61177}{11520}\right)\frac{\npii}{m_2},
\\[2ex]
N_{413}(\xa,\pa) &=
-\frac{7073 \npii}{1200 m_2}.
\end{align}
\end{subequations}
\end{widetext}

The fulfillment of the Poincar\'e algebra does not imply a complete check of the Hamiltonian,
but rather a check of all terms besides the purely static ones.
Of course, the Poincar\'e algebra is invariant against canonical transformations,
particularly those induced by coordinate-gauge transformations,
so the single terms in the Hamiltonian have no direct physical meaning.
The reader interested in a representation of a higher-order PN Hamiltonian
through center-of-mass and relative coordinates is referred to \cite{GS15}.

\begin{acknowledgments}

The authors thank Thibault Damour for useful discussions.
We also thank Jan Steinhoff for bringing to our attention the review by H.\ Georgi.
The work of P.J.\ was supported in part by the Polish National Science Center Grant
No.\ 2014/15/Z/ST9/00038 (\textit{Networking and R\&D for the Einstein Telescope}).
P.J.\ also acknowledges support through the Transregional Collaborative Research Center 7
(SFB/TR7) of the German Research Foundation (DFG) entitled ``Gravitational Wave Astronomy: Methods--Sources--Observation''
when performing computations in Jena before mid 2013. 

\end{acknowledgments}

\appendix

\section{Regularization}
\label{Regularization}

In this appendix we describe techniques which we have used to regularize
divergent integrals which appear in our paper.

\subsection{Three-dimensional Riesz-implemented Hadamard regularization}
\label{3RHregularization}

The Riesz-implemented Hadamard (RH) regularization
was developed in the context of deriving PN equations of motion of binary systems
in Refs.\ \cite{J1997,JS98} (see also \cite{DJS2000b})
to deal with locally divergent integrals computed in 3 dimensions.
The RH regularization relies on multiplying the full integrand, say $i(\x)$,
of the divergent integral by a regularization factor,
\be
\label{regfactor}
i(\x) \longrightarrow i(\x)\Big(\frac{r_1}{s_1}\Big)^{\epsilon_1} \Big(\frac{r_2}{s_2}\Big)^{\epsilon_2},
\ee
and studying the double limit $\epsilon_1\to0$, $\epsilon_2\to0$
(here $s_1$ and $s_2$ are arbitrary three-dimensional UV regularization scales).
Let us thus consider such an integral performed over the whole space ${\mathbb R}^3$
and let us assume that it develops \emph{only local poles} (so it is convergent at spatial infinity).
The value of the integral, after performing the RH regularization in 3 dimensions,
has the structure
\begin{align}
\label{IRHuv}
I^{\text{RH}}&(3;\epsilon_1,\epsilon_2) :=
\int_{{\mathbb R}^3} i({\mathbf x})
\Big(\frac{r_1}{s_1}\Big)^{\epsilon_1} \Big(\frac{r_2}{s_2}\Big)^{\epsilon_2}\,\md^3x
\nonumber\\[1ex]
&= A + c_1 \Big(\frac{1}{\epsilon_1} + \ln\frac{r_{12}}{s_1} \Big)
+ c_2 \Big(\frac{1}{\epsilon_2} + \ln\frac{r_{12}}{s_2} \Big)
\nonumber\\[1ex]
&\quad + \mathcal{O}(\epsilon_1,\epsilon_2).
\end{align}

In the case of an integral over ${\mathbb R}^3$ developing poles \emph{only at spatial infinity}
(so it is locally integrable)
it would be enough to use a regularization factor of the form $(r/r_0)^\epsilon$
(where $r_0$ is an IR regularization scale), but it is more convenient to use the factor
\be
\Big(\frac{r_1}{r_0}\Big)^{a\epsilon} \Big(\frac{r_2}{r_0}\Big)^{b\epsilon}
\ee
and study the limit $\epsilon\to0$.
Let us denote the integrand again by $i(\mathbf{x})$. The value of the integral,
after performing the RH regularization in 3~dimensions,
has the structure
\begin{align}
\label{inf1}
I^{\text{RH}}(3;a,b,\epsilon)
&:= \int_{{\mathbb R}^3} i({\mathbf x}) \Big(\frac{r_1}{r_0}\Big)^{a\epsilon} \Big(\frac{r_2}{r_0}\Big)^{b\epsilon}\,\md^3x
\nonumber\\[1ex]
&\phantom{:}= A - c_\infty \bigg(\frac{1}{(a+b)\epsilon} + \ln\frac{r_{12}}{r_0} \bigg)
+ \mathcal{O}(\epsilon).
\end{align}

Many integrals appearing in Eqs.\ \eqref{IRHuv} and \eqref{inf1}
we compute using three-dimensional form of the following formula,
first derived in $d$ dimensions probably by Riesz \cite{Riesz49}:
\begin{multline}
\label{riesz}
\int r_1^\alpha\,r_2^\beta\,\md^dx = \pi^{d/2}\,r_{12}^{\alpha+\beta+d}
\\[1ex]
\times \frac{\Gamma\left(\frac{\alpha+d}{2}\right)
\Gamma\left(\frac{\beta+d}{2}\right)
\Gamma\left(-\frac{\alpha+\beta+d}{2}\right)}
{\Gamma\left(-\frac{\alpha}{2}\right)
\Gamma\left(-\frac{\beta}{2}\right)
\Gamma\left(\frac{\alpha+\beta+2d}{2}\right)}.
\end{multline}
To compute the 4PN-accurate Hamiltonian one needs to employ
a generalization of the three-dimensional version of the Riesz formula \eqref{riesz}
for integrands of the form $r_1^\alpha\,r_2^\beta\,\ms^\gamma$, where
\be
\ms := r_1 + r_2 + r_{12}.
\ee
Such formula was derived in Ref.\ \cite{JS98} and it reads
\begin{subequations}
\label{griesz}
\be
\label{r2a}
\int r_1^\alpha\,r_2^\beta\,\ms^\gamma\,\md^3x
= R(\alpha,\beta,\gamma)\,r_{12}^{\alpha+\beta+\gamma+3},
\ee
where [let us note that the formula given below is invariant under the permutation $(a\leftrightarrow b)$]
\begin{widetext}
\begin{align}
\label{r2b}
R(\alpha,\beta,\gamma)&:=2\pi
\frac{\Gamma\left(\alpha+2\right)\Gamma\left(\beta+2\right)
\Gamma\left(-\alpha-\beta-\gamma-4\right)}{\Gamma\left(-\gamma\right)}
\nonumber\\[2ex]&
\times\left[
I_{1/2}\left(\alpha+2,-\alpha-\gamma-2\right)
+I_{1/2}\left(\beta+2,-\beta-\gamma-2\right)
-I_{1/2}\left(\alpha+\beta+4,-\alpha-\beta-\gamma-4\right)-1
\right].
\end{align}
\end{widetext}
\end{subequations}
The function $I_{1/2}$ in Eq.\ (\ref{r2b}) is defined as follows:
\be
I_{1/2}\left(x,y\right):=\frac{B_{1/2}\left(x,y\right)}{B\left(x,y\right)},
\ee 
where $B$ stands for the beta function (Euler's integral of the first kind) and
$B_{1/2}$ is the incomplete beta function; it can be expressed in terms of the
Gauss hypergeometric function $_2F_1$:
\be
B_{1/2}\left(x,y\right)=\frac{1}{2^x x}\,
{_2F_1}\!\!\left(1-y,x;x+1;\frac{1}{2}\right).
\ee

The regularization procedure based on the formulas \eqref{riesz} and \eqref{griesz}
consists of several steps. We enumerate them now.
The most general integrand we have to consider has the form
\be
\label{gintegrand1}
\nipi^{q_1} \niipi^{q_2} \nipii^{q_3} \niipii^{q_4} r_1^\alpha r_2^\beta \ms^\gamma,
\ee
where $q_1$, \ldots, $q_4$ are non-negative integers
and $\gamma$ is a negative integer.
We first eliminate the unit vector ${\bf n}_2$ by the identity
\be
\label{n2byn1&n12}
{\bf n}_2=\frac{r_1}{r_2}{\bf n}_1+\frac{r_{12}}{r_2}{\bf n}_{12}.
\ee
We thus plug Eq.\ \eqref{n2byn1&n12} into \eqref{gintegrand1}
and expand the scalar product ${\bf n}_1\cdot{\bf n}_{12}$
by means of the relation
\be
\label{n1n12}
{\bf n}_1\cdot{\bf n}_{12}=\frac{r_2^2-r_1^2-r_{12}^2}{2r_1 r_{12}}.
\ee
After this the most general integrand reduces to
\be
\label{gintegrand2}
\nipi^{q_1} \nipii^{q_2} r_1^\alpha r_2^\beta \ms^\gamma,
\ee
where again $q_1$ and $q_2$ are non-negative integers.

We perform integration in \emph{prolate spheroidal coordinates}.
By using these coordinates it is possible to represent integrand \eqref{gintegrand2}
as a linear combination of integrands of the type $r_1^\alpha r_2^\beta \ms^\gamma$.
To show this let us locate the particles in the focal points of the prolate spheroidal coordinates
[they lie along the $z$ axis of the Cartesian coordinate system $(x,y,z)$],
so the particles' position vectors have the following Cartesian coordinates:
\be
\mathbf{x}_1 = (0,0,-r_{12}/2),
\quad
\mathbf{x}_2 = (0,0,r_{12}/2).
\ee
The Cartesian components of the unit vector ${\mathbf n}_1:=(\x-\x_1)/r_1$
can be expressed by $r_1$, $r_2$ and the azimuthal angle $\phi$,
\begin{subequations}
\label{n1co}
\begin{align}
n_1^x &= \frac{\sqrt{\left[\left(r_1+r_2\right)^2-r_{12}^2\right]
\left[r_{12}^2-\left(r_1-r_2\right)^2\right]}}{2r_1 r_{12}}\cos\phi,
\\
n_1^y &= \frac{\sqrt{\left[\left(r_1+r_2\right)^2-r_{12}^2\right]
\left[r_{12}^2-\left(r_1-r_2\right)^2\right]}}{2r_1 r_{12}}\sin\phi,
\\
n_1^z &= \frac{r_{12}^2+r_1^2-r_2^2}{2r_1 r_{12}}.
\end{align}
\end{subequations}
Without loss of generality we can place the vector $\mathbf{p}_1$ in the $(x,z)$ plane
(we can also assume that $p_{1x}>0$).
One can then show that the Cartesian components of the vector $\mathbf{p}_1$
are as follows:
\begin{subequations}
\label{p1co}
\begin{align}
p_{1x} &= \sqrt{\pipi - \npi^2},
\\[1ex]
p_{1y} &= 0,
\\[1ex]
p_{1z} &= -\npi.
\end{align}
\end{subequations}
The $x$ and $z$ Cartesian components of the vector $\mathbf{p}_2$ read
\begin{subequations}
\label{p2co}
\begin{align}
p_{2x} &= \frac{\pipii - \npi\npii}{\sqrt{\pipi - \npi^2}},
\\[1ex]
p_{2z} &= -\npii;
\end{align}
the $y$ component of the vector $\mathbf{p}_2$ can be computed from
\be
\label{p2y}
p_{2y} = \pm \sqrt{\piipii - p_{2x}^2 - p_{2z}^2}.
\ee
\end{subequations}
We have checked that the result of the procedure described below
does not depend on the $\pm$ ambiguity in Eq.\ \eqref{p2y}.

Making use of Eqs.\ \eqref{n1co}--\eqref{p2co} we compute scalar products $\nipi$ and $\nipii$
in Eq.\ \eqref{gintegrand2}. After this the integrand
\eqref{gintegrand2} becomes a sum of terms of the form $A(r_1,r_2)B(\phi)$,
where $B(\phi)$ is a polynomial in $\sin\phi$ and $\cos\phi$.
Each of these terms we integrate in the following way:
\begin{align}
\int A(r_1,r_2)B(\phi)\,\md^3x &= \int A(r_1,r_2)\,\md^2x \int_0^{2\pi}B(\phi)\,\md\phi
\nonumber\\[1ex]
&= \left<B\right> \int A(r_1,r_2)\,\md^3x,
\end{align}
where $\left<B\right>$ is the average over the angle $\phi$,
\be
\left<B\right> := \frac{1}{2\pi} \int_0^{2\pi}B(\phi)\,\md\phi.
\ee
After this step the integrand \eqref{gintegrand2} becomes the linear combination of the type
\be
\label{r01}
\sum_I c_I\,r_1^{\alpha_I}r_2^{\beta_I}\ms^{\gamma_I},
\ee
where the constant coefficients $c_I$ may depend only on $r_{12}$, $\pipi$, $\pipii$, $\piipii$, $\npi$, and $\npii$.
The integral of \eqref{r01} is computed by means of Eq.\ \eqref{IRHuv} or \eqref{inf1}
and with the usage of formulas \eqref{riesz} and \eqref{griesz}.

Appendix A6 of Ref.\ \cite{HSS13} contains generalization of the above presented procedure
to $d$ space dimensions (and it employs prolate spheroidal coordinates in $d$ dimensions).

\subsection{UV corrections}
\label{UVcorr}

Reference \cite{JS98} showed that the three-dimensional RH regularization
described above used to derive the 3PN two-point-mass Hamiltonian gave ambiguous results.
Namely, by means of integration by parts (assuming that all involved integrals are convergent
at infinity, so all surface terms can be neglected) one can replace one form of Hamiltonian density (or its part)
by some other form. Integration of both equivalent densities should give the same result, but it did not.
To correct the result of the three-dimensional RH regularization (i.e., to remove ambiguity),
Ref.\ \cite{DJS2001} (see Secs.\ 3 and 4 there) developed dimensional-regularization (DR) technique,\footnote{
The presentation of this technique given below is an improved and more complete version
of the explanations contained in Sec.\ III of \cite{JS13}.}
which we have also used to make the results of the three-dimensional RH regularization
of the locally divergent part of the 4PN-accurate Hamiltonian unique.

The technique of Ref.\ \cite{DJS2001} boils down to the computation of the difference
\be
\lim_{d\to3} H_\text{4PN}^\text{loc}(d) - H_\text{4PN}^\text{RH loc}(3),
\ee
where $H_\text{4PN}^\text{RH loc}(3)$ is the ``local part'' of the Hamiltonian
obtained by means of the three-dimensional RH regularization
[it is the sum of all integrals of the type $I^{\text{RH}}(3;\epsilon_1,\epsilon_2)$
introduced in Eq.\ \eqref{IRHuv}];
$H_\text{4PN}^\text{loc}(d)$ is its $d$-dimensional counterpart.

Reference \cite{DJS2001} showed that to find the DR correction to the integral $I^{\text{RH}}(3;\epsilon_1,\epsilon_2)$
related with the local pole at, say, $\mathbf{x}=\mathbf{x}_1$,
it is enough to consider only this part of the integrand $i(\x)$ which develops
logarithmic singularities, i.e.\ which locally behaves like $1/r_1^3$,
\be
\label{UVexp3}
i(\mathbf{x}) = \cdots + \tilde{c}_1(\mathbf{n}_1)\,r_1^{-3} + \cdots,
\quad \text{when}\ \mathbf{x}\to\mathbf{x}_1.
\ee
Then the pole part of the integral \eqref{IRHuv}
(related with the singularity at $\mathbf{x}=\mathbf{x}_1$)
we recover by RH regularization of the integral
of $\tilde{c}_1(\mathbf{n}_1)\,r_1^{-3}$
over the ball $B(\mathbf{x}_1,{\ell_1})$ of radius $\ell_1$ surrounding the particle $\mathbf{x}_1$.
The RH regularized value of this integral reads
\begin{align}
I_1^{\text{RH}}(3;\epsilon_1) &:=
\int_{B(\mathbf{x}_1,{\ell_1})} \tilde{c}_1(\mathbf{n}_1) \, r_1^{-3}
\Big(\frac{r_1}{s_1}\Big)^{\epsilon_1} \, \md^3 {\mathbf r}_1
\nonumber\\
&\phantom{:}= c_1 \int_0^{\ell_1} r_1^{-1} \Big(\frac{r_1}{s_1}\Big)^{\epsilon_1}\,\md r_1,
\end{align}
where $c_1$ is the angle-averaged value of the coefficient $\tilde{c}_1(\mathbf{n}_1)$.
The expansion of the integral $I_1^{\text{RH}}(3;\epsilon_1)$ around $\epsilon_1=0$ equals
\be
I_1^{\text{RH}}(3;\epsilon_1) = c_1\Big(\frac{1}{\epsilon_1}
+ \ln\frac{\ell_1}{s_1}\Big) + \mathcal{O}(\epsilon_1).
\ee

The idea of the technique developed in \cite{DJS2001}
relies on replacing the RH-regularized value of the three-dimensional integral $I_1^{\text{RH}}(3;\epsilon_1)$
by the value of its $d$-dimensional version $I_1(d)$.
We thus consider the $d$-dimensional counterpart of the expansion \eqref{UVexp3}.
It reads
\be
i(\mathbf{x}) = \cdots + \ell_0^{k(d-3)}\tilde{\mc}_1(d;{\mathbf n}_1) \, r_1^{6-3d} + \cdots,
\quad \text{when}\ \mathbf{x}\to\mathbf{x}_1,
\ee
where $\ell_0$ is the scale which relates
the Newtonian $G_\textrm{N}$ and the $D$-dimensional ($D=d+1$) $G_D$ gravitational constants,
\be
\label{ell0}
G_D = G_\textrm{N}\,\ell_0^{d-3}.
\ee
The number $k$ in the exponent of $\ell_0^{k(d-3)}$ is related with the momentum-order of the considered term
[the term with $k$ is of the order of $\mathcal{O}(p^{10-2k})$, where $k=1,\ldots,5$].
The integral $I_1(d)$ we define as
\begin{align}
\label{I1def}
I_1(d) &:= \ell_0^{k(d-3)} \int_{B(\mathbf{x}_1,{\ell_1})}
\tilde{\mc}_1(d;{\mathbf n}_1) \, r_1^{6-3d}\, \md^d {\mathbf r}_1
\nonumber\\
&\phantom{:}= \ell_0^{k(d-3)} \mc_1(d) \int_0^{\ell_1} r_1^{5-2d}\,\md r_1,
\end{align}
where $\mc_1(d)$ is the angle-averaged value
of the coefficient $\tilde{\mc}_1(d;{\mathbf n}_1)$,
\be
\mc_1(d) := \oint_{\mathbb{S}^{d-1}(\mathbf{0},1)}
\tilde{\mc}_1(d;{\mathbf n}_1)\,\md\Omega_{d-1}.
\ee
One checks that always
\be
\label{limitc1IR}
\lim_{d\to3} \mc_1(d) = \mc_1(3) = c_1.
\ee
The radial integral in Eq.\ \eqref{I1def} is convergent
if the real part $\Re(d)$ of $d$ fulfills $\Re(d)<3$.
Let us introduce
$$
\varepsilon := d-3
$$
and let us expand $\mc_1(d)$ around $\varepsilon=0$,
\be
\label{UVcorr10}
\mc_1(d) = \mc_1(3+\varepsilon) = c_1 + \mc'_1(3)\varepsilon + \mathcal{O}(\varepsilon^2).
\ee
Then the expansion of the integral $I_1(d)$ around $\varepsilon=0$ reads
\be
\label{UVcorr11}
I_1(d) = -\frac{c_1}{2\varepsilon} -\frac{1}{2} \mc'_1(3)
+ c_1 \ln\frac{\ell_1}{\ell_0} + \mathcal{O}(\varepsilon).
\ee
In Eqs.\ \eqref{UVcorr10}--\eqref{UVcorr11} we have used \eqref{limitc1IR}.
Let us note that the coefficient $\mc'_1(3)$ usually depends on $\ln r_{12}$
and it has the structure
\be
\mc'_1(3) = \mc'_{11}(3) + \mc'_{12}(3) \ln\frac{r_{12}}{\ell_0}.
\ee
Therefore the DR correction also changes the terms $\propto\ln{r_{12}}$.

The DR correction to the RH-regularized value of the integral
$I^{\text{RH}}(3;\epsilon_1,\epsilon_2)$ relies on replacing this integral by
\be
I^{\text{RH}}(3;\epsilon_1,\epsilon_2) + \Delta I_1 + \Delta I_2,
\ee
where
\be
\Delta I_a := I_a(d) - I_1^{\text{RH}}(3;\epsilon_1),
\quad a=1,2.
\ee
Then one computes the double limit
\begin{multline}
\label{IRHcorr}
\lim_{\substack{\epsilon_1\to3 \\ \epsilon_2\to3}}
\Big(I^{\text{RH}}(3;\epsilon_1,\epsilon_2) + \Delta I_1 + \Delta I_2\Big)
= A -\frac{c_1+c_2}{2\varepsilon}
\\[1ex]
- \frac{1}{2} \big(\mc'_1(3) + \mc'_2(3)\big)
+ \big(c_1 + c_2\big) \ln\frac{r_{12}}{\ell_0} + \mathcal{O}(\varepsilon)
\\[1ex]
= A -\frac{c_1+c_2}{2\varepsilon}
- \frac{1}{2} \big(\mc'_{11}(3) + \mc'_{21}(3)\big)
\\[1ex]
+ \big(c_1 - \frac{1}{2}\mc'_{12}(3) + c_2
- \frac{1}{2}\mc'_{22}(3)\big) \ln\frac{r_{12}}{\ell_0} + \mathcal{O}(\varepsilon).
\end{multline}
Note that all poles $\propto1/\epsilon_1,1/\epsilon_2$
and all terms depending on radii $\ell_1$, $\ell_2$ or scales $s_1$, $s_2$ cancel each other.
The result \eqref{IRHcorr} is as if all computations were fully done in $d$ dimensions.

In the DR correcting UV divergences of the 3PN two-point-mass Hamiltonian performed in Ref.\ \cite{DJS2001},
after collecting all terms of the type \eqref{IRHcorr} together, all poles $\propto 1/(d-3)$ cancel each other.
This is not the case for the UV divergences of the 4PN two-point-mass Hamiltonian considered in the present paper.
As explained in Sec.\ \ref{removeUVpoles} of our paper after collecting all terms of the type \eqref{IRHcorr},
one has to add to the Hamiltonian a \emph{unique} total time derivative [given in Eq.\ \eqref{ttdD}]
to eliminate all poles $\propto 1/(d-3)$ (together with $\ell_0$-dependent logarithms).

\subsection{IR corrections}
\label{IRcorr}

To regularize IR-divergent integrals which appear in the derivation of the 4PN two-point-mass Hamiltonian,
we have originally developed a technique analogous to the one described above
and used to compute DR corrections to UV-divergent integrals regularized in 3 dimensions.
After completing tedious computations we have obtained IR terms analogous to UV terms described by Eq.\ \eqref{IRHcorr}.
After adding all these terms we have the expression with poles $\propto 1/(d-3)$.
Then we have checked that there exists \emph{no} total time derivative by means of which one can eliminate these poles.
The conclusion was that even in $d$-dimensional computation of IR-divergent integrals
one has to introduce an additional IR regularization factor $(r/s)^B$ with a new scale $s$.

To be consistent with DR correction of UV divergences performed in $d$ dimensions,
we have developed a $d$-dimensional version of IR regularization.
We have devised two different regularization schemes.
We will however see that the results of these regularizations
are identical with the results achieved by means of purely three-dimensional computations.

There is a crucial difference between the results of application of UV and IR regularizations:
the results of IR regularizations depend on an arbitrary IR regularization scale which we had to introduce,
whereas the result of UV regularization does not depend on any scale.
Moreover the results of two different IR regularizations developed below are different;
therefore, we have to conclude that the result of IR regularization is \emph{ambiguous}.
In Appendix \ref{IRamb} we show that the ambiguity can be expressed in terms of only one unknown dimensionless parameter.

The basic idea of both IR regularizations is, similarly to what we have done for UV divergences,
to replace this part of the three-dimensional integral $I_\infty^{\text{RH}}(3;a,b,\epsilon)$ from Eq.\ \eqref{inf1}
which is responsible for IR divergences, by its $d$-dimensional counterpart.
Let us thus consider the three-dimensional integral \eqref{inf1} which is IR divergent.
In all considered cases one checks that its IR pole term proportional to $1/\epsilon$
is related to this part of the integrand $i(\x)$ of \eqref{inf1},
which, after expansion around $r=\infty$ ($r:=|\mathbf{x}|$),
is proportional to $1/r^3$,
\be
\label{e3IR}
i(\mathbf{x}) = \cdots + \infctiii({\mathbf n}) \, r^{-3} + \cdots,
\quad \text{when}\ r\to\infty,
\ee
where $\mathbf{n}:=\mathbf{x}/r$.
It means that one can reproduce the pole term of Eq.\ \eqref{inf1}
by means of integration of $\infctiii({\mathbf n})\,r^{-3}$ taken over the exterior of the ball $\mathbb{B}(\mathbf{0},R)$
(with the center at the origin $\mathbf{0}$ of the coordinate system and the radius $R$).
We thus define
\begin{align}
I_\infty^{\text{RH}}(3;a,b,\epsilon) &:=
\int_{\mathbb{R}^3\setminus \mathbb{B}(\mathbf{0},R)} \infctiii({\mathbf n})\,r^{-3}
\Big(\frac{r}{r_0}\Big)^{(a+b)\epsilon} \, \md^3x
\nonumber\\[1ex]
&\phantom{:}= \infciii \int_R^\infty r^{-3} \Big(\frac{r}{r_0}\Big)^{(a+b)\epsilon}\,r^2 \,\md r
\nonumber\\[1ex]
&\phantom{:}= -\infciii \left(\frac{1}{(a+b)\epsilon} + \ln\frac{R}{r_0}\right)
+ \mathcal{O}(\epsilon),
\end{align}
where $\infciii$ is the angle-averaged value of the coefficient $\infctiii({\mathbf n})$,
\be
\infciii := \oint\limits_{\mathbb{S}^2(\mathbf{0},1)}\infctiii({\mathbf n})\,\md\Omega_2.
\ee

\subsubsection{Modifying behavior of the $\hTTvi ij$ at infinity}
\label{IRcorr1}

All terms contributing to poles at spatial infinity
are collected in Eq.\ \eqref{h12-2-2}.
They (with the exception of the first term) have the structure
\be
\label{infdi2}
f_{ij}(\mathbf{x})\,{\ilhTTivddot ij},
\ee
and the first term in Eq.\ \eqref{h12-2-2} we treat as
$$
\frac{1}{2(d-1)}\fii{\hTTiv ij}\Deltad({\ilhTTivddot ij}).
$$
To regularize all these terms properly we have made the replacement
\be
\label{infdi3}
{\ilhTTivddot ij} \longrightarrow \Deltad^{-1}\bigg[\left(\frac{r}{s}\right)^B\,{\hTTivddot ij}\bigg],
\ee
where $(r/s)^B$ is an IR regularization factor
with $s$ being a new constant (needed to make the regularization factor dimensionless).

After making the replacement \eqref{infdi3} in each term in Eq.\ \eqref{h12-2-2},
we have found in $d$-dimensions for each term this part of its expansion around $r=\infty$
which contributes to the IR divergence.
It is proportional to $r^{6-3d+B}$ and has the structure
\begin{multline}
\label{infdi4}
f_{ij}(\mathbf{x})\,\Deltad^{-1}\Big[\left(\frac{r}{s}\right)^B\,{\hTTivddot ij}\Big]
= \cdots + \tilde{\mc}^{\,1}_\infty(d,B;{\mathbf n})
\\[1ex]
\times s^{-B} \, r^{6-3d+B} + \cdots,
\quad \text{when}\ r\to\infty.
\end{multline}
One can make the replacement \eqref{infdi3} directly in $d=3$ dimensions,
then instead of the expansion \eqref{infdi4} one obtains
\begin{multline}
f_{ij}(\mathbf{x})\,\Delta^{-1}\Big[\left(\frac{r}{s}\right)^B\,{\hTTivddot ij}\Big]
= \cdots + \tilde{c}^{\,0}_\infty(B;{\mathbf n})
\\[1ex]
\times s^{-B} \, r^{B-3} + \cdots,
\quad \text{when}\ r\to\infty.
\end{multline}
We have checked that always
\begin{subequations}
\label{inftyCnd3}
\begin{align}
\lim_{d\to 3} \tilde{\mc}^{\,1}_\infty(d,B;{\mathbf n}) &= \tilde{c}^{\,0}_\infty(B;{\mathbf n}),
\\[1ex]
\lim_{B\to 0} \tilde{c}^{\,0}_\infty(B;{\mathbf n}) &= \infctiii({\mathbf n}),
\end{align}
\end{subequations}
where $\infctiii({\mathbf n})$ is the coefficient in the three-dimensional expansion \eqref{e3IR}.

For all terms from Eq.\ \eqref{h12-2-2} we have computed the integral
\begin{align}
\label{inftyI1}
I^1_\infty(d,B) &:= s^{-B} \int_{\mathbb{R}^d\setminus \mathbb{B}(\mathbf{0},R)}
\tilde{\mc}^{\,1}_\infty(d,B;{\mathbf n}) \, r^{6-3d+B}\, \md^dx
\nonumber\\[1ex]
&\phantom{:}= \mc^{1}_\infty(d,B) \,s^{-B} \int_R^\infty r^{5-2d+B}\,\md r
\nonumber\\[1ex]
&\phantom{:}= -\mc^{1}_\infty(d,B) \,s^{-B} \frac{R^{6-2d+B}}{6-2d+B},
\end{align}
where $\mc^1_\infty(d,B)$ is the angle-averaged value
of the coefficient $\tilde{\mc}^{\,1}_\infty(d,B;{\mathbf n})$,
\be
\mc^1_\infty(d,B) := \oint_{\mathbb{S}^{d-1}(\mathbf{0},1)}
\tilde{\mc}^{\,1}_\infty(d,B;{\mathbf n})\,\md\Omega_{d-1}.
\ee
One easily checks that the equalities analogous to \eqref{inftyCnd3} are also fulfilled
for the angle-averaged values of the coefficients,
\begin{subequations}
\label{inftyCd3}
\begin{align}
\lim_{d\to 3} \mc^1_\infty(d,B) &= c^0_\infty(B),
\\[1ex]
\lim_{B\to 0} c^0_\infty(B) &= \infciii,
\end{align}
\end{subequations}
where the three-dimensional coefficient $c^0_\infty(B)$ is the angle-averaged value
of the three-dimensional coefficient $\tilde{c}^{\,0}_\infty(B;{\mathbf n})$,
\be
c^0_\infty(B) := \oint_{\mathbb{S}^{2}(\mathbf{0},1)}
\tilde{c}^{\,0}_\infty(B;{\mathbf n})\,\md\Omega_{2}.
\ee
The integral $I^1_\infty(d,B)$ can be shortly written as
(let us recall that $\varepsilon:=d-3$)
\be
\label{inftyI2}
I^1_\infty(d,B) = I^1_\infty(3+\varepsilon,B)
= \frac{N(\varepsilon,B)}{B-2\varepsilon},
\ee
where we have defined
\be
N(\varepsilon,B):= -\mc^1_\infty(3+\varepsilon,B) s^{-B} R^{B-2\varepsilon}.
\ee

As the regularized value of the integral \eqref{inftyI2} for $\varepsilon\to0$ and $B\to0$
we take the finite part (FP) of the pole occurring at $B=2\varepsilon$ in $d$ dimensions
(we follow here Ref.\ \cite{Blanchet&others2005}; see especially Sec.\ VIII there).
We thus define
\begin{align}
\label{FP-B}
\mathrm{FP}\,I^1_\infty
&:= \underset{\varepsilon\to0}{\mathrm{FP}}\,\lim_{B\to0} \frac{N(\varepsilon,B)-N(\varepsilon,2\varepsilon)}{B-2\varepsilon}
\nonumber\\[1ex]&
\phantom{:}= \underset{\varepsilon\to0}{\mathrm{FP}}\,\frac{N(\varepsilon,0)-N(\varepsilon,2\varepsilon)}{-2\varepsilon}
\nonumber\\[1ex]&
\phantom{:}= \underset{\varepsilon\to0}{\mathrm{FP}}\,\left(\frac{\partial N}{\partial B}(0,0) + \mathcal{O}(\varepsilon)\right)
= \frac{\partial N}{\partial B}(0,0)
\nonumber\\[1ex]&
\phantom{:}= -\frac{\partial \mc^1_\infty}{\partial B}(3,0) - \mc^1_\infty(3,0) \ln\frac{R}{s}.
\end{align}
As the correction to  integral $I^{\text{RH}}(3;a,b,\epsilon)$
we define the difference
\be
\Delta I^1_\infty := \mathrm{FP}\,I^1_\infty - I_\infty^{\text{RH}}(3;a,b,\epsilon),
\ee
so the regularized value of the three-dimensional IR-divergent integral over $i({\mathbf x})$ reads
\begin{align}
\label{infRHcorr1a}
\bigg(\int_{{\mathbb R}^3} i({\mathbf x})\,\md^3x\bigg)_\textrm{reg\,1}
&:= \lim_{\epsilon\to0} \big(I^{\text{RH}}(3;a,b,\epsilon) + \Delta I^1_\infty\big)
\nonumber\\[1ex]
&\phantom{:}= A - \frac{\partial \mc^1_\infty}{\partial B}(3,0) - \mc^1_\infty(3,0) \ln\frac{r_{12}}{s}.
\end{align}

In view of the equalities \eqref{inftyCd3} it is clear that the value of the right-hand side of Eq.\ \eqref{infRHcorr1a}
would be the same if all computations leading to it were performed in 3 dimensions.
We thus have
\begin{align}
\label{infRHcorr1}
\bigg(\int_{{\mathbb R}^3} i({\mathbf x})\,\md^3x\bigg)_\textrm{reg\,1}
= A - \frac{\partial c^0_\infty}{\partial B}(0) - c_\infty\,\ln\frac{r_{12}}{s}.
\end{align}

\subsubsection{$d$-dimensional RH regularization}
\label{IRcorr2}

We have also considered another way of regularizng IR divergences.
Namely, before integrating an IR-divergent integral over $d$-dimensional space,
we multiply the full integrand by a factor
\be
\label{dRH1}
\Big(\frac{r_1}{s}\Big)^{\alpha_1} \Big(\frac{r_2}{s}\Big)^{\alpha_2}
\ee
and after evaluating it we take the finite part of the IR pole occurring at $\alpha_1+\alpha_2=2(d-3)$.
This recipe means nothing more than the $d$-dimensional version
of the Riesz-implemented Hadamard regularization we performed in 3 dimensions.

In this approach instead of the expansion \eqref{infdi4} one has
(we introduce here $\beta:=\alpha_1+\alpha_2$)
\begin{align}
\label{dRH2}
f_{ij}(\mathbf{x})\,{\ilhTTivddot ij} &= \cdots + \tilde{\mc}^{\,2}_\infty(d;{\mathbf n}) \, s^{-\beta} \, r^{6-3d+\beta} + \cdots,
\nonumber\\
&\qquad \text{when}\ r\to\infty,
\end{align}
and instead of the integral \eqref{inftyI1} one considers the integral
\begin{align}
\label{inftyIR1}
I^2_\infty(d,\beta) &:= s^{-\beta} \int_{\mathbb{R}^d\setminus \mathbb{B}(\mathbf{0},R)}
\tilde{\mc}^{\,2}_\infty(d;{\mathbf n}) \, r^{6-3d+\beta}\, \md^dx
\nonumber\\[1ex]
&\phantom{:}= \mc^2_\infty(d) \,s^{-\beta} \int_R^\infty r^{5-2d+\beta}\,\md r
\nonumber\\[1ex]
&\phantom{:}= -\mc^2_\infty(d) \,s^{-\beta} \frac{R^{6-2d+\beta}}{6-2d+\beta},
\end{align}
where
\be
\mc^2_\infty(d) := \oint_{\mathbb{S}^{d-1}(\mathbf{0},1)}
\tilde{\mc}^{\,2}_\infty(d;{\mathbf n})\,\md\Omega_{d-1}.
\ee
One checks that always
\be
\label{infc2limit}
\lim_{d\to 3} \mc^2_\infty(d) = \mc^2_\infty(3) = c_\infty.
\ee

The crucial difference between the integral $I^2_\infty(d,\beta)$ and the integral $I^1_\infty(d,B)$ of \eqref{inftyI1} is such
that in $I^2_\infty(d,\beta)$ the coefficient $\tilde{\mc}^{\,2}_\infty(d;{\mathbf n})$
[and its angle-averaged value $\mc^2_\infty(d)$] does not depend on $\beta$,
whereas in $I^1_\infty(d,B)$ the coefficient $\tilde{\mc}^{\,1}_\infty(d,B;{\mathbf n})$
[and its angle-averaged value $\mc^1_\infty(d;B)$] does depend on $B$.
The integral \eqref{inftyIR1} can be written as
\begin{subequations}
\begin{align}
\label{inftyIR2}
I^2_\infty(d,\beta) &\phantom{;}= \frac{\eta(\varepsilon,\beta)}{\beta-2\varepsilon},
\\[1ex]
\eta(\varepsilon,\beta) &:= -\mc^2_\infty(3+\varepsilon) s^{-\beta} R^{\beta-2\varepsilon},
\end{align}
\end{subequations}
and its regularized value for $\varepsilon\to0$ and $\beta\to0$
we define as the finite part
\begin{align}
\label{FP-beta}
\mathrm{FP}\,I^2_\infty
&:= \underset{\varepsilon\to0}{\mathrm{FP}}\,\lim_{\beta\to0} \frac{\eta(\varepsilon,\beta)-\eta(\varepsilon,2\varepsilon)}{\beta-2\varepsilon}
\nonumber\\[1ex]&
\phantom{:}= \frac{\partial\eta}{\partial\beta}(0,0)
= -\mc^2_\infty(3) \ln\frac{R}{s}.
\end{align}
The correction to the integral $I^{\text{RH}}(3;a,b,\epsilon)$
is again the difference
\be
\Delta I^2_\infty := \mathrm{FP}\,I^2_\infty - I_\infty^{\text{RH}}(3;a,b,\epsilon),
\ee
so the regularized value of the three-dimensional IR-divergent integral over $i({\mathbf x})$ reads
\begin{align}
\label{infRHcorr2a}
\bigg(\int_{{\mathbb R}^3} i({\mathbf x})\,\md^3x\bigg)_\textrm{reg\,2}
&:=  \lim_{\epsilon\to0} \big(I^{\text{RH}}(3;a,b,\epsilon) + \Delta I^2_\infty\big)
\nonumber\\[1ex]
&\phantom{:}= A - \mc^2_\infty(3) \ln\frac{r_{12}}{s}.
\end{align}

By virtue of Eq.\ \eqref{infc2limit} it is clear that the value of the right-hand side of Eq.\ \eqref{infRHcorr2}
would be the same if all computations were performed in 3 dimensions.
We thus have
\begin{align}
\label{infRHcorr2}
\bigg(\int_{{\mathbb R}^3} i({\mathbf x})\,\md^3x\bigg)_\textrm{reg\,2}
= A - c_\infty\,\ln\frac{r_{12}}{s}.
\end{align}
Let us finally note that the above result can be immediately read off from Eq.\ \eqref{inf1}
after dropping the pole part and identifying the scales $r_0$ and $s$.

\subsubsection{IR ambiguity}
\label{IRamb}

Comparison of the two IR regularization schemes considered above
immediately leads to the conclusion that the results of their
application to computation of the integral of $h_{(12)}^{2,2}$, Eq.\ \eqref{h12-2-2}, are different.
By virtue of Eqs.\ \eqref{infRHcorr1} and \eqref{infRHcorr2}
one gets
\begin{align}
\label{ambIR}
\Delta H^{\textrm{reg\,2,2}}_{\textrm{4PN}}
&:= \Bigg(\int\limits_{{\mathbb R}^3}h_{(12)}^{2,2}\,\md^3x\Bigg)_\textrm{\!\!reg\,1}
- \Bigg(\int\limits_{{\mathbb R}^3} h_{(12)}^{2,2}\,\md^3x\Bigg)_\textrm{\!\!reg\,2}
\nonumber\\[1ex]
&\phantom{:}= -\sum \frac{\partial c^0_\infty}{\partial B}(0),
\end{align}
where the summation is over all terms of the integrand $h_{(12)}^{2,2}$.
We have computed the difference $\Delta H^{\textrm{reg\,2,2}}_{\textrm{4PN}}$.
It turns out that it can be written as
\be
\label{ambIR2}
\Delta H^{\textrm{reg\,2,2}}_{\textrm{4PN}} = \textrm{(total time derivative)}
+ \frac{137}{768} F,
\ee
where $F$ is defined in Eq.\ \eqref{defF}.

To take into account ambiguity \eqref{ambIR2} of IR regularization
we have introduced in Eq.\ \eqref{h12-2-2-a} a dimensionless ambiguity parameter $C$.
According to Eq.\ \eqref{ambIR2} this ambiguity can be written (up to adding a total time derivative)
as a multiple of the term $F$, and this is the content of Eq.\ \eqref{h12-2-2-e}.

\subsection{``Partie finie" value of singular function}
\label{AHPF}

The concept of ``partie finie" value of function at its singular point
was previously used e.g.\ in the calculation of ADM point-particle Hamiltonians
at the 2PN and 2.5PN levels (see Appendix B in \cite{Schafer85}),
and also at the 3PN and 3.5PN orders (see also \cite{J1997}).

Let $f$ be a smooth real-valued function defined in an open ball
$\mathbb{B}(\x_0,E)\subset{\mathbb R}^3$ of radius $E>0$ and origin at
$\x_0\in{\mathbb R}^3$, excluding the point $\x_0$, i.e.,
$f\in C^\infty(\mathbb{B}({\bf x}_0,E)\setminus\{\x_0\})$.
At $\x_0$ the function $f$ is assumed to be singular.
It is enough to consider functions $f$ which have only rational singularities.
We thus assume that there exists positive integer $N$ such that the limit
\be
\label{hpf1}
\lim_{{\bf x}\to{\bf x}_0} f({\bf x})|{\bf x}-{\bf x}_0|^N
\ee
exists and is finite. Let us denote by $N_\text{min}$ the smallest positive integer $N$ 
for which the limit (\ref{hpf1}) exists and is finite. We define
\be
\label{hpf2}
g({\bf x}) := \left\{ \begin{array}{ll}
f({\bf x})|{\bf x}-\x_0|^{N_\text{min}},
& \text{for}\quad {\bf x}\ne\x_0,
\\[1ex]
\lim_{{\bf x}\to{\bf x}_0} f({\bf x})|{\bf x}-\x_0|^{N_\text{min}},
& \text{for}\quad {\bf x}=\x_0.
\end{array} \right.
\ee
We also define two families of auxiliary functions $f_{\bf n}$ and $g_{\bf n}$ 
(labeled by unit vectors ${\bf n}$, ${\bf n}\cdot{\bf n}=1$):
\begin{subequations}
\label{hpf3ab}
\begin{align}
\label{hpf3a}
\left(0;E\right)\ni\epsilon\mapsto f_{\bf n}(\epsilon)
&:= f\left(\x_0+\epsilon{\bf n}\right) \in {\mathbb R},
\\[1ex]
\label{hpf3b}
\left\langle0;E\right)\ni\epsilon\mapsto g_{\bf n}(\epsilon)
&:= g\left(\x_0+\epsilon{\bf n}\right) \in {\mathbb R}.
\end{align}
\end{subequations}
Our final assumption is that {\em for all unit vectors ${\bf n}$ the function
$g_{\bf n}$ is smooth in the half-closed interval $\left\langle0;E\right)$}.
Using Eqs.\ (\ref{hpf2}) and (\ref{hpf3ab}) we obtain
\be
\label{hpf3}
f_{\bf n}(\epsilon) = \frac{g_{\bf n}(\epsilon)}{\epsilon^{N_\text{min}}}.
\ee
From Eq.\ (\ref{hpf3}), after expanding $g_{\bf n}$ into Taylor series around 
$\epsilon=0$, the {\em formal} expansion of $f_{\bf n}$ into Laurent series around 
$\epsilon=0$ follows:
\begin{align}
\label{laurent}
f_{\bf n}(\epsilon) &= \sum\limits_{m=-N_\text{min}}^{\infty}a_m({\bf n})\,\epsilon^m
\nonumber\\[1ex]
&= \frac{g_{\bf n}(0)}{\epsilon^{N_\text{min}}}
+ \frac{g'_{\bf n}(0)}{\epsilon^{N_\text{min}-1}}
+ \cdots
+ \frac{g^{(N_\text{min}-1)}_{\bf n}(0)}{(N_\text{min}-1)!\,\epsilon}
\nonumber\\[1ex]&\qquad
+ \frac{1}{N_\text{min}!}g^{(N_\text{min})}_{\bf n}(0) + {\cal O}(\epsilon).
\end{align}
We define the regularized ``partie finie" value of the function $f$ at $\x_0$
as the coefficient at $\epsilon^0$ in the expansion \eqref{laurent}
averaged over all unit vectors ${\bf n}$:
\begin{align}
\label{HPFdef}
f_{\rm reg}(\x_0)
&:= \frac{1}{4\pi}\oint\!\md\Omega_2\,a_0({\bf n})
\nonumber\\[1ex]
&\phantom{:}= \frac{1}{4\pi N_\text{min}!}\oint\!\md\Omega_2\,g^{(N_\text{min})}_{\bf n}(0).
\end{align}
Let us consider the function
$$
{\widetilde g}({\bf x}) := f({\bf x})|{\bf x}-\x_0|^{\widetilde N},
$$
where $\widetilde{N}$ is the positive integer ${\widetilde N}>N_\text{min}$
Then one can define $f_{\rm reg}(\x_0)$ using the function ${\widetilde g}$ 
instead of $g$. It is easy to show that the value of $f_{\rm reg}(\x_0)$ 
will not change, so the definition (\ref{HPFdef}) does not depend on the choice 
of the number ${\widetilde N}$ provided ${\widetilde N}\ge N_\text{min}$.

Definition \eqref{HPFdef} is used to give the meaning to integrals of $f(\x)\,\da$,
where the function $f$ is assumed to be singular at $\x=\xa$.
Namely, we define
\be
\label{DDdef}
\int\md^3x\,f(\x)\,\da := f_{\textrm{reg}}(\xa).
\ee
The definition \eqref{DDdef} is an extension of the notion of ``good" Dirac $\delta$-functions
introduced by Infeld and Pleba\'nski \cite{IP60}.
Their definition assumes that ``good" $\delta$-function,
besides having the properties of ordinary Dirac $\delta$ distributions,
also satisfies the condition (cf.\ Appendix 1 in \cite{IP60})
\be
\label{IPdef}
\int\md^3x\, \frac{\da}{r_a^k} = 0,\quad
\mbox{for $k=1,2,\ldots,p$.}
\ee
Obviously the definition (\ref{DDdef})
entails the fulfillment of the condition (\ref{IPdef}).

The important feature of the definition (\ref{HPFdef}) is 
that the regularized value of the product of functions is not, in general,
equal to the product of the regularized values of the individual functions:
\be
\label{prHPH1}
\left(f_1 f_2 \cdots \right)_{\text{reg}}(\x_0) \neq f_{1\,\text{reg}}(\x_0)\, f_{2\,\text{reg}}(\x_0) \cdots
\ee
(the above property but with the equality sign was called ``tweedling of products''
by Infeld and Pleba\'nski; see, e.g., Appendix 2 in \cite{IP60}).
The property \eqref{prHPH1} of three-dimensional ``partie finie" operation leads to problems.
Let us consider some function $S$ singular at $\x=\xa$.
It is natural to demand that
\be
\label{prHPH2}
S(\x)\,\da = S_{\textrm{reg}}(\xa)\,\da.
\ee
The above rule is always used when solving Poisson equations
with singular source terms of the form $S(\x)\,\da$
[see Eq.\ \eqref{eqPoisson} below].
Then, multiplying both sides of Eq.\ \eqref{prHPH2} by another function $T$
which is also singular at $\x=\xa$, one gets
\be
\label{prHPH3}
T({\bf x})\,S({\bf x})\,\da = T({\bf x})\,S_{\textrm{reg}}(\xa)\,\da.
\ee
The rule \eqref{prHPH2} applied to Eq.\ \eqref{prHPH3} implies that
\be
\label{prHPH4}
(TS)_{\textrm{reg}}(\xa) = T_{\textrm{reg}}(\xa)\,S_{\textrm{reg}}(\xa),
\ee
what in general contradicts Eq.\ \eqref{prHPH1}.

To avoid this kind of problem one should employ dimensional regularization.
One can check that in the generic $d$-dimensional case the distributivity is always satisfied,
\be
\label{dHPF1}
\big(f^{(d)}_1 f^{(d)}_2 \cdots\big)_{\textrm{reg}}(\xa)
= f^{(d)}_{1\,\textrm{reg}}(\xa) f^{(d)}_{2\,\textrm{reg}}(\xa) \cdots,
\ee
where $f^{(d)}_{\textrm{reg}}(\xa)$ is rather not defined
by the $d$-dimensional analog of the definition \eqref{HPFdef},
but it is directly computed by employing the existence of such region
in the complex $d$-plane where the function $f^{(d)}(\x)$
is finite for $\x=\xa$ (see the example at the end of this appendix).
In $d$ dimensions one can thus always use
\be
\label{dHPF2}
f^{(d)}(\x)\,\delta^{(d)}(\x-\x_a)
= f^{(d)}_{\textrm{reg}}(\xa)\,\delta^{(d)}(\x-\x_a).
\ee
Therefore one defines the dimensional-regularization rule,
\be
\label{dHPFdef}
f_{\textrm{reg}}(\xa) := \lim_{d\to3}\big(f^{(d)}_{\textrm{reg}}(\xa)\big),
\ee
where $f^{(d)}$ is the $d$-dimensional version of $f$.
In the computation of the 4PN Hamiltonian the usage of the rule \eqref{dHPFdef}
boils down in practice to the usage of the three-dimensional definition \eqref{HPFdef}
\emph{after} the usage of distributivity \eqref{dHPF1}
(in the case of computing regularized value of the product of functions).

As an example of justifying the distributivity \eqref{dHPF1}
let us consider the following 4PN-related contact integral
[here we denote by $\fii^{(d)}$ the function,
given by the right-hand side of Eq.\ \eqref{phi2-d},
which is the $d$-dimensional version of the three-dimensional potential denoted by $\fii$]:
\be
\label{prHPF8}
\int\md^dx\Big(\phi_{(2)}^{(d)}(\mathbf{x})\Big)^5\delta_1
= \int\md^dx\,\kappa^5 \big(m_1 r_1^{2-d} + m_2 r_2^{2-d}\big)^5\,\delta_1.
\ee
Because
$\Re(d)<2 \,\Rightarrow\,\lim_{\x\to\x_1}r_1^{2-d}= 0$, then
\begin{align}
\label{prHPF9}
\int\md^dx\Big(\phi_{(2)}^{(d)}(\mathbf{x})\Big)^5\delta_1
= \kappa^5 \big(m_2 r_{12}^{2-d}\big)^5
= \big[\big(\fii^{(d)}\big)_{\textrm{reg}}(\x_1)\big]^5.
\end{align}
Therefore the three-dimensional value of the integral \eqref{prHPF8} equals
\begin{align}
\label{prHPF10}
\int\md^3x\,\big(\fii(\x)\big)^5\,\delta_1
&= \lim_{d\to3}\big[\big(\fii^{(d)}\big)_{\textrm{reg}}(\x_1)\big]^5
\nonumber\\[1ex]
&= \big[\big(\fii\big)_{\textrm{reg}}(\x_1)\big]^5.
\end{align}
Computation of the three-dimensional integral of $\big(\fii(\x)\big)^5\,\delta_1$
directly by means of Eq.\ \eqref{HPFdef} leads to the result different from the result of Eq.\ \eqref{prHPF10}.

\subsection{Distributional differentiation of homogeneous functions}
\label{gendiff}

Appearance of UV divergences is not the only cost of employing Dirac-delta sources.
Another consequence is that in our computations we have to differentiate homogeneous
functions using an enhanced (or distributional) rule,
which comes from standard distribution theory
(see Sec.\ 3.3 in Chapter III of Ref.\ \cite{GS64}).

Let $f$ be a real-valued function defined in a neighborhood of the origin of $\mathbb{R}^3$.
$f$ is said to be a \emph{positively homogeneous function of degree $\lambda$},
if for any number $a>0$
\be
\label{ed1}
f(a\,\x) = a^\lambda\,f(\x).
\ee
Let $k:=-\lambda-2$. If $\lambda$ is an integer and if $\lambda\le-2$
(i.e.\ $k$ is a non-negative integer), then the partial derivative of $f$ with
respect to the coordinate $x^i$ has to be calculated by means of the formula
[cf.\ Eq.\ (5.15) in \cite{K85}]
\begin{multline}
\label{ed2}
\partial_i f(\x) = \partial_{\underline i}f(\x)
+ \frac{(-1)^k}{k!} \frac{\partial^k\delta({\bf x})}{\partial x^{i_1}\cdots\partial x^{i_k}}
\\[1ex]
\times \oint_\Sigma \md\sigma_i\,f(\x')\,x'^{i_1}\cdots x'^{i_k},
\end{multline}
where $\partial_{\underline i}f$ means the derivative computed using
the standard rules of differentiations, $\Sigma$ is any smooth close surface
surrounding the origin and $\md\sigma_i$ is the surface element on $\Sigma$.

The rule \eqref{ed2} should also be applied to differentiation of 
functions which are homogeneous not with respect to ${\bf x}$, but with respect 
to ${\bf x}-{\bf x}_0$, for some constant ${\bf x}_0\in{\mathbb R}^3$.
Let $f$ be such function, then there exists another function $\phi$ for which the relation
\be
\label{ed3}
f(\x) = \phi(\x-\x_0)
\ee
is fulfilled and the function $\bxi\mapsto\phi(\bxi)$ is a positively homogeneous 
function of degree $\lambda$, i.e., for any number $a>0$
\be
\label{ed4}
\phi(a\,\bxi) = a^\lambda\,\phi(\bxi).
\ee
Obviously (here $\bxi:={\bf x}-{\bf x}_0$)
\be
\label{ed5}
\frac{\partial f({\bf x})}{\partial x^i} = \frac{\partial \phi(\bxi)}{\partial \xi^i}
\ee
and for the derivative $\partial\phi/\partial\xi^i$ the rule \eqref{ed1} is directly 
applicable. Therefore the partial derivative with respect to the coordinate 
$x^i$ of the function $f$ satisfying conditions (\ref{ed3}) and (\ref{ed4}) 
should, by virtue of (\ref{ed5}) and (\ref{ed2}), be calculated by means of the 
formula
\begin{multline}
\label{erddef}
\partial_i f(\x) = \partial_{\underline i}f(\x)
+ \frac{(-1)^k}{k!}
\frac{\partial^k\delta({\bf x}-{\bf x}_0)}{\partial x^{i_1}\cdots\partial x^{i_k}}
\\[1ex]
\times \oint_\Sigma \md\sigma_i\,\phi(\bxi')\,\xi'^{i_1}\cdots\xi'^{i_k},
\end{multline}
where $\partial_{\underline i}f$ means the derivative computed using
the standard rules of differentiations, $\Sigma$ is any smooth close surface
surrounding the point ${\bf x}_0$ and $\md\sigma_i$ is the surface element on~$\Sigma$.

As an example let us employ the formula (\ref{erddef})
to calculate first and second partial derivatives of $1/r_a$, $1/r_a^2$, and $1/r_a^3$.
For the first partial derivatives we obtain
\begin{subequations}
\begin{align}
\label{e41a1}
\partial_i\frac{1}{r_a} &= \partial_{\underline i}\frac{1}{r_a},
\\[2ex]
\label{e41a2}
\partial_i\frac{1}{r_a^2} &= \partial_{\underline i}\frac{1}{r_a^2},
\\[2ex]
\label{e41a3}
\partial_i\frac{1}{r_a^3} &= \partial_{\underline i}\frac{1}{r_a^3} - \frac{4\pi}{3}\partial_i\da.
\end{align}
\end{subequations}
Let us note that in Eq.\ \eqref{e41a1} there is no need to use the rule \eqref{erddef},
and in Eq.\ \eqref{e41a2} the term proportional to $\da$ [obtained from the usage of (\ref{erddef})], vanishes.
The second partial derivatives read
\begin{subequations}
\begin{align}
\label{e41b1}
\partial_i\partial_j\frac{1}{r_a} &= \partial_{\underline i}\partial_{\underline j}\frac{1}{r_a}
- \frac{4\pi}{3}\delta_{ij}\da,
\\[2ex]
\label{e41b2}
\partial_i\partial_j\frac{1}{r_a^2}
&= \partial_{\underline i}\partial_{\underline j}\frac{1}{r_a^2},
\\[2ex]
\label{e41b3}
\partial_i\partial_j\frac{1}{r_a^3}
&= \partial_{\underline i}\partial_{\underline j}\frac{1}{r_a^3}
- \frac{2\pi}{15} \left( 16\partial_i\partial_j\da + 3\delta_{ij}\Delta\da \right).
\end{align}
\end{subequations}
Making use of Eqs.\ (\ref{e41b1}), (\ref{e41b2}), and (\ref{e41b3}), one gets
\begin{subequations}
\begin{align}
\label{e41c1}
\Delta\frac{1}{r_a} &= -4\pi\da,
\\[2ex]
\label{e41c2}
\Delta\frac{1}{r_a^2} &= \frac{2}{r_a^4},
\\[2ex]
\label{e41c3}
\Delta\frac{1}{r_a^3} &= \frac{6}{r_a^5}
- \frac{10\pi}{3}\Delta\da.
\end{align}
\end{subequations}

The distributional derivative (\ref{erddef}) does not obey the Leibniz rule. It 
can easily be seen by considering the distributional partial derivative of the 
product $1/r_a$ and $1/r_a^2$. Let us suppose that the Leibniz rule is 
applicable here:
\be
\label{leibniz}
\partial_i{\frac{1}{r_a^3}}
= \partial_i{\left(\frac{1}{r_a}\frac{1}{r_a^2}\right)}
= \frac{1}{r_a^2}\, \partial_i{\frac{1}{r_a}}
+ \frac{1}{r_a}\, \partial_i{\frac{1}{r_a^2}}.
\ee
By virtue of Eqs.\ (\ref{e41a1}) and (\ref{e41a2}) the right-hand side of Eq.\ 
(\ref{leibniz}) can be computed using standard differential calculus (no terms 
with Dirac deltas), whereas computing the left-hand side of (\ref{leibniz}) by 
means of (\ref{e41a3}) one obtains some term proportional to $\partial_i\da$.

The distributional differentiation is necessary when one differentiates
homogeneous functions under the integral sign.
Let us consider the following locally divergent integral (here $a\ne b$):
\be
\label{e42a}
p_{ai}\,p_{aj}\int\md^3x\left(\partial_i\partial_j\frac{1}{r_a}\right)\frac{1}{r_b^4}.
\ee
We shall regularize this integral in two different ways.
We first replace in \eqref{e42a} differentiations with respect to $x^i$
by those with respect to $x^i_a$ (obviously $\partial_i r_a = -\partial_{ai}r_a$).
Then we shift the differentiations before the integral sign
and apply directly the Riesz formula \eqref{riesz}. The result is
\begin{align}
\label{e42b}
p_{ai}\,p_{aj}\int\md^3x &\left(\partial_i\partial_j\frac{1}{r_a}\right)\frac{1}{r_b^4}
= p_{ai}\,p_{aj}\,\partial_{ai}\partial_{aj} \int \frac{\md^3x}{r_a r_b^4}
\nonumber\\
&= p_{ai}\,p_{aj}\,\partial_{ai}\partial_{aj}
\left(-\frac{2\pi}{r_{ab}^2}\right)
\nonumber\\[1ex]
&= \frac{4\pi\big[\papa-4\nabpa^2\big]}{r_{ab}^4}.
\end{align}
We have obtained \eqref{e42b} performing integration first and then differentiation.
Now we shall regularize the integral (\ref{e42a}) doing differentiation first.
To do it we have to use the rule (\ref{erddef}), which gives [cf.\ Eq.\ (\ref{e41b1})]
\be
\label{e42c}
\partial_i\partial_j\frac{1}{r_a} = \left(3n_a^i n_a^j-\delta_{ij}\right)\frac{1}{r_a^3}
- \frac{4\pi}{3}\delta_{ij}\da.
\ee
We substitute (\ref{e42c}) into (\ref{e42a}):
\begin{multline}
\label{e42d}
p_{ai}\,p_{aj}\int\md^3x\left(\partial_i\partial_j\frac{1}{r_a}\right)\frac{1}{r_b^4}
\\[1ex]
= \int\md^3x\frac{3\napa^2-\papa}{r_a^3 r_b^4}
- \frac{4\pi}{3}\papa\int\md^3x\frac{\da}{r_b^4}.
\end{multline}
The second integral on the right-hand side of (\ref{e42d}) is calculated
by means of the definition (\ref{DDdef}). The obvious result is
\be
\label{e42e}
\int\md^3x\frac{\da}{r_b^4} = \frac{1}{r_{ab}^4}.
\ee
To calculate the first integral on the right-hand side of (\ref{e42d})
we apply the procedure described in Appendix \ref{3RHregularization}.
We obtain
\be
\label{e42f}
\int\md^3x\frac{3\napa^2-\papa}{r_a^3 r_b^4}
= \frac{16\pi\big[\papa-3\nabpa^2\big]}{3r_{ab}^4}.
\ee
Collecting Eqs.\ \eqref{e42d}--\eqref{e42f} together
we get the result which coincides with the result (\ref{e42b}) obtained before.
The two ways of regularizing the integral (\ref{e42a}), described above,
coincide only if we apply formula (\ref{erddef}) when we perform differentiation before integration.

It is not difficult to show that the formula \eqref{ed2}
is also valid (without any change) in the $d$-dimensional case,
i.e.\ when $f$ is a real-valued and positively homogeneous function of degree $\lambda$
defined in a neighborhood of the origin of $\mathbb{R}^d$.
The formula \eqref{ed2} can be applied in $d$ dimensions when the number $k:=-\lambda+d-1$ is a non-negative integer.
For example, the $d$-dimensional versions of Eqs.\ \eqref{e41a1} and \eqref{e41b1} read
\begin{subequations}
\begin{align}
\label{e41a1d}
\partial_i\frac{1}{r_a^{d-2}} &= \partial_{\underline i}\frac{1}{r_a^{d-2}},
\\[1ex]
\label{e41b1d}
\partial_i\partial_j\frac{1}{r_a^{d-2}} &= \partial_{\underline i}\partial_{\underline j}\frac{1}{r_a^{d-2}}
- \frac{1}{\kappa d}\delta_{ij}\da.
\end{align}
\end{subequations}

Using the formula \eqref{ed2} [or \eqref{erddef}] in $d$ (or 3) dimensions requires
averaging of products of unit vectors over the unit sphere.
This can be done by means of the following formulas
(see Appendix A 2 of \cite{HSS13} for $d$-dimensional expression
and Appendix A 5 of \cite{BD1986} for its three-dimensional version):
\begin{subequations}
\begin{align}
\oint_{\mathbb{S}^{d-1}}\md\Omega_{d-1}\,n^{i_1}\cdots n^{i_{2k+1}} &= 0, \quad k=0,1,2,\ldots\,,
\\
\oint_{\mathbb{S}^{d-1}}\md\Omega_{d-1}\,n^{i_1}\cdots n^{i_{2k}}&= \frac{(d-2)!!}{(d+2(k-1))!!}\Omega_{d-1}
\nonumber\\[1ex]&\kern-12ex
\times\delta_{\{i_1 i_2}\cdots\delta_{i_{2k-1}i_{2k}\}},
\quad k=0,1,2,\ldots\,.
\end{align}
\end{subequations}
Here, for positive integer $k$, $k!!:=k(k-2)\cdots1$ for $k$ odd and $k!!:=k(k-2)\cdots2$ for $k$ even,
$\Omega_{d-1}$ is the area of the unit sphere in $\mathbb{R}^d$,
\be
\Omega_{d-1} = \frac{2\pi^{d/2}}{\Gamma(d/2)},
\ee
and $A_{\{i_1 i_2 \cdots i_k\}}:=\sum_{\sigma\in S}A_{\sigma(i_1)\cdots\sigma(i_k)}$,
where $S$ is the smallest set of permutations of $\{1,\ldots,k\}$
making $A_{\{i_1 i_2 \cdots i_k\}}$ fully symmetric in $i_1,i_2,\cdots,i_k$.
Let us note some simple cases:
\begin{subequations}
\begin{align}
\oint_{\mathbb{S}^{d-1}}\md\Omega_{d-1}\,n^{i_1}n^{i_2} &= \frac{1}{d}\Omega_{d-1}\delta_{i_1i_2},
\\
\oint_{\mathbb{S}^{d-1}}\md\Omega_{d-1}\,n^{i_1}n^{i_2}n^{i_3}n^{i_4} &= \frac{1}{d(d+2)}\Omega_{d-1}
\nonumber\\&\kern-12ex
\times\big(\delta_{i_1i_2}\delta_{i_3i_4}+\delta_{i_1i_3}\delta_{i_2i_4}+\delta_{i_1i_4}\delta_{i_2i_3}\big).
\end{align}
\end{subequations}

\subsection{Riesz kernel}

To avoid problems related with applying distributional derivatives (even in $d$ dimensions)
one can replace Dirac delta distribution $\delta$ by some functional representation (``delta sequence'').
In $d$ dimensions one can e.g.\ employ the \emph{Riesz kernel} $K_a(\epsilon_a)$:
\be
\label{kernelR}
\da = \lim_{\ve_a\to0} K_a(\epsilon_a),
\quad
K_a(\epsilon_a) := \frac{\Gamma\big((d-\epsilon_a)/2\big)}{\pi^{d/2}\,2^{\epsilon_a}\,\Gamma(\epsilon_a/2)} r_a^{\epsilon_a-d}.
\ee
Then one should replace in the constraint equations \eqref{CE}
Dirac-delta sources by Riesz kernels \eqref{kernelR},
solve the constraints perturbatively and develop the whole PN scheme
(let us stress that then no distributional differentiation is needed).
At the end of the calculation, one takes the limits $\epsilon_1\to0$, $\epsilon_2\to0$,
and only after this one computes $d\to3$ limit.

This procedure was applied by us in \cite{DJS2008} to recompute all UV divergent integrals at the 3PN level.
It is however too difficult to be performed fully at the 4PN level,
so it has not been applied in the main part of the present paper,
but it was used for checking some results.
It should be mentioned that the Riesz-kernel method of regularization has been applied by Damour in Refs.\ \cite{TD1980,TD1983}.
It also may be pointed out that the dimensional regularization calculations of Ref.\ \cite{tHV1972}
have been performed in momentum representation, where also only ordinary (i.e., nondistributional) space-time derivatives show up.

As an example let us first compute the potential $\fii$ for Riesz-kernel sources.
Instead of Eq.\ \eqref{delta-fii} we thus solve
\be
\Deltad\fii = -\sum_a m_a K_a.
\ee
Making use of Eq.\ \eqref{di02} one gets
\be
\fii = -\sum_a \frac{2^{-\epsilon_a}\pi^{-d/2}\Gamma((d - \epsilon_a)/2)}{\epsilon_a(2 - d + \epsilon_a)\Gamma(\epsilon_a/2)}
m_a r_a^{2 - d + \epsilon_a}.
\ee
Let us next consider the integral of Eq.\ \eqref{prHPF8},
which now takes the form
\be
\int\md^dx\,\fii^5 \,K_1.
\ee
To compute this integral it is enough to use Eq.\ \eqref{riesz}.
After computing the double limit $\epsilon_1\to0$, $\epsilon_2\to0$
(the order of the limits does not matter) one obtains
\be
\frac{\pi^{-5 d/2}\Gamma(d/2)^5}{32(d-2)^5} m_2^5 r_{12}^{10-5 d}.
\ee
The value of the above formula in the limit $d\to3$ 
coincides with the result \eqref{prHPF10}.

Let us finally mention that the usage of the Riesz kernel directly in 3 dimensions
does not resolve ambiguities. For example, three-dimensional computations
corresponding to $d$-dimensional ones presented in the above example
would lead to the result which is different from \eqref{prHPF10}
(so the ``tweedling property'' of the product would be violated).

\section{Inverse Laplacians}
\label{invLap}

In the present paper we have to consider
(both in $d$ and in 3 dimensions)
numerous Poisson equations with distributional source terms,
\be
\Deltad f = \sum_a g(\x)\,\da,
\ee
where usually the function $g$ is singular at $\x=\x_a$.
Equation of this type we solve as follows:
\begin{align}
\label{eqPoisson}
f(\x) &= \Deltad^{-1}\,\sum_a g(\x)\,\da
= \Deltad^{-1}\,\sum_a g_{\rm reg}({\bf x}_a)\,\da
\nonumber\\[1ex]
&= \sum_a g_{\rm reg}({\bf x}_a)\Deltad^{-1}\delta_a,
\end{align}
where the regularized value $g_{\rm reg}$ of the function $g$ is defined in
Eq.\ \eqref{HPFdef} (in 3 dimensions)
or in Eq.\ \eqref{dHPFdef} (when, to avoid ambiguities,
one has to employ the $d$-dimensional version of computing of $g_{\rm reg}$).
The function $\Deltad^{-1}\delta_a$ is defined below in Eq.\ \eqref{di06-1} (in $d$ dimensions)
or in Eq.\ \eqref{i06-1} (in 3 dimensions).

\subsection{$d$-dimensional inverse Laplacians}
\label{invLapd}

We start from considering some solutions of equation
\be
\label{di01}
\Deltad^n f = r_a^\alpha,\quad n=1,2,\ldots\,,
\ee
where $\Deltad^n$ denotes $n$-folded superposition of the $d$-dimensional flat Laplacian $\Deltad$.
By direct computation one checks that the function
\be
\label{di02}
\Deltad^{-n}r_a^\alpha := \frac{\Gamma(\alpha/2 + 1)\Gamma((\alpha+d)/2)}{2^{2n}\Gamma(\alpha/2 + n + 1)\Gamma((\alpha+d)/2 + n)} r_a^{\alpha+2n}
\ee
is the solution of Eq.\ \eqref{di01}.
The function $\Deltad^{-n}r_a^\alpha$ we call {\em $n$th inverse Laplacian of $r_a^\alpha$}.
Let us note the formulas for the first, second, and third inverse Laplacians of $r_a^\alpha$:
\begin{widetext}
\begin{subequations}
\label{di03}
\begin{align}
\Deltad^{-1}r_a^\alpha &:= \frac{r_a^{\alpha+2}}{(\alpha+2)(\alpha+d)},
\\[1ex]
\Deltad^{-2}r_a^\alpha &:= \frac{r_a^{4 + \alpha}}{(2 + \alpha)(4 + \alpha)(d + \alpha)(2 + d + \alpha)},
\\[1ex]
\Deltad^{-3}r_a^\alpha &:= \frac{r_a^{6 + \alpha}}{(2 + \alpha)(4 + \alpha)(6 + \alpha)(d + \alpha)(2 + d + \alpha)(4 + d + \alpha)}.
\end{align}
\end{subequations}
\end{widetext}

We also have to consider solutions of equation
\be
\label{di04}
\Deltad^n f = \da,\quad n=1,2,\ldots\,.
\ee
Making use of Eq.\ \eqref{di02} and the formula
\be
\label{di05}
\Deltad r_a^{2-d} = -\kappa^{-1}\da,
\ee
one checks that the function
\be
\label{di06}
\Deltad^{-n}\da := -\frac{\kappa\,\Gamma(2-d/2)}{4^{n-1}(n-1)!\,\Gamma(n+1-d/2)}r_a^{2n-d}
\ee
solves Eq.\ \eqref{di04}.
The function $\Deltad^{-n}\da$ we call the \emph{$n$th inverse Laplacian of $\da$}.
The first inverse Laplacian of $\da$ thus reads
\be
\label{di06-1}
\Deltad^{-1}\da := -\kappa\,r_a^{2-d}.
\ee

We also need solutions of equations of a more general type than Eq.\ \eqref{di01}.
It reads
\be
\label{di07}
\Deltad^n f = r_a^\alpha n_a^{i_1}\cdots n_a^{i_k},\quad n=1,2,\ldots\,.
\ee
More precisely, we need to consider Eq.\ \eqref{di07} for $n=1,2,3$
and for each of these values of $n$ we need to take $k=1,\ldots,6$.
To save space we list below solutions (inverse Laplacians) related to Eq.\ \eqref{di07} only for $n=1,2,3$ and $k=1,2,3$:
\begin{widetext}
\begin{subequations}
\begin{align}
\Deltad^{-1} n_a^i r_a^\alpha &:= \frac{n_a^i r_a^{2 + \alpha}}{(\alpha + 1) (\alpha + d + 1)},
\\[2ex]
\Deltad^{-1} n_a^i n_a^j r_a^\alpha &:= \frac{[(2 + \alpha)(d + \alpha)n_a^i n_a^j - 2 \delta_{ij}]r_a^{2 + \alpha}}{\alpha(2 + \alpha)(d + \alpha)(2 + d + \alpha)},
\\[2ex]
\Deltad^{-1} n_a^i n_a^j n_a^k r_a^\alpha &:= \frac{[2 (\delta_{jk} n_a^i + \delta_{ik} n_a^j + \delta_{ij} n_a^k)
- (1 + \alpha) (1 + d + \alpha)n_a^i n_a^j n_a^k] r_a^{2 + \alpha}}{(1 - \alpha)(1 + \alpha)(1 + d + \alpha)(3 + d + \alpha)},
\\[2ex]
\Deltad^{-2} n_a^i r_a^\alpha &:= \frac{n_a^i r_a^{4 + \alpha}}{(1 + \alpha) (3 + \alpha) (1 + d + \alpha) (3 + d + \alpha)},
\\[2ex]
\Deltad^{-2} n_a^i n_a^j r_a^\alpha &:= \frac{[(4 + \alpha) (d + \alpha) n_a^i n_a^j - 4 \delta_{ij}]r_a^{4 + \alpha}}
{\alpha (2 + \alpha) (4 + \alpha) (d + \alpha) (2 + d + \alpha) (4 + d + \alpha)},
\\[2ex]
\Deltad^{-2} n_a^i n_a^j n_a^k r_a^\alpha &:=
\frac{[(3 + \alpha) (1 + d + \alpha) n_a^i n_a^j n_a^k - 4 (\delta_{ij} n_a^k + \delta_{jk} n_a^i + \delta_{ik} n_a^j)]r_a^{4 + \alpha}}
{(\alpha - 1) (1 + \alpha) (3 + \alpha) (1 + d + \alpha) (3 + d + \alpha) (5 + d + \alpha)},
\\[2ex]
\Deltad^{-3} n_a^i r_a^\alpha &:= \frac{n_a^i r_a^{6 + \alpha}}
{(1 + \alpha) (3 + \alpha) (5 + \alpha) (1 + d + \alpha) (3 + d + \alpha) (5 + d + \alpha)},
\\[2ex]
\Deltad^{-3} n_a^i n_a^j r_a^\alpha &:= \frac{[(6 + \alpha) (d + \alpha) n_a^i n_a^j - 6 \delta_{ij}]r_a^{6 + \alpha} }
{\alpha (2 + \alpha) (4 + \alpha) (6 + \alpha) (d + \alpha) (2 + d + \alpha) (4 + d + \alpha) (6 + d + \alpha)},
\\[2ex]
\Deltad^{-3} n_a^i n_a^j n_a^k r_a^\alpha &:=
\frac{[(5 + \alpha) (1 + d + \alpha) n_a^i n_a^j n_a^k - 6 (\delta_{ij} n_a^k + \delta_{jk} n_a^i + \delta_{ik} n_a^j)]r_a^{6 + \alpha}}
{(\alpha - 1) (1 + \alpha) (3 + \alpha) (5 + \alpha) (1 + d + \alpha) (3 + d + \alpha) (5 + d + \alpha) (7 + d + \alpha)}.
\end{align}
\end{subequations}
\end{widetext}

\subsection{Three-dimensional inverse Laplacians}
\label{invLap3}

For convenience let us start from rewriting
some $d$-dimensional results of Appendix \ref{invLapd} in $d=3$ dimensions.
The function
\be
\label{i01}
\Delta^{-n}r_a^\alpha := \frac{\Gamma(\alpha+2)}{\Gamma(\alpha+2n+2)}r_a^{\alpha+2n}
\ee
is the $n$th inverse Laplacian of $r_a^\alpha$
[it is thus the solution of Eq.\ \eqref{di01} in $d=3$ dimensions],
so the first inverse Laplacian of $r_a^\alpha$ reads
\be
\label{i02}
\Delta^{-1}r_a^\alpha := \frac{r_a^{\alpha+2}}{(\alpha+2)(\alpha+3)}.
\ee
The function
\be
\label{i06}
\Delta^{-n}\da := -\frac{1}{4\pi(2n-2)!}r_a^{2n-3}
\ee
solves Eq.\ \eqref{di04} in $d=3$ dimensions
and is the $n$th inverse Laplacian of $\da$.
The first inverse Laplacian of $\da$ thus equals
\be
\label{i06-1}
\Delta^{-1}\da := -\frac{1}{4\pi r_a}.
\ee

In the derivations of the field functions needed to calculate the 4PN-accurate two-point-mass Hamiltonian
we have used some special solutions to the partial differential equations of the form
\be
\label{i1}
\Delta^n f = r_a^\alpha r_b^\beta \quad (a\ne b),\quad n=1,2,\ldots\,.
\ee
This special solution to Eq.\ (\ref{i1}) is defined below, we denote it by $I(n;\alpha,\beta)$ and 
call the {\em $n$th inverse Laplacian of $r_a^\alpha r_b^\beta$}.

Let us mention that in Ref.\ \cite{Chu2009} one can find formulas [see Eq.\ (B13) there and equations below it],
which describe the formal solution to Eq.\ \eqref{i1} in $d$ dimensions
for any complex values of $n$, $\alpha$, and $\beta$.
The solution is expressed in terms of the Appell hypergeometric function $F_4$ of two variables.
We have tried to use it in our 4PN-related computations but, in general,
this has led to calculations too complicated to be useful.

The large family of three-dimensional inverse Laplacians of $r_a^\alpha r_b^\beta$ is based on the 
solution to the Poisson equation
\be
\label{i2}
\Delta f = \frac{1}{r_a r_b}.
\ee
One immediately checks that $\ln(r_a+r_b\pm r_{ab})$ solves Eq.\ (\ref{i2}). 
Because $\ln(r_a+r_b-r_{ab})$ is singular along the segment joining the points
${\bf x}_a$ and ${\bf x}_b$, the first inverse Laplacian of $r_a^{-1}r_b^{-1}$ 
we define to be
\be
\label{i3}
I(1;-1,-1) := \ln(r_a+r_b+r_{ab}).
\ee
Let us apply the operator $\Delta_a$ ($\Delta_a$ contains differentiations with 
respect to $x_a^i$ only) to the both sides of Eq.\ (\ref{i2}).
Assuming commutativity of the operators $\Delta_a$ and $\Delta$ one obtains
\be
\label{i3a}
\Delta_a(\Delta f) = \Delta(\Delta_a f) = \frac{1}{r_b}\left(\Delta_a\frac{1}{r_a}\right)
= -4\pi \frac{\delta_a}{r_b},
\ee
where we have used
\be
\label{i3b}
\Delta_a\frac{1}{r_a} = \Delta\frac{1}{r_a} = -4\pi\da.
\ee
After applying the rule [see Eq.\ \eqref{prHPH2} and the discussion around this equation]
\be
\label{i3c}
f(\x)\da = f_{\rm reg}(\x_a)\da
\ee
to the right-hand side of Eq.\ (\ref{i3a}), one gets
\be
\label{i3d}
\Delta(\Delta_a f) = -4\pi \frac{\da}{r_{ab}}.
\ee
Using Eq.\ \eqref{i06-1} from (\ref{i3d}) one obtains
\be
\label{i3e}
\Delta_a f = -\frac{4\pi}{r_{ab}}\Delta^{-1}\da
= \frac{1}{r_a r_{ab}}.
\ee
Analogously one can derive the equation
\be
\label{i3f}
\Delta_b f = \frac{1}{r_b r_{ab}}.
\ee
One checks that the solution (\ref{i3}) fulfills Eqs.\ (\ref{i3e}) and 
(\ref{i3f}), whereas the function $\ln(r_a+r_b-r_{ab})$ does not.

Using the result (\ref{i3}) it is possible to calculate all multiple inverse 
Laplacians of the type\footnote{
The method presented below was devised in Ref.\ \cite{JS1997}; see Appendix C there.}
\be
\label{i4}
I(n;2k-1,2l-1),\quad n=1,2,\ldots,\quad k,l=0,1,2,\ldots\,.
\ee
To obtain $I(n;2k-1,2l-1)$ (for $n=1,2,\ldots$, $k,l=0,1,2,\ldots$) one assumes that
it has the following structure:
\begin{align}
\label{i6}
I(n;2k-1,2l-1) &:= W_1(r_a,r_b,r_{ab})
\nonumber\\[1ex]&\quad
+ W_2(r_a,r_b,r_{ab})\ln(r_a+r_b+r_{ab}),
\end{align}
where $W_1$ and $W_2$ are polynomials of variables $r_a$, $r_b$, and $r_{ab}$, 
consisting of terms of only $2(k+l+n-1)$-order in these variables.
The form of Eq.\ \eqref{i6} can be inferred from dimensional analysis
which takes into account the first inverse Laplacian of $r_a^{-1}r_b^{-1}$ given in Eq.\ (\ref{i3}).
To fix the coefficients of $W_1$ and $W_2$ {\em uniquely} one requires that the 
following conditions are fulfilled:
\begin{subequations}
\label{i7}
\begin{align}
\label{i7a}
\Delta I(n;2k-1,2l-1) &= I(n-1;2k-1,2l-1),
\\[2ex]
\label{i7b}
\Delta_a I(n;2k-1,2l-1) &= 2k(2k-1)I(n;2k-3,2l-1), 
\\[2ex]
\label{i7c}
\Delta_b I(n;2k-1,2l-1) &= 2l(2l-1)I(n;2k-1,2l-3).
\end{align}
Equation (\ref{i7b}) and (\ref{i7c}) corresponds to Eq.\ (\ref{i3e}) and (\ref{i3f}), respectively.
In the case of $k=0$ Eq.\ (\ref{i7b}) should be replaced by
\be
\label{i7bi}
\Delta_a I(n;-1,2l-1) = \frac{r_{ab}^{2l-1}}{(2n-2)!}r_a^{2n-3};
\ee
analogously for $l=0$ instead of Eq.\ (\ref{i7c}) one uses
\be
\label{i7ci}
\Delta_b I(n;2k-1,-1) = \frac{r_{ab}^{2k-1}}{(2n-2)!}r_b^{2n-3}.
\ee
\end{subequations}

To illustrate the above procedure let us consider the second inverse Laplacian $I(2;-1,-1)$.
We are thus looking for a solution $f$ of the partial differential equation $\Delta^2f=1/(r_ar_b)$. 
We assume that the solution $f$ is of the form $W_1(r_a,r_b,r_{ab})+W_2(r_a,r_b,r_{ab})\ln(r_a+r_b+r_{ab})$,
where $W_1$ and $W_2$ are polynomials of indicated variables consisting of only quadratic terms in these variables
(i.e.\ the most general polynomial of this type is $a_1r_a^2+a_2r_b^2+a_3r_{ab}^2+a_4r_ar_b+a_5r_ar_{ab}+a_6r_br_{ab}$). 
We need to fix the values of 12 coefficients
(in fact less because of symmetry with respect to interchanging the labels $a$ and $b$ of the particles)
defining the polynomials $W_1$ and $W_2$. 
To do this we employ Eqs.\ \eqref{i7a}, \eqref{i7bi}, and \eqref{i7ci}, which take the form:
(1)~$\Delta f=\ln(r_a+r_b+r_{ab})$; (2)~$\Delta_a f=r_a/(2r_{ab})$; (3)~$\Delta_b f=r_b/(2r_{ab})$.
The equations 1--3 fix the 12 coefficients of the polynomials $W_1$ and $W_2$ uniquely. 
The result is given in Eq.\ \eqref{il2-1-1} below.

Below we list explicit formulas for the inverse Laplacians which have been used
throughout this paper (here $a \ne b$ and $s_{ab}:=r_a+r_b+r_{ab}$):
\begin{widetext}
\begin{subequations}
\label{il}
\begin{align}
\label{il1}
\Delta^{-1}\frac{1}{r_a r_b} &:= \ln s_{ab},
\\[2ex]
\label{il2-1-1}
\Delta^{-2}\frac{1}{r_a r_b} &:= \frac{1}{36} \left(-r_a^2 + 3 r_a r_{ab} + r_{ab}^2 - 3 r_a r_b + 3 r_{ab} r_b - r_b^2 \right)
+ \frac{1}{12} \left( r_a^2 - r_{ab}^2 + r_b^2 \right)\ln s_{ab},
\\[2ex]
\Delta^{-3}\frac{1}{r_a r_b} &:=
\frac{1}{28800} \big( -63 r_a^4 + 150 r_a^3 r_{ab} + 126 r_a^2 r_{ab}^2
- 90 r_a r_{ab}^3 - 63 r_{ab}^4
- 90 r_a^3 r_b + 90 r_a^2 r_{ab} r_b + 90 r_a r_{ab}^2 r_b - 90 r_{ab}^3 r_b
\nonumber\\[1ex]&\quad
- 2 r_a^2 r_b^2
+ 90 r_a r_{ab} r_b^2 + 126 r_{ab}^2 r_b^2 - 90 r_a r_b^3 + 150 r_{ab} r_b^3
- 63 r_b^4 \big)
\nonumber\\[1ex]&\quad
+\frac{1}{960} \left( 3 r_a^4 - 6 r_a^2 r_{ab}^2 + 3 r_{ab}^4
+ 2 r_a^2 r_b^2 - 6 r_{ab}^2 r_b^2 + 3 r_b^4 \right) \ln s_{ab},
\\[2ex]
\Delta^{-1}\frac{r_a}{r_b} &:=
\frac{1}{18}\left(
-r_a^2 - 3r_ar_{ab} - r_{ab}^2 + 3r_ar_b + 3r_{ab}r_b + r_b^2\right)
+\frac{1}{6}\left(r_a^2 + r_{ab}^2 - r_b^2\right)\ln s_{ab},
\\[2ex]
\Delta^{-2}\frac{r_a}{r_b} &:=
\frac{1}{7200}\left(
-r_a^4 - 30r_a^3r_{ab} - 62r_a^2r_{ab}^2 + 90r_ar_{ab}^3 + 63r_{ab}^4 - 30r_a^3r_b + 30r_a^2r_{ab}r_b
\right.\nonumber\\[1ex]&\quad
- 90r_ar_{ab}^2r_b + 90r_{ab}^3r_b - 62r_a^2r_b^2 - 90r_ar_{ab}r_b^2 - 126r_{ab}^2r_b^2 + 90r_ar_b^3
\nonumber\\[1ex]&\quad\left.
+ 90r_{ab} r_b^3+63r_b^4\right)
+\frac{1}{240}\left(r_a^4 + 2r_a^2r_{ab}^2 - 3r_{ab}^4 + 2r_a^2r_b^2 + 6r_{ab}^2r_b^2 - 3r_b^4\right)\ln s_{ab},
\\[2ex]
\Delta^{-3}\frac{r_a}{r_b} &:= \frac{1}{2822400} \big(-37 r_a^6 - 210 r_a^5 r_ab - 951 r_a^4 r_{ab}^2
+ 1540 r_a^3 r_{ab}^3 + 2013 r_a^2 r_{ab}^4 - 1050 r_a r_{ab}^5 - 1025 r_{ab}^6 - 210 r_a^5 r_b
\nonumber\\[1ex]&\quad
+ 210 r_a^4 r_{ab} r_b - 840 r_a^3 r_{ab}^2 r_b + 840 r_a^2 r_{ab}^3 r_b + 1050 r_a r_{ab}^4 r_b - 1050 r_{ab}^5 r_b
- 37 r_a^4 r_b^2 - 420 r_a^3 r_{ab} r_b^2 - 1062 r_a^2 r_{ab}^2 r_b^2
\nonumber\\[1ex]&\quad
+ 2100 r_a r_{ab}^3 r_b^2 + 3075 r_{ab}^4 r_b^2 - 280 r_a^3 r_b^3 + 420 r_a^2 r_{ab} r_b^3 - 2100 r_a r_{ab}^2 r_b^3
+ 2800 r_{ab}^3 r_b^3 - 951 r_a^2 r_b^4 - 1050 r_a r_{ab} r_b^4
\nonumber\\[1ex]&\quad
- 3075 r_{ab}^2 r_b^4 + 1050 r_a r_b^5 + 1050 r_{ab} r_b^5 + 1025 r_b^6 \big)
+ \frac{1}{13440} \big( r_a^6 + 3 r_a^4 r_{ab}^2 - 9 r_a^2 r_{ab}^4 + 5 r_{ab}^6 + r_a^4 r_b^2 + 6 r_a^2 r_{ab}^2 r_b^2
\nonumber\\[1ex]&\quad
- 15 r_{ab}^4 r_b^2 + 3 r_a^2 r_b^4 + 15 r_{ab}^2 r_b^4 - 5 r_b^6 \big) \ln s_{ab},
\\[2ex]
\Delta^{-1}r_a r_b &:=
\frac{1}{3600}\left(63r_a^4 + 90r_a^3r_{ab} - 62r_a^2r_{ab}^2 - 30r_ar_{ab}^3 -
r_{ab}^4 +  90r_a^3r_b - 90r_a^2r_{ab}r_b+ 30r_ar_{ab}^2r_b
\right.\nonumber\\[1ex]&\quad\left.
- 30r_{ab}^3r_b - 
126r_a^2r_b^2 - 90r_ar_{ab}r_b^2 - 62r_{ab}^2r_b^2 + 90r_ar_b^3 + 
90r_{ab}r_b^3 + 63r_b^4\right)
\nonumber\\[1ex]&\quad
+\frac{1}{120}\left(-3r_a^4 + 2r_a^2r_{ab}^2 +
r_{ab}^4 + 6r_a^2r_b^2 + 2r_{ab}^2r_b^2 - 3r_b^4\right)\ln s_{ab},
\\[2ex]
\Delta^{-2}r_a r_b &:= \frac{1}{705600} \big( 531 r_a^6 + 630 r_a^5 r_{ab} - 531 r_a^4 r_{ab}^2 - 420 r_a^3 r_{ab}^3
- 531 r_a^2 r_{ab}^4 + 630 r_a r_{ab}^5 + 531 r_{ab}^6 + 630 r_a^5 r_b
\nonumber\\[1ex]&\quad
- 630 r_a^4 r_{ab} r_b - 630 r_a r_{ab}^4 r_b + 630 r_{ab}^5 r_b - 531 r_a^4 r_b^2 - 914 r_a^2 r_{ab}^2 r_b^2
- 531 r_{ab}^4 r_b^2 - 420 r_a^3 r_b^3 - 420 r_{ab}^3 r_b^3
\nonumber\\[1ex]&\quad
- 531 r_a^2 r_b^4 - 630 r_a r_{ab} r_b^4 - 531 r_{ab}^2 r_b^4 + 630 r_a r_b^5 + 630 r_{ab} r_b^5 + 531 r_b^6 \big)
\nonumber\\[1ex]&\quad
+ \frac{1}{3360} \big(-3 r_a^6 + 3 r_a^4 r_{ab}^2 + 3 r_a^2 r_{ab}^4
- 3 r_{ab}^6 + 3 r_a^4 r_b^2
+ 2 r_a^2 r_{ab}^2 r_b^2 + 3 r_{ab}^4 r_b^2 + 3 r_a^2 r_b^4 + 3 r_{ab}^2 r_b^4 - 3 r_b^6 \big) \ln s_{ab},
\\[2ex]
\Delta^{-3}r_a r_b &:= \frac{1}{1219276800} \big( 18975 r_a^8 + 18900 r_a^7 r_{ab} - 21100 r_a^6 r_{ab}^2 - 
    18900 r_a^5 r_{ab}^3 - 50550 r_a^4 r_{ab}^4 + 65100 r_a^3 r_{ab}^5 + 
    88500 r_a^2 r_{ab}^6
\nonumber\\[1ex]&\quad
- 31500 r_a r_{ab}^7 - 35825 r_{ab}^8 + 18900 r_a^7 r_b - 
    18900 r_a^6 r_{ab} r_b - 6300 r_a^5 r_{ab}^2 r_b + 6300 r_a^4 r_{ab}^3 r_b - 
    44100 r_a^3 r_{ab}^4 r_b
\nonumber\\[1ex]&\quad
+ 44100 r_a^2 r_{ab}^5 r_b + 31500 r_a r_{ab}^6 r_b - 
    31500 r_{ab}^7 r_b - 18204 r_a^6 r_b^2 + 3780 r_a^5 r_{ab} r_b^2 - 
    30804 r_a^4 r_{ab}^2 r_b^2 - 7560 r_a^3 r_{ab}^3 r_b^2
\nonumber\\[1ex]&\quad
- 39492 r_a^2 r_{ab}^4 r_b^2 + 44100 r_a r_{ab}^5 r_b^2 + 88500 r_{ab}^6 r_b^2 - 
    8820 r_a^5 r_b^3 - 3780 r_a^4 r_{ab} r_b^3 - 840 r_a^3 r_{ab}^2 r_b^3 - 
    7560 r_a^2 r_{ab}^3 r_b^3
\nonumber\\[1ex]&\quad
- 44100 r_a r_{ab}^4 r_b^3 + 65100 r_{ab}^5 r_b^3 - 
    1542 r_a^4 r_b^4 - 3780 r_a^3 r_{ab} r_b^4 - 30804 r_a^2 r_{ab}^2 r_b^4 + 
    6300 r_a r_{ab}^3 r_b^4 - 50550 r_{ab}^4 r_b^4
\nonumber\\[1ex]&\quad
- 8820 r_a^3 r_b^5 + 
    3780 r_a^2 r_{ab} r_b^5 - 6300 r_a r_{ab}^2 r_b^5 - 18900 r_{ab}^3 r_b^5 - 
    18204 r_a^2 r_b^6
- 18900 r_a r_{ab} r_b^6 - 21100 r_{ab}^2 r_b^6
\nonumber\\[1ex]&\quad
+ 18900 r_a r_b^7 + 18900 r_{ab} r_b^7 + 18975 r_b^8 \big)
+ \frac{1}{967680} \big( -15 r_a^8 + 20 r_a^6 r_{ab}^2 + 30 r_a^4 r_{ab}^4 - 60 r_a^2 r_{ab}^6 + 
    25 r_{ab}^8 + 12 r_a^6 r_b^2
\nonumber\\[1ex]&\quad
+ 12 r_a^4 r_{ab}^2 r_b^2
+ 36 r_a^2 r_{ab}^4 r_b^2 - 60 r_{ab}^6 r_b^2 + 6 r_a^4 r_b^4 + 
    12 r_a^2 r_{ab}^2 r_b^4 + 30 r_{ab}^4 r_b^4 + 12 r_a^2 r_b^6 + 
    20 r_{ab}^2 r_b^6 - 15 r_b^8 \big) \ln s_{ab},
\\[2ex]
\Delta^{-1}\frac{r_a^3}{r_b} &:= \frac{1}{1200} \big( -63r_a^4 - 90r_a^3r_{ab} - 2r_a^2r_{ab}^2 - 90r_ar_{ab}^3 -
63r_{ab}^4 +  150r_a^3r_b + 90r_a^2r_{ab}r_b+ 90r_ar_{ab}^2r_b
+150r_{ab}^3r_b
\nonumber\\[1ex]&\quad
+ 126r_a^2r_b^2 + 90r_ar_{ab}r_b^2 
+ 126r_{ab}^2r_b^2 - 90r_ar_b^3 -  90r_{ab}r_b^3 - 63r_b^4 \big)
\nonumber\\[1ex]&\quad
+\frac{1}{40}\big(3r_a^4+2r_a^2r_{ab}^2+3r_{ab}^4-6r_a^2 r_b^2-6r_{ab}^2 r_b^2 +3r_b^4 \big)\ln s_{ab}.
\end{align}
\end{subequations}
\end{widetext}

\section{Explicit results for the field functions}
\label{functions}

In this appendix we give the explicit formulas for the PN approximate solutions
of both the constraint \eqref{CE} and the field \eqref{FE} equations
which can be found in the literature or were computed by us for the first time.
The different inverse Laplacians needed to compute field functions
presented below are given in Appendix \ref{invLap}.

\subsection{Potentials $\phi_{(n)}$}

The solutions of Eqs.\ \eqref{delta-fii-fiv} for the potentials $\phi_{(2)}$ and $\phi_{(4)}$ are known in $d$ dimensions.
They can be obtained by means of Eqs.\ \eqref{eqPoisson} and \eqref{di06-1}
using the result $(\phi_{(2)})_\textrm{reg}(\xa)=\kappa\sum_{b\ne a}m_b r_{ab}^{2-d}$.
They read
\begin{subequations}
\begin{align}
\label{phi2-d}
\phi_{(2)} &= \kappa\sum_a m_a r_a^{2-d},
\\[2ex]
\phi_{(4)} &= -\frac{1}{2}\SivI + \frac{d-2}{4(d-1)}\SivII,
\\[1ex]
\SivI &= -\kappa \sum_a \frac{\papa}{m_a} r_a^{2-d},
\\[1ex]
\SivI &= -\kappa^2  \sum_a\sum_{b\ne a} m_a m_b r_{ab}^{2-d}r_a^{2-d},
\end{align}
\end{subequations}
where the coefficient $\kappa$ is defined in Eq.\ \eqref{kappa}.
These solutions are valid also for general $n$-body, i.e., not only for two-body, point-mass systems.

The potential $\phi_{(6)}$ is split into two parts, $\phi_{(6)}=\phi_{(6)1}+\phi_{(6)2}$,
where $\phi_{(6)1}$ is the solution of the Poisson equation \eqref{lapphi61}
and $\phi_{(6)2}$ fulfills the equation \eqref{lapphi62}.
The solutions to these equations are fully known only for two-body point-mass systems and in $d=3$ dimensions
(they can be found in an implicit form e.g.\ in Refs.\ \cite{OOKH74,OKH75} and \cite{Schafer85}).
They can be symbolically written as
\begin{widetext}
\begin{subequations}
\begin{align}
\label{phi61}
\phi_{(6)1} &= \Delta^{-1}\left( \sum_a\left(
-\frac{1}{64}\phi_{(2)}^2+\frac{1}{8}\phi_{(4)}
+\frac{5}{16}\phi_{(2)}\frac{{\bf p}_a^2}{m_a^2}
+\frac{1}{8}\frac{({\bf p}_a^2)^2}{m_a^4}\right)m_a\delta_a
- \big({\pitiii ij}\big)^2 \right),
\\[2ex]
\label{phi62}
\phi_{(6)2} &= \frac{1}{2}\,\Delta^{-1}\Big(\phi_{(2),ij}{\hTTiv ij}\Big)
= \frac{1}{8\pi} \sum_a m_a \, \Delta^{-1}\bigg(\left(\frac{1}{r_a}\right)_{\!\!,ij}{\hTTiv ij}\bigg).
\end{align}
\end{subequations}
A more explicit form of the function $\phi_{(6)1}$ can be written as\footnote{
In Appendix \ref{functions} we employ the following notation:
$\nabla=(\partial_i)$, $\nabla_a=(\partial_{ai})$,
$\pan=p_{ai}\partial_i$, $\panb=p_{ai}\partial_{bi}$, $\nanb=\partial_{ai}\partial_{bi}$.}
\begin{align}
\phi_{(6)1} &=-\frac{1}{32\pi}\sum_a\frac{\papap^2}{m_a^3 r_a}
-\frac{1}{\left(16\pi\right)^2}\sum_a\sum_{b\ne a}\frac{m_b\papa}{m_a r_{ab}}
\left(\frac{5}{r_a}+\frac{1}{r_b}\right)
+\frac{1}{\left(16\pi\right)^3}\sum_a\sum_{b\ne a}\frac{m_a^2 m_b}{r_{ab}^2}
\left(\frac{1}{r_a}+\frac{1}{r_b}\right)
\nonumber\\[1ex]&\quad
-\frac{9}{4}\frac{1}{\left(16\pi\right)^2} \sum_a \frac{2\papa-\napa^2}{r_a^2}
+ \frac{1}{\left(16\pi\right)^2}\sum_a\sum_{b\ne a}\left\{
-\frac{1}{4}\pana\pbnb\nanb^2\left(\Delta^{-1}r_a r_b\right)
\right.\nonumber\\[1ex]&\quad
+\left[-8\papb\nanb+3\pana\pbnb-8\panb\pbna\right]
\left(\Delta^{-1}\frac{1}{r_a r_b}\right)
\nonumber\\[1ex]&\quad\left.
+4\pana\pbna\nanb\left(\Delta^{-1}\frac{r_a}{r_b}\right)\right\},
\end{align}
where the inverse Laplacians $\Delta^{-1}(r_a\,r_b)$, $\Delta^{-1}(r_a^{-1}\,r_b^{-1})$, and $\Delta^{-1}(r_a\,r_b^{-1})$
can be found in Eqs.\ \eqref{il}.
The Poisson integral needed to calculate the function $\phi_{(6)2}$ reads (here $b\ne a$)
\begin{multline}
\label{phi62a}
\Delta^{-1}\bigg(\left(\frac{1}{r_a}\right)_{\!\!,ij}{\hTTiv ij}\bigg)
= \frac{1}{32\left(16\pi\right)^2}
\frac{m_a m_b}{r_a^3 r_b r_{ab}^3}\left(
3 r_a^4 - 12 r_a^3 r_{ab} + 18 r_a^2 r_{ab}^2 - 12 r_a r_{ab}^3 + 3 r_{ab}^4
\right.\\[1ex]\left.
+ 28 r_a^3 r_b
- 12 r_a^2 r_{ab} r_b - 12 r_a r_{ab}^2 r_b + 28 r_{ab}^3 r_b
- 30 r_a^2 r_b^2 + 60 r_a r_{ab} r_b^2
- 30 r_{ab}^2 r_b^2 - 36 r_a r_b^3
- 36 r_{ab} r_b^3 + 35 r_b^4 \right)
\\[2ex]
+ \frac{3}{64\pi} \frac{\papa-3\uapa^2}{m_a r_a^2}
+ \frac{1}{32\pi}\frac{1}{m_b}\left\{
2\pbnb^2\frac{r_b}{r_a}
-4\frac{\pbpb}{r_a r_b}
+\left[2\pbpb\nnb^2
-4\pbn\pbnb\nnb
\right.\right.\\[2ex]\left.\left.
-4\pbn\pbnb\nnb\right]
\left(\Delta^{-1}\frac{r_b}{r_a}\right)
+8\pbn^2\left(\Delta^{-1}\frac{1}{r_a r_b}\right)
+\frac{1}{6}\pbnb^2\nnb^2\left(\Delta^{-1}\frac{r_b^3}{r_a}\right)\right\},
\end{multline}
where the inverse Laplacians $\Delta^{-1}(r_b\,r_a^{-1})$, $\Delta^{-1}(r_a^{-1}\,r_b^{-1})$, and $\Delta^{-1}(r_b^3\,r_a^{-1})$
are given in Eqs.\ \eqref{il}.
\end{widetext}

\subsection{Longitudinal field momenta ${\pit ij}_{(n)}$}

Equations \eqref{grad-pitiii-ixb} fulfilled by the longitudinal field momenta ${\pit ij}_{(n)}$
can be written in the form
\be
\label{eqpit}
{\pit ij}_{(n),j} = S_{(n)}^i,
\ee
where $S_{(n)}^i$ stands for the source term.
Making use of the decomposition \eqref{pitbyv} one rewrites
Eq.\ \eqref{eqpit} in terms of the vectorial function $V^i_{(n)}$,
\be
\label{eqV}
\Deltad V^i_{(n)} + \frac{d-2}{d}\partial_{ij} V^j_{(n)} = S_{(n)}^i.
\ee
It is not difficult to find the formal solution of Eq.\ \eqref{eqV}
in terms of the first and the second inverse Laplacian of the source $S_{(n)}^i$.
It reads
\be
\label{solV}
V^i_{(n)} = \Deltad^{-1}S_{(n)}^i
- \frac{d-2}{2(d-1)}\partial_{ij}\Deltad^{-2}S_{(n)}^j.
\ee
Let us now assume that the source term in Eq.\ \eqref{eqpit}
has the form of a divergence of a symmetric and trace-free quantity,
\be
S_{(n)}^i = \partial_j T_{(n)}^{ij},
\ee
where $T_{(n)}^{ij}=T_{(n)}^{ji}$ and $T_{(n)}^{ii}=0$.
Then, making use of Eq.\ \eqref{pitbyv} and the definition \eqref{defTT} of the TT operator in $d$ dimensions,
one can show that
\be
\label{pitbyTT}
{\pit ij}_{(n)} = T_{(n)}^{ij} - \big(T_{(n)}^{ij}\big)^\mathrm{TT}.
\ee

Using the source term from Eq.\ \eqref{grad-pitiii} we have obtained
the explicit solution \eqref{solV} for the leading-order function ${\Viii i}$
in $d$ dimensions,
\begin{multline}
{\Viii i} = \frac{\kappa}{8(d-1)}
\sum_a \Big((3d-2)p_{ai}
\\[1ex]
+ (d-2)^2 \uapa n_a^i\Big) r_a^{2-d}.
\end{multline}
This solution is also valid for $n$-body point-mass systems.
The leading-order longitudinal field momentum $\pitiii ij$
can be constructed from the function $\Viii i$ by means of the formula \eqref{pitbyv}.
Let us also quote the following useful form of the field momentum ${\pitiii ij}$ in $d=3$ dimensions,
\begin{align}
{\pitiii ij} &= \frac{1}{16\pi}\sum_a p_{ak} \bigg\{
2\bigg[\delta_{ik}\left(\frac{1}{r_a}\right)_{\!\!,j}
+\delta_{jk}\left(\frac{1}{r_a}\right)_{\!\!,i}\bigg]
\nonumber\\[1ex]&\qquad
- \delta_{ij}\left(\frac{1}{r_a}\right)_{\!\!,k}
- \frac{1}{2}r_{a,ijk}
\bigg\}.
\end{align}

The next-to-leading-order function $\Vv i$ has been calculated
in $d=3$ dimensions and for two-point-mass systems only. It reads
\begin{align}
{\Vv i} &= -\frac{3}{16(16\pi)^2} \sum_a \frac{m_a}{r_a^2} \big(9 p_{ai} + 2 \napa n_a^i\big)
\nonumber\\[1ex]&\quad
+ \frac{1}{(16\pi)^2} \sum_a\sum_{b\ne a} m_b \bigg\{
\Big[4\partial_{ai}\pan
\nonumber\\[1ex]&\quad
- \frac{3}{2}\partial_i\pana
+ 4 p_{ai}\nna\Big] \left(\Delta^{-1}\frac{1}{r_a r_b}\right)
\nonumber\\[1ex]&\quad
- 2 \partial_i \pan \nna \left(\Delta^{-2}\frac{1}{r_a r_b}\right)
\nonumber\\[1ex]&\quad
- \partial_{ai} \pana \nna \left(\Delta^{-1}\frac{r_a}{r_b}\right)
\nonumber\\[1ex]&\quad
+ \frac{1}{4} \partial_i \pana \nna^2 \left(\Delta^{-2}\frac{r_a}{r_b}\right) \bigg\},
\end{align}
where the inverse Laplacians $\Delta^{-1}(r_a^{-1}\,r_b^{-1})$, $\Delta^{-2}(r_a^{-1}\,r_b^{-1})$,
$\Delta^{-1}(r_a\,r_b^{-1})$, and $\Delta^{-2}(r_a\,r_b^{-1})$ are given in Eqs.\ \eqref{il}.
The field momentum ${\pitv ij}$ can be constructed from the vectorial function $\Vv i$ by means of the formula \eqref{pitbyv}.

\subsection{${\hTTiv ij}$ and ${\hTTvi ij}$ related terms}

The leading-order TT part ${\hTTiv ij}$ of the metric is the solution of Eq.\ \eqref{feqhTT4}.
It is the sum
\be
{\hTTiv ij} = {\hTTivpo ij} + {\hTTivpii ij},
\ee
where ${\hTTivpo ij}$ does not depend on particles' momenta and ${\hTTivpii ij}$ is quadratic in momenta.
The quadratic-in-momenta part of ${\hTTivpii ij}$ can be computed in $d$ dimensions
and for general $n$-body systems. It reads
\begin{align}
{\hTTivpii ij} &= \frac{\kappa(d-2)}{8(d-1)}
\sum_a \frac{1}{m_a} \Big\{\big[\papa - (d + 2)\napa^2\big]\delta_{ij}
\nonumber\\[1ex]&\kern-3ex
+ \big[d (d - 2) \napa^2 - (d + 2) \papa\big] n_a^i n_a^j
\nonumber\\[1ex]&\kern-3ex
+ 2 d \napa \big(n_a^i p_{aj} + n_a^j p_{ai}\big) + 2p_{ai}p_{aj}\Big\} r_a^{2-d}.
\end{align}
The momentum independent part $\hTTivpo ij$ is known only in $d=3$ dimensions
and for two-body systems.
Its fully explicit form reads (here $s_{ab}:=r_{a}+r_{b}+r_{ab}$ for $a\ne b$; see, e.g., Ref.\ \cite{JS98})
\begin{widetext}
\begin{align}
\label{h4TTp0}
{\hTTivpo ij} &= \frac{1}{8}\frac{1}{\left(16\pi\right)^2}
\sum\limits_a\sum\limits_{b\ne a} m_a m_b
\Bigg\{ - \frac{32}{s_{ab}}
\left(\frac{1}{r_{ab}} + \frac{1}{s_{ab}}\right) n_{ab}^i n_{ab}^j
+ 2 \left(\frac{r_a + r_b}{r_{ab}^3} + \frac{12}{s_{ab}^2}\right) n_a^i n_{b}^j
\nonumber\\[1ex]&\qquad
+ 16 \left( \frac{2}{s_{ab}^2} - \frac{1}{r_{ab}^2} \right)
\left( n_a^i n_{ab}^j +  n_a^j n_{ab}^i \right)
+ \left[ \frac{5}{r_{ab}r_a}
- \frac{1}{r_{ab}^3}\left(\frac{r_b^2}{r_a}+3r_a\right)
- \frac{8}{s_{ab}}
\left(\frac{1}{r_a} + \frac{1}{s_{ab}}\right) \right] n_a^i n_{a}^j
\nonumber\\[1ex]&\qquad
+ \left[ 5 \frac{r_a}{r_{ab}^3} \left( \frac{r_a}{r_b} - 1 \right)
- \frac{17}{r_{ab}r_a} + \frac{4}{r_a r_b}
+ \frac{8}{s_{ab}} \left( \frac{1}{r_a} + \frac{4}{r_{ab}} \right) \right]
\delta_{ij} \Bigg\}.
\end{align}

The next-to-leading-order conservative TT part ${\hTTvi ij}$ of the metric function ${\hTT ij}$
is very complicated. It fulfills equation \eqref{feqhTT6}.
The piece of ${\hTTvi ij}$ which diverges linearly at infinity in $d=3$ dimensions equals $\Deltad^{-1}{\hTTivddot ij}$.
The part of this inverse Laplacian which corresponds to the function ${\hTTivpii ij}$ can be computed in $d$ dimensions
(and for general $n$-point-mass systems). It reads
\begin{align}
\Deltad^{-1}{\hTTivpiiddot ij} &=
\frac{\kappa}{48(4-d)(d-1)}\frac{\partial^2}{\partial t^2}
\sum\limits_a \frac{r_a^{4-d}}{m_a} \left\{
\Big[(16-d^2)\napa^2-(14-d)\papa\Big]\delta_{ij}
+ 2(7d-8) p_{ai} p_{aj}
\right.\nonumber\\[1ex]&\left.\qquad
+ (4-d) \Big[(4+d)\papa -(d-2)^2\napa^2\Big] n_{a}^i n_{a}^j
- 2 (4-d) (2d-1) \napa (n_a^i p_{aj} + n_a^j p_{ai}) \right\}.
\end{align}
The part of the inverse Laplacian $\Deltad^{-1}{\hTTivddot ij}$ corresponding to the function ${\hTTivpo ij}$
can be computed only in 3 dimensions and for two-point-mass systems.
It reads
\begin{align}
\Delta^{-1}{\hTTivpoddot ij} &= -\frac{1}{\left(16\pi\right)^2} \frac{\partial^2}{\partial t^2}
\sum\limits_a\sum\limits_{b\ne a} m_a m_b \left\{
\frac{1}{8}\left[ \frac{1}{r_{ab}^3}\left(r_a^3-r_a^2 r_b\right) + \frac{5r_a}{r_{ab}} \right]\delta_{ij}
+ \frac{1}{12}\left( \partial_i\partial_{bj}+\partial_j\partial_{bi}-\partial_i\partial_{aj}-\partial_j\partial_{ai} \right)\frac{r_a^3}{r_{ab}}
\right.\nonumber\\[1ex]&\left.\qquad
+\frac{1}{144r_{ab}^3}\partial_i\partial_j\left[r_a^5-r_a^3(r_b^2+17r_{ab}^2)\right]
- \frac{1}{2}\delta_{ij}\left(\Delta^{-1}\frac{1}{r_a r_b}\right)
+ \frac{1}{2}\left(8\partial_{ai}\partial_{bj}-\partial_i\partial_j\right)
\left(\Delta^{-2}\frac{1}{r_a r_b}\right) \right\}.
\end{align}
The inverse Laplacians $\Delta^{-1}(r_a^{-1}\,r_b^{-1})$ and $\Delta^{-2}(r_a^{-1}\,r_b^{-1})$ can be found in Eqs.\ \eqref{il}.

Other parts of the function ${\hTTvi ij}$ are determined by $\Deltad^{-1}{\CTTvi ij}$
and they enter the density $h_{(12)}^{2,1}$ from Eq.\ \eqref{h12-21}.
We list below formulas by which, together with the inverse Laplacians
enumerated in Appendix \ref{invLap3}, one can compute the density $h_{(12)}^{2,1}$ explicitly in 3 dimensions.
These formulas read
\begin{subequations}
\label{h12-21a}
\begin{align}
\big(\fii{\pitiii ij}\big)^\mathrm{TT} &= \frac{4m_2}{(16\pi)^2} \Big(
-2(p_{1i}\partial_{1j}+p_{1j}\partial_{1i})\frac{1}{r_1r_2}
+\frac{1}{2}\partial_{1i}\partial_{1j}\pinai\frac{r_1}{r_2} \Big)^\mathrm{TT}
+ \big(1\leftrightarrow 2\big),
\\[2ex]
\Big(\fii\Delta{\hTTiv ij}\Big)^\mathrm{TT} &= \frac{4m_1^2m_2}{(16\pi)^3}
\bigg(\frac{3}{2r_{12}}\partial_{1i}\partial_{2j}\frac{1}{r_1r_2}
+ \frac{1}{2r_{12}^3}\Big(13\partial_{1i}\partial_{1j}\frac{r_1}{r_2}+15\partial_{2i}\partial_{2j}\frac{r_2}{r_1}\Big)
+ \frac{35}{8} \partial_{1i}\partial_{1j}\frac{1}{r_1^2r_2} \bigg)^\mathrm{TT}
\nonumber\\[1ex]&\quad
+ \frac{4}{(16\pi)^2}\frac{m_1}{m_2} \bigg( \Big(2\piipii\partial_{2i}\partial_{2j}
- 4\big(p_{2j}\partial_{2i} + p_{2i}\partial_{2j}\big)\piinaii \Big)\frac{1}{r_1r_2}
+ \partial_{2i}\partial_{2j}\piinaii^2\frac{r_2}{r_1} \bigg)^\mathrm{TT}
\nonumber\\[1ex]&\quad
+ \frac{1}{12\pi} \frac{m_1}{m_2} \bigg(\frac{\piipii\delta_{ij} - 3p_{2i}p_{2j}}{r_{12}}\delta_2\bigg)^\mathrm{TT}
+ \big(1\leftrightarrow 2\big),
\\[2ex]
{\BTTvi ij} &= \frac{m_1^2 m_2}{(16\pi)^3} \bigg( \frac{3}{r_{12}}\partial_{1i}\partial_{2j}\frac{1}{r_1r_2}
- \frac{35}{8} \partial_{1i}\partial_{1j}\frac{1}{r_1^2r_2} \bigg)^\mathrm{TT}
- \frac{4}{(16\pi)^2} \frac{m_2}{m_1} \pipi \bigg(\partial_{1j}\partial_{2i}\frac{1}{r_1r_2}\bigg)^\mathrm{TT}
\nonumber\\[1ex]&\quad
+ \frac{1}{(16\pi)^2} \bigg(
\Biglb(8\pipii\partial_{1i}\partial_{2j} + 16 p_{1i}\Big(\frac{1}{3}\partial_{1j}\piinaii - \partial_{2j}\piinai\Big)
- 8 p_{1i}p_{2j} \nainaii\Bigrb)\frac{1}{r_1r_2}
\nonumber\\[1ex]&\quad
+ \Biglb( \partial_{1i}\pinai\Big(\partial_{2j}\piinai-\frac{2}{3}\partial_{1j}\piinaii\Big)
+ p_{2j}\partial_{1i}\pinai\nainaii \Bigrb) \frac{r_1}{r_2}
\nonumber\\[1ex]&\quad
+ \Biglb( p_{1i}\piinaii\Big(3\partial_{2j}\nainaii-\frac{2}{3}\partial_{1j}\Delta_2\Big)
- \partial_{1i}\partial_{2j}\pinaii\piinaii \Bigrb) \frac{r_2}{r_1}
\nonumber\\[1ex]&\quad
+ \frac{1}{4}\partial_{1i}\pinai\piinaii\Big(\frac{1}{3}\partial_{1j}\Delta_2-\partial_{2j}\nainaii\Big) r_1r_2
\bigg)^\mathrm{TT}
\nonumber\\[1ex]&\quad
+ \bigg(\frac{5}{32\pi}\frac{m_2}{m_1}\frac{p_{1i}p_{1j}}{r_{12}}\delta_1 + \frac{1}{4}\frac{\pipi p_{1i}p_{1j}}{m_1^3}\delta_1
\bigg)^\mathrm{TT}+ \big(1\leftrightarrow 2\big).
\end{align}
\end{subequations}
\end{widetext}
To compute the density $h_{(12)}^{2,1}$ explicitly one has to employ Eqs.\ \eqref{h12-21a}
together with the following inverse Laplacians,
which can be found in Eqs.\ \eqref{i06} and \eqref{il}:
\begin{subequations}
\begin{align*}
&\Delta^{-1}\da, \quad \Delta^{-2}\da, \quad \Delta^{-3}\da,
\\[1ex]
&\Delta^{-1}\frac{1}{r_1r_2},\quad \Delta^{-2}\frac{1}{r_1r_2},\quad \Delta^{-3}\frac{1}{r_1r_2},
\\[1ex]
&\Delta^{-1}\frac{r_1}{r_2},\quad \Delta^{-2}\frac{r_1}{r_2},\quad \Delta^{-3}\frac{r_1}{r_2},
\\[1ex]
&\Delta^{-1}r_1r_2,\quad \Delta^{-2}r_1r_2,\quad \Delta^{-3}r_1r_2.
\end{align*}
\end{subequations}
Let us also note that after combining ${\BTTvi ij}$ and $\big(\fii\Delta{\hTTiv ij}\big)^\mathrm{TT}$
into $\big({\Bvi ij}+(1/4)\,\fii\Delta{\hTTiv ij}\big)^\mathrm{TT}$,
see Eq.\ \eqref{h12-211}, there is no need to compute any inverse Laplacians of $(r_1^2r_2)^{-1}$.
The inverse Laplacian $\Delta^{-1}\big({\ggradfii kl}{\hTTiv kl}\big)$ also needed to compute $h_{(12)}^{2,1}$ explicitly,
can be inferred from Eqs.\ \eqref{phi62} and \eqref{phi62a}.

\subsection{Local $d$-dimensional UV analysis}

In Sec.\ 3 of Ref.\ \cite{DJS2001} a method of local analysis of UV divergences in $d$ dimensions was proposed.
This method was essentially used in \cite{DJS2001} for studying the local behavior
of only one specific function, namely, the momentum-independent part $\hTTivpo ij$ of the field function ${\hTTiv ij}$.
At the 4PN level we have had to use it in many more cases;
therefore, we present this method here in more detail.

The problem is to find the local behavior in $d$ dimensions and, say, around $\x=\x_1$, of the solution
of equation of the type (for some cases we have to consider the Poisson equation without TT projection)
\be
\label{localUV1}
(\Deltad f)^\mathrm{TT} = g^\mathrm{TT}.
\ee
We assume that the full analytic solution of this equation is not available.
The source function $g$ depends on the field point $\x$ only through $\x-\x_1$ and $\x-\x_2$;
therefore, one defines the auxiliary function $\bar{g}$ as follows:
\be
\label{localUV2}
\bar{g}(r_1) := g(\x-\x_1,\x-\x_2) = g(r_1\n_1,r_1\n_1+r_{12}\n_{12}),
\ee
where we have used $\x-\x_1=r_1\n_1$ and $\x-\x_2=r_1\n_1+r_{12}\n_{12}$.
The method of Ref.\ \cite{DJS2001} relies on expansion of the source function $\bar{g}$
around $r_1=0$ and applying the operator $\Deltad^{-1}$ to each term of the expansion.
This expansion has the form
\be
\label{localUV3}
\bar{g}(r_1) = \sum_{k=-m}^\infty \frac{\bar{g}^{(k)}(\n_1)}{k!}r_1^k,
\ee
where $m\ge0$ is some nonnegative integer.
With the expansion \eqref{localUV3} of the source term one can relate
the following expansion of the solution to Eq.\ \eqref{localUV1} near $\x=\x_1$:
\be
\label{localUV4}
f_\textrm{nonhom}^\mathrm{TT}(\x) = \sum_{k=-m}^\infty \frac{1}{k!} \Deltad^{-1}\big(\bar{g}^{(k)}(\n_1)r_1^k\big)^\mathrm{TT}.
\ee
The above formula does not contain all terms of the expansion of the physically acceptable solution of Eq.\ \eqref{localUV1}.
The missing terms are solutions of homogeneous Poisson equation. We have devised a method to compute these terms.

Let us write the formal solution of Eq.\ \eqref{localUV1}
in the form of the integral
\be
\label{localUV5}
f^\mathrm{TT}(\x) = -\kappa \int\md^dx'\, g(\x')\big(|\x-\x'|^{2-d}\big)^\mathrm{TT}.
\ee
The crucial element of the method is expansion of the kernel of the integral \eqref{localUV5} around $\x=\x_1$.
To do this one introduces the auxiliary function
\begin{align}
\label{localUV6a}
K(r_1) &= |\x-\x'|^{2-d} = |(\x - \x_1) - (\x' - \x_1)|^{2-d}
\nonumber\\[1ex]
&= |r_1\n_1 - r'_1\n'_1|^{2-d},
\end{align}
and expands it around $r_1=0$:
\be
\label{localUV6}
K(r_1) = \sum_{\ell=0}^\infty \frac{K^{(\ell)}(\n_1;\n'_1,r'_1)}{\ell!} r_1^\ell.
\ee
One then substitutes the expansion \eqref{localUV6} into the integral \eqref{localUV5}
and integrates the sum term by term. Consequently one gets the expansion of the form
\be
\label{localUV7}
f_\textrm{hom}^\mathrm{TT}(\x) = -\kappa \sum_{\ell=0}^\infty \frac{1}{\ell!}
 \int\md^dx'\, g(\x') \big(K^{(\ell)}(\n_1;\n'_1,r'_1)r_1^\ell\big)^\mathrm{TT}.
\ee

As an example of applying the above formula let us compute the leading-order term
in the expansion of the momentum-independent part $\hTTivpo ij$ of the function $\hTTiv ij$ near $\x=\x_1$.
Using Eq.\ \eqref{h4TTp0} one easily checks that the function $\hTTivpo ij$ is finite for $\x=\x_1$ in $d=3$ dimensions.
This finite value cannot be obtained in $d$ dimensions from the expansion \eqref{localUV4},
but it follows from Eq.\ \eqref{localUV7} (with $\ell=0$).
After making use of Eq.\ \eqref{Siv}
[which is the source term for the function $\hTTiv ij$, see Eq.\ \eqref{feqhTT4}]
one gets
\begin{multline}
\label{localUV8}
{\hTTivpo ij}(\x=\x_1) = \frac{(d + 1)d(d - 2)^3\kappa^2}{16 (d - 1)^3 (4 - d)} m_1 m_2
\\[1ex]
\times (\delta_{ij} - d n_{12}^i n_{12}^j) r_{12}^{4 - 2 d}.
\end{multline}

\end{document}